\documentclass[reprint,amsmath,amssymb,aps,]{revtex4-2}

\usepackage{graphicx}
\usepackage{dcolumn}
\usepackage{bm}
\usepackage{color} % color
\usepackage{amsmath}
\usepackage{mathrsfs}
\usepackage{hyperref} %---link
\hypersetup{
	colorlinks=true,
	linkcolor=blue,
    filecolor=blue,      
    citecolor=blue,
	urlcolor=blue  %---weblink
}

\DeclareUnicodeCharacter{2009}{\,}

\usepackage{placeins}
\begin{document}

\title{Microscopic Origin of Superradiant Biphoton Emission in Atomic Ensembles}

\author{Zi-Yu Liu,$^{1,2\dagger}$ Jiun-Shiuan Shiu,$^{1,2}$ Wei-Lin Chen,$^{1}$ and Yong-Fan Chen$^{1,2}$}

\email{yfchen@mail.ncku.edu.tw}
\email{$\textcolor{black}{^{\dagger\;}}$liuziyu.ncku@gmail.com}

\affiliation{$^1$Department of Physics, National Cheng Kung University, Tainan 70101, Taiwan \\
$^2$Center for Quantum Frontiers of Research $\&$ Technology, Tainan 70101, Taiwan}

\date{February 11, 2026}

%%%%%%%%%%%%%%%%%%%%%%%%%%%%%%%%%%%%%%%%%%%%%%%%%%%%%%%%%%%%%%%%%%%%%%%%%%%%%%%%%%%%%%%%%%%%%%%%%%%%%
%%%%%%%%%%%%%%%%%%%%%%%%%%%%%%%%%%%%%%%%%%%%%%%%%%%%%%%%%%%%%%%%%%%%%%%%%%%%%%%%%%%%%%%%%%%%%%%%%%%%%

\begin{abstract}

Superradiant biphoton emission from atomic ensembles provides a powerful route to generating highly correlated quantum light, yet its microscopic physical origin has remained incompletely understood. In particular, it is often unclear how collective enhancement, spontaneous emission, and vacuum fluctuations jointly give rise to both paired biphoton generation and unavoidable unpaired background within a single, self-consistent framework. Here we present a fully quantum microscopic theory within a unified Heisenberg--Langevin--Maxwell framework that explicitly incorporates dissipation and quantum noise, thereby revealing the microscopic origin of superradiant biphoton emission in atomic ensembles. The theory provides a consistent description of parametric gain and unpaired noise within the same open-quantum-system framework and applies to both Doppler-free cold atomic ensembles and Doppler-broadened warm vapors. In the high-optical-depth regime, the coupled propagation equations admit analytical solutions, under which the biphoton dynamics rigorously reduce to an effective collective two-level emission process. Within this limit, the biphoton correlation time and spectral properties are shown to obey closed-form scaling relations governed by optical depth and excited-state decoherence. Our results establish a unified microscopic picture of superradiant biphoton generation and clarify the fundamental role of vacuum fluctuations and dissipation in setting the brightness, pairing efficiency, and temporal structure of atomic biphoton sources, with direct relevance to quantum networking and atomic quantum interfaces.

\end{abstract}

%%%%%%%%%%%%%%%%%%%%%%%%%%%%%%%%%%%%%%%%%%%%%%%%%%%%%%%%%%%%%%%%%%%%%%%%%%%%%%%%%%%%%%%%%%%%%%%%%%%%%
%%%%%%%%%%%%%%%%%%%%%%%%%%%%%%%%%%%%%%%%%%%%%%%%%%%%%%%%%%%%%%%%%%%%%%%%%%%%%%%%%%%%%%%%%%%%%%%%%%%%%

\maketitle

%%%%%%%%%%%%%%%%%%%%%%%%%%%%%%%%%%%%%%%%%%%%%%%%%%%%%%%%%%%%%%%%%%%%%%%%%%%%%%%%%%%%%%%%%%%%%%%%%%%%%
%%%%%%%%%%%%%%%%%%%%%%%%%%%%%%%%%%%%%%%%%%%%%%%%%%%%%%%%%%%%%%%%%%%%%%%%%%%%%%%%%%%%%%%%%%%%%%%%%%%%%

\newcommand{\figone}{
    \begin{figure}[t]
    \centering
    \includegraphics[width = 8.7 cm]{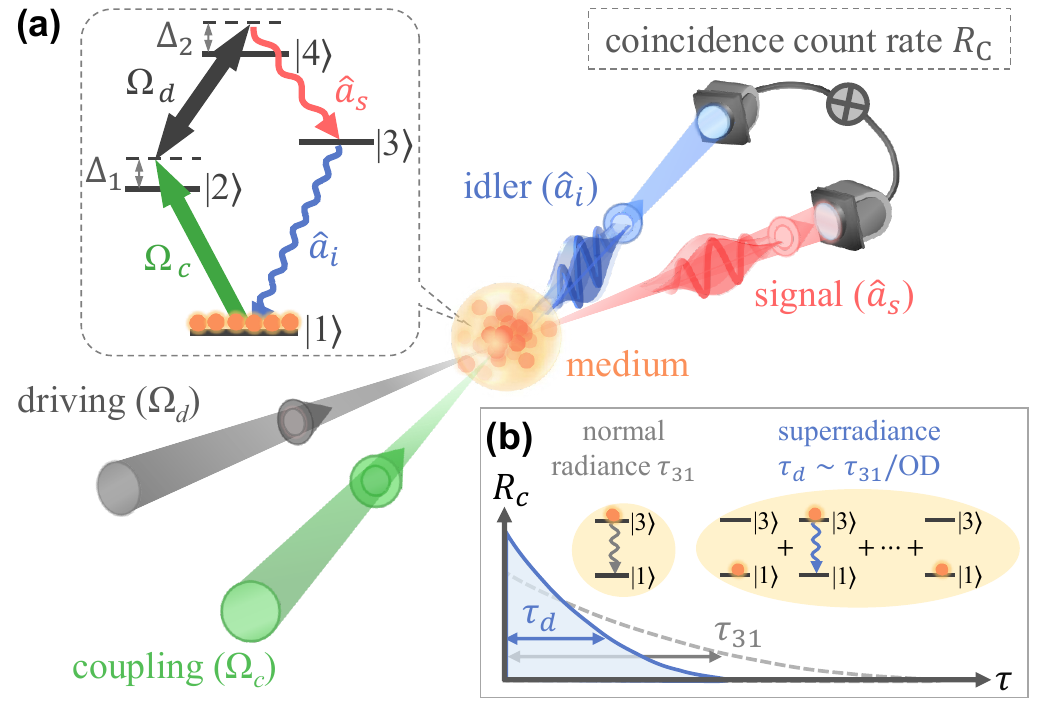}
    \caption{    	
Schematic illustration of biphoton generation and superradiant emission in a diamond-type atomic ensemble. (a) Four-level diamond-type configuration for biphoton generation. Co-propagating coupling ($\Omega_c$) and driving ($\Omega_d$) fields sequentially excite atoms via the $|1\rangle \rightarrow |2\rangle \rightarrow |4\rangle$ transitions. Vacuum-fluctuation–seeded spontaneous emission through the $|4\rangle \rightarrow |3\rangle \rightarrow |1\rangle$ cascade is isotropic, while directional temporal correlations between the signal ($\hat{a}^{\dagger}_s$) and idler ($\hat{a}^{\dagger}_i$) photons emerge from collectively enhanced, phase-matched four-wave mixing in the ensemble, as characterized by the coincidence rate $R_\mathrm{C}$. (b) Illustration of $R_\mathrm{C}$ as a function of the idler-photon delay time. In the low-OD limit, the $1/e$ decay time approaches the single-atom spontaneous lifetime $\tau_{31}$ on $|3\rangle \rightarrow |1\rangle $ transition. As OD increases, cooperative emission produces a brighter and temporally narrower $R_C$, indicating superradiant shortening of the decay time $\tau_d$.
}
    \label{fig1}
    \end{figure}
}

\newcommand{\figtwo}{
	\begin{figure}[t]
		\centering
		\includegraphics[width = 8.6 cm]{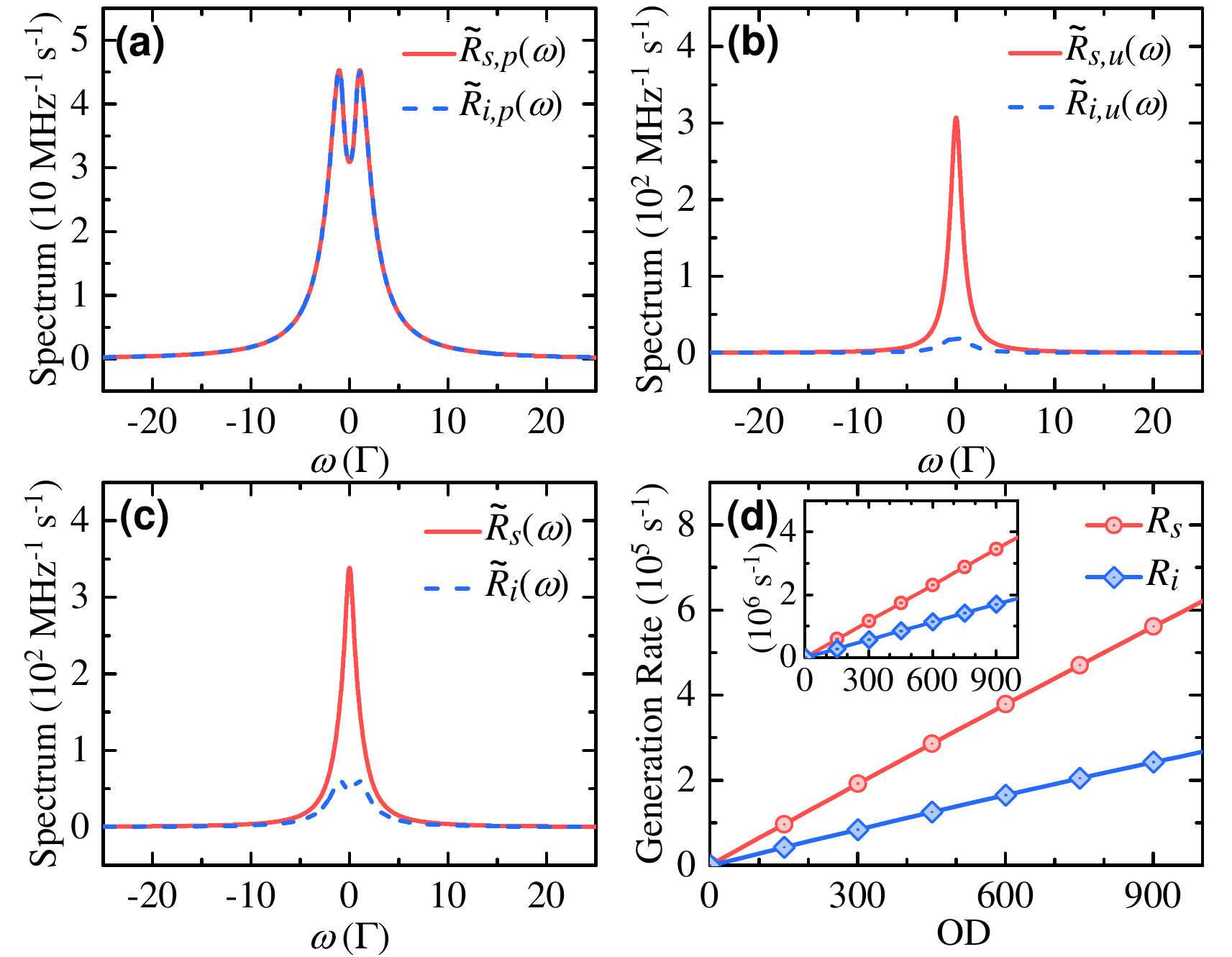}
		\caption{
Spectral characteristics of signal and idler photons and their dependence on OD. (a) Pairing spectra $R_{s,p}$ and $R_{i,p}$, (b) unpairing spectra $R_{s,u}$ and $R_{i,u}$, and (c) total spectra $R_s$ and $R_i$ of the signal and idler fields for $\mathrm{OD}=10$. In (a), the pairing spectra coincide, $R_{s,p}=R_{i,p}$. (d) Total generation rates $R_s$ and $R_i$ as functions of OD. The inset shows an alternative biphoton-generation configuration with 1367~nm (signal) and 780~nm (idler). Other parameters are $\Omega_c=\Omega_d=1\Gamma$, $\Delta_1=-50\Gamma$, and $\Delta_2=0$.
}
	\label{fig2}
	\end{figure}
}

\newcommand{\figthree}{
	\begin{figure}[t]
		\centering
		\includegraphics[width = 8.6 cm]{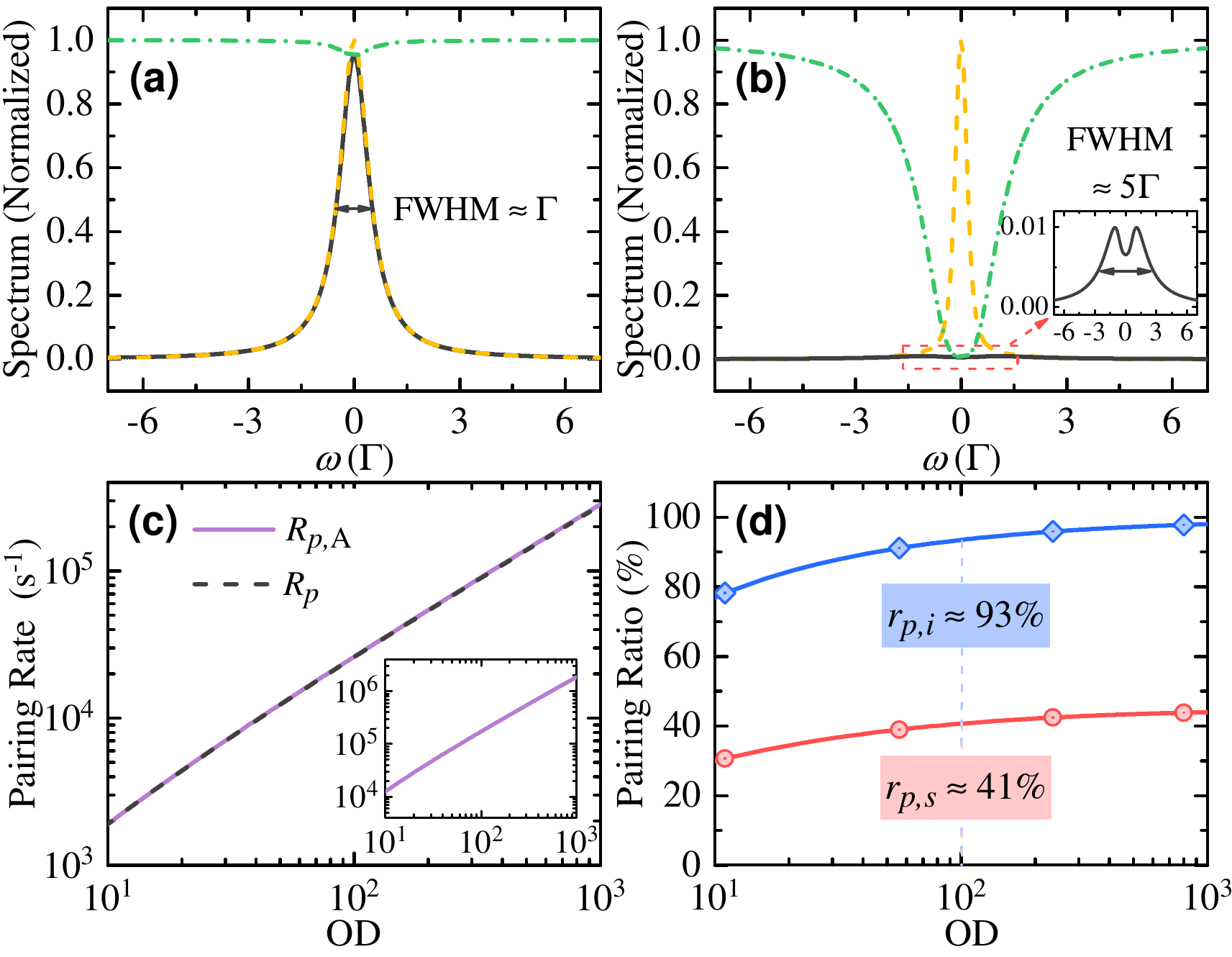}
		\caption{			
Spectral characteristics of paired photons and their dependence on OD. (a), (b) Spectral profiles of the pairing rate $R_p$ (black solid) together with its absorption (green dash-dotted) and emission (yellow dashed) components for $\mathrm{OD}=0.1$ and $\mathrm{OD}=10$, respectively. The absorption and emission gain spectra are normalized independently, and each $R_p$ spectrum is normalized to the maximum of its corresponding emission spectrum. The inset highlights the OD-induced spectral broadening of $R_p$. (c) Pairing rate as a function of OD obtained from the general ($R_p$) and analytical-solution ($R_{p,\mathrm{A}}$) results. The two curves nearly overlap for $\mathrm{OD}>10$. The inset shows the pairing rate for biphoton generation at 1367~nm (signal) and 780~nm (idler), which yields a higher generation rate under otherwise identical driving conditions. (d) Pairing ratios of the signal ($r_{p,s}$) and idler ($r_{p,i}$) photons as functions of OD. All panels use $\Omega_c=\Omega_d=1\Gamma$, $\Delta_1=-50\Gamma$, and $\Delta_2=0$.
}
	\label{fig3}
	\end{figure}
}

\newcommand{\figfour}{
	\begin{figure}[t]
		\centering
		\includegraphics[width = 8.8 cm]{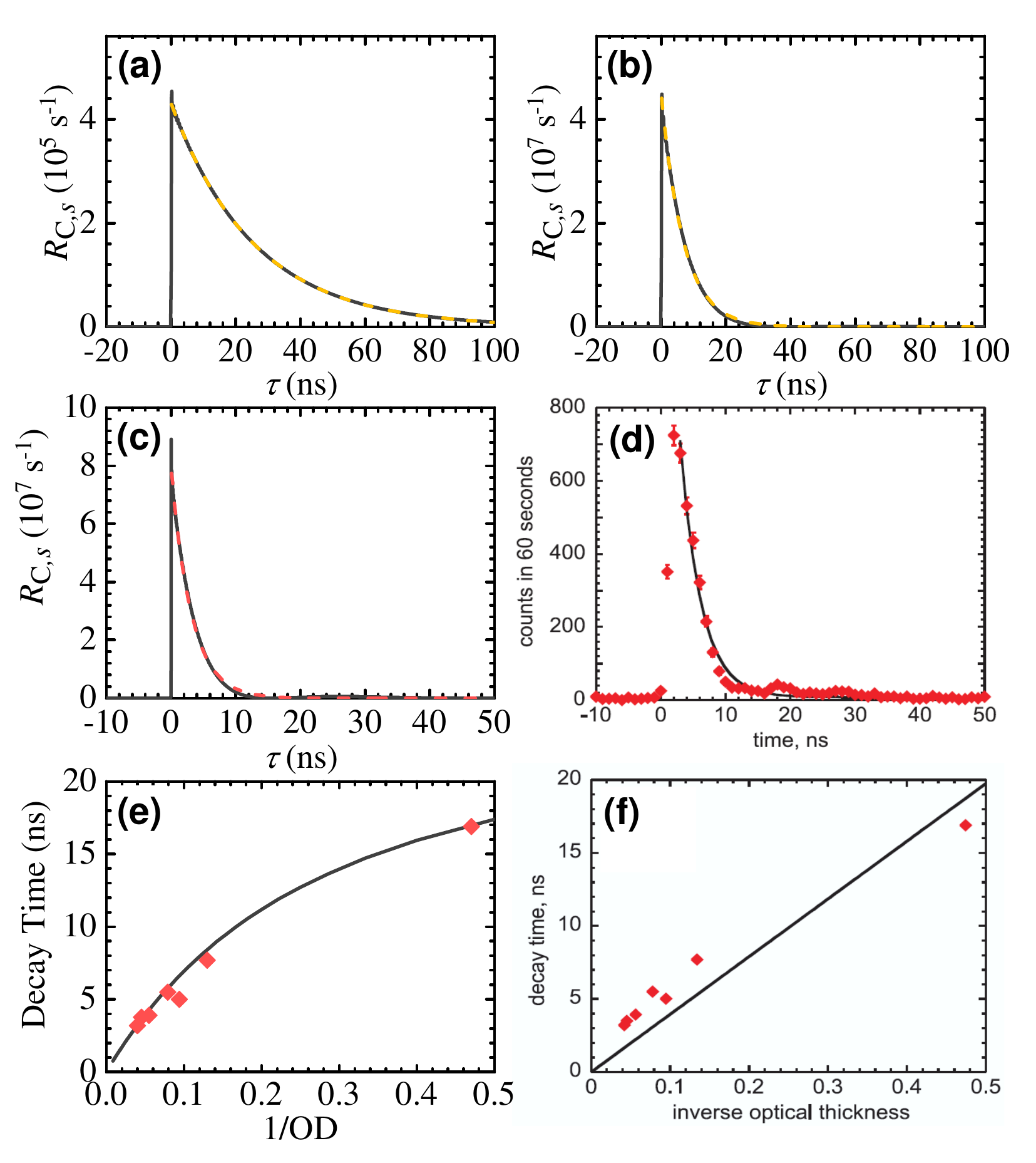}
		\caption{			
Superradiant behavior of biphoton temporal correlations and quantitative comparison with experimental data.  
(a), (b) Signal-side coincidence rate $R_{\mathrm{C},s}$ as a function of the idler-photon delay time for $\mathrm{OD}=0.1$ and $\mathrm{OD}=10$, respectively, with $\Omega_c=\Omega_d=1\Gamma$, $\Delta_1=-50\Gamma$, and $\Delta_2=0$. Exponential fits (yellow dashed curves) yield $1/e$ decay times of approximately $26~\mathrm{ns}$ and $7~\mathrm{ns}$. 
(c) Theoretical prediction of $R_{\mathrm{C},s}$ at $\mathrm{OD}=25$. 
(d) Experimental trace (red dots) reproduced from Ref.~\cite{Chaneliere} for direct quantitative comparison. The fitted decay times in (c) and (d) are 3.3~ns and 3.2~ns, respectively, demonstrating close agreement between theory and experiment. 
(e) Theoretical decay time (black curve) as a function of $1/\mathrm{OD}$, with red diamonds indicating experimental values extracted from the same reference. 
Panels (c) and (e) are calculated using the backward-field scheme and the corresponding experimental energy-level configuration. Experimental data are reproduced with permission.
}
	\label{fig4}
	\end{figure}
}

\newcommand{\figfive}{
	\begin{figure}[t]
		\centering
		\includegraphics[width = 8.6 cm]{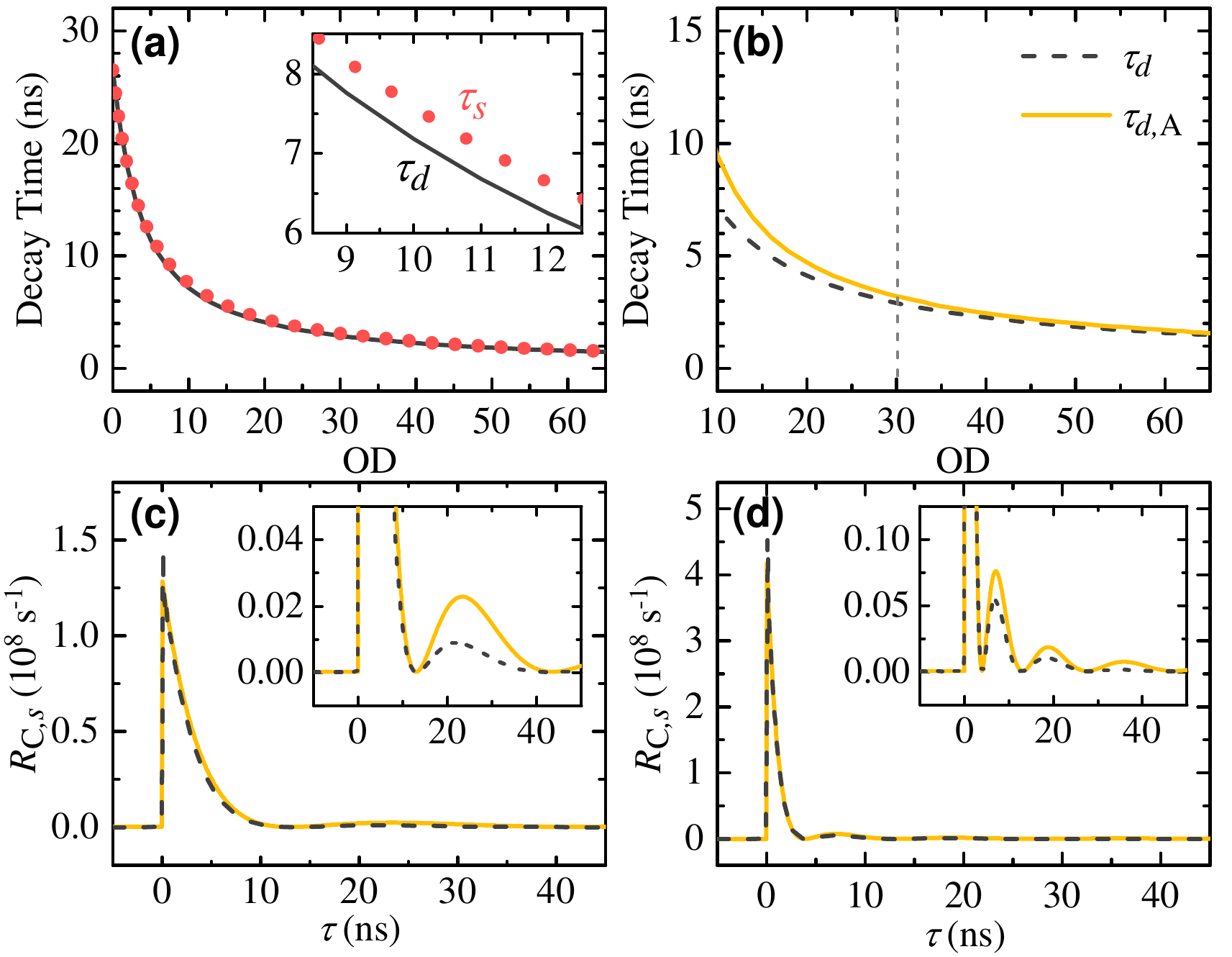}
		\caption{			
Comparison between the superradiant scaling time and the biphoton correlation decay time, and validation of the analytical approximation. (a) Decay time extracted from the full correlation function $\tau_d$ (black solid, from $G^{(2)}_{s\textrm{-}i}(\tau)$) compared with the superradiant scaling time $\tau_s$ (red dots). The inset highlights the small deviation between the two curves. (b) $1/e$ decay time as a function of OD obtained from the general ($\tau_d$) and analytical ($\tau_{d,\mathrm{A}}$) expressions. The discrepancy remains below $25\%$ for $10\lesssim\alpha\lesssim30$, and the two results converge for $\alpha\gtrsim30$, where $\tau_d\approx\tau_{d,\mathrm{A}}\approx4/(\alpha\Gamma_{31})$. (c), (d) Signal-side coincidence rate as a function of idler delay $\tau$ obtained from the full solution [$G^{(2)}_{s\textrm{-}i}(\tau)$, black dashed] and the analytical approximation [$G^{(2)}_{s\textrm{-}i,\mathrm{A}}(\tau)$, yellow solid] for $\mathrm{OD}=30$ and $\mathrm{OD}=100$. The insets highlight oscillatory structures that become more pronounced at higher OD. All panels use $\Omega_c=\Omega_d=1\Gamma$, $\Delta_1=-50\Gamma$, and $\Delta_2=0$.
}
	\label{fig5}
	\end{figure}
}

\newcommand{\figsix}{
	\begin{figure}[t]
		\centering
		\includegraphics[width = 8.6 cm]{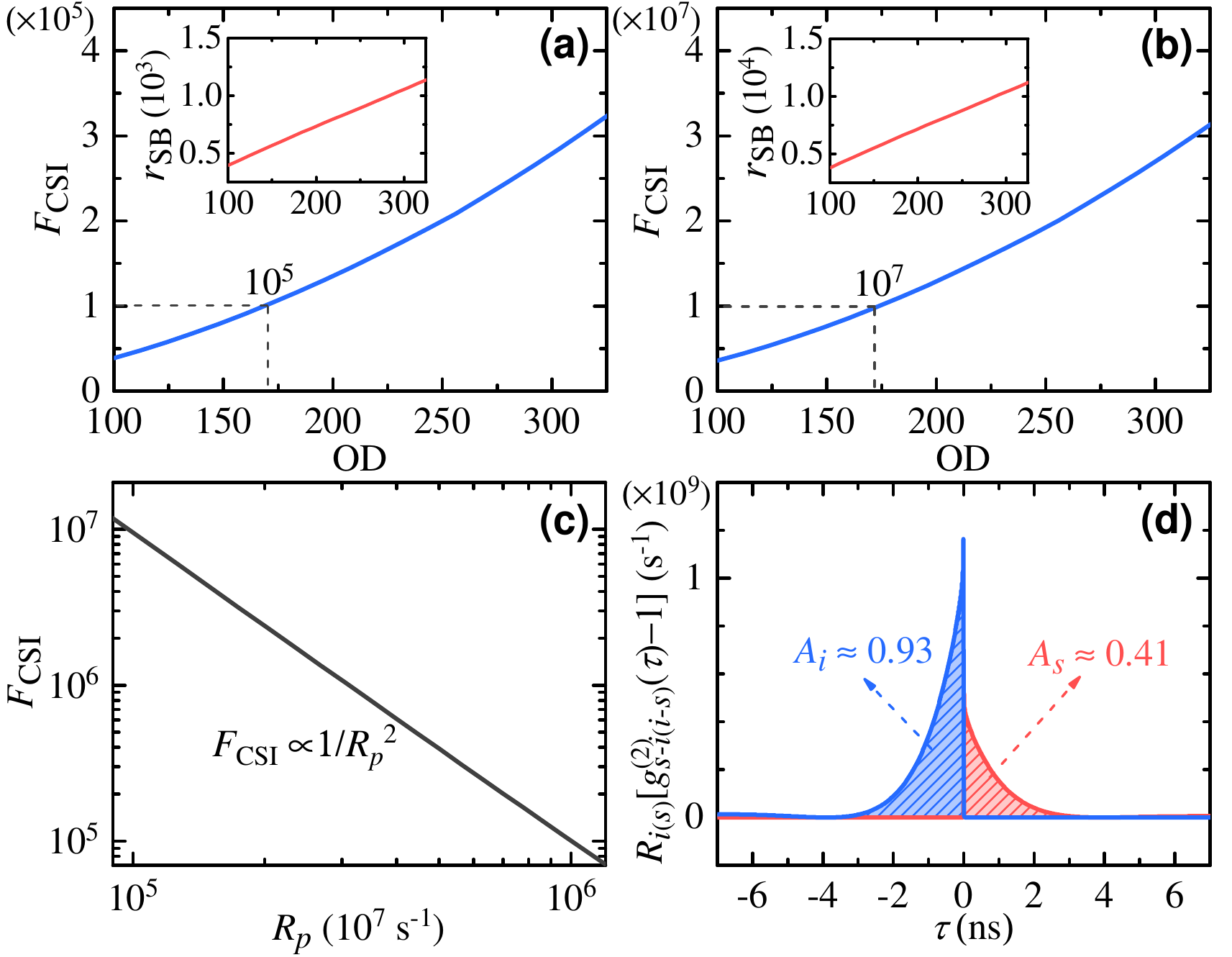}
		\caption{			
Nonclassicality enhancement characterized by the Cauchy--Schwarz surpassing factor $F_{\textrm{CSI}}$. (a), (b) $F_{\textrm{CSI}}$ as a function of OD for fixed pairing rates $R_p=1.0\times10^{6}\,\mathrm{s}^{-1}$ and $1.0\times10^{5}\,\mathrm{s}^{-1}$, respectively. In this OD range, $R_p$ scales as $R_p\propto\mathrm{OD}$, $R_p\propto\Omega_c^{\,2}$, and $R_p\propto\Omega_d^{\,2}$. To maintain a constant $R_p$, the Rabi frequencies are adjusted to $\Omega_c=\Omega_d=10^{1/4}(\mathrm{OD}/370)^{-1/4}\Gamma$ in (a) and $\Omega_c=\Omega_d=(\mathrm{OD}/370)^{-1/4}\Gamma$ in (b). The insets show the corresponding peak signal-to-background ratios $r_{\textrm{SB}}$. (c) $F_{\textrm{CSI}}$ as a function of $R_p$ at $\mathrm{OD}=170$, obtained by sweeping $\Omega_c$ and $\Omega_d$ between the settings used in (a) and (b). The overall scaling $F_{\textrm{CSI}}\propto1/R_p^{2}$ is evident. (d) $R_{i(s)}[g^{(2)}_{s\textrm{-}i(i\textrm{-}s)}(\tau)-1]$ as a function of the idler delay $\tau$ for $\mathrm{OD}=100$. The integrated areas yield $A_s=0.41$ and $A_i=0.93$, corresponding to the average numbers of correlated idler (signal) photons conditioned on detecting a signal (idler) photon. These values agree with the pairing ratios $r_{p,s}$ and $r_{p,i}$ shown in Fig.~\ref{fig3}(d) at the same OD. Unless otherwise stated, the remaining parameters are $\Omega_c=\Omega_d=1\Gamma$, $\Delta_1=-50\Gamma$, and $\Delta_2=0$.
}
	\label{fig6}
	\end{figure}
}

\newcommand{\figseven}{
	\begin{figure}[t]
		\centering
		\includegraphics[width = 8.6 cm]{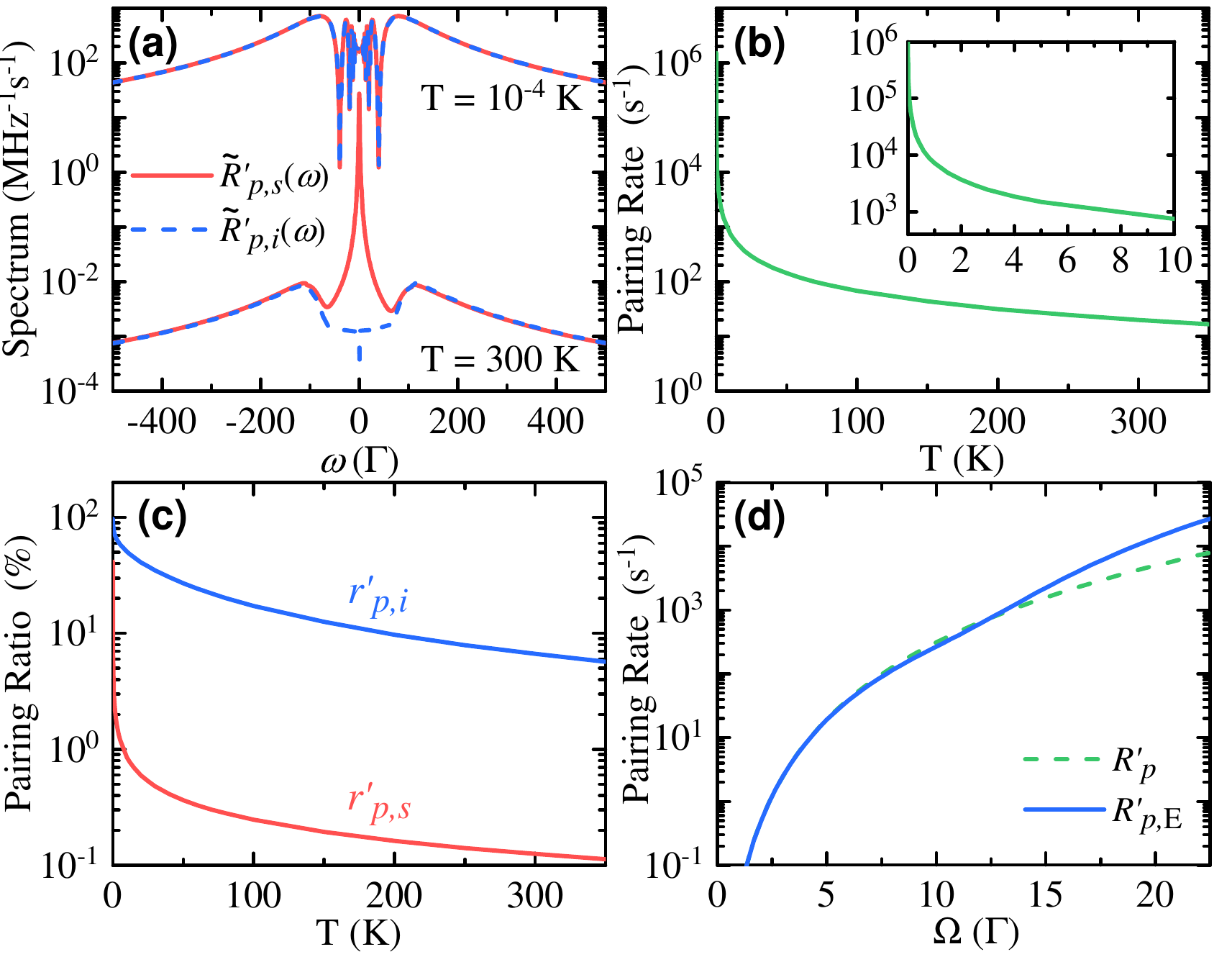}
		\caption{			
Doppler-broadened pairing spectra, pairing rates, and pairing ratios in warm atomic ensembles. (a) Pairing spectra of the signal (red solid) and idler (blue dashed) fields at different temperatures. (b) Pairing rate as a function of temperature. The inset enlarges the low-temperature region from 0 to 10 K. (c) Corresponding pairing ratios as functions of temperature. The parameters in (a)–(c) are $\mathrm{OD}=1000$, $\Omega_c=\Omega_d=5\Gamma$, $\Delta_1=-500\Gamma$, and $\Delta_2=0$. (d) Pairing rate as a function of the laser Rabi frequency $\Omega$ in the Doppler-broadened regime. Results obtained from the exact-population model are shown by blue solid curve, while the GSA prediction is shown by the green dashed curve. The parameters are $\mathrm{OD}=1000$, $\Omega_c=\Omega_d\equiv\Omega$, $\Delta_1=-500\Gamma$, and $\Delta_2=0$.
}
	\label{fig7}
	\end{figure}
}

\newcommand{\figeight}{
	\begin{figure}[t]
		\centering
		\includegraphics[width = 9.2 cm]{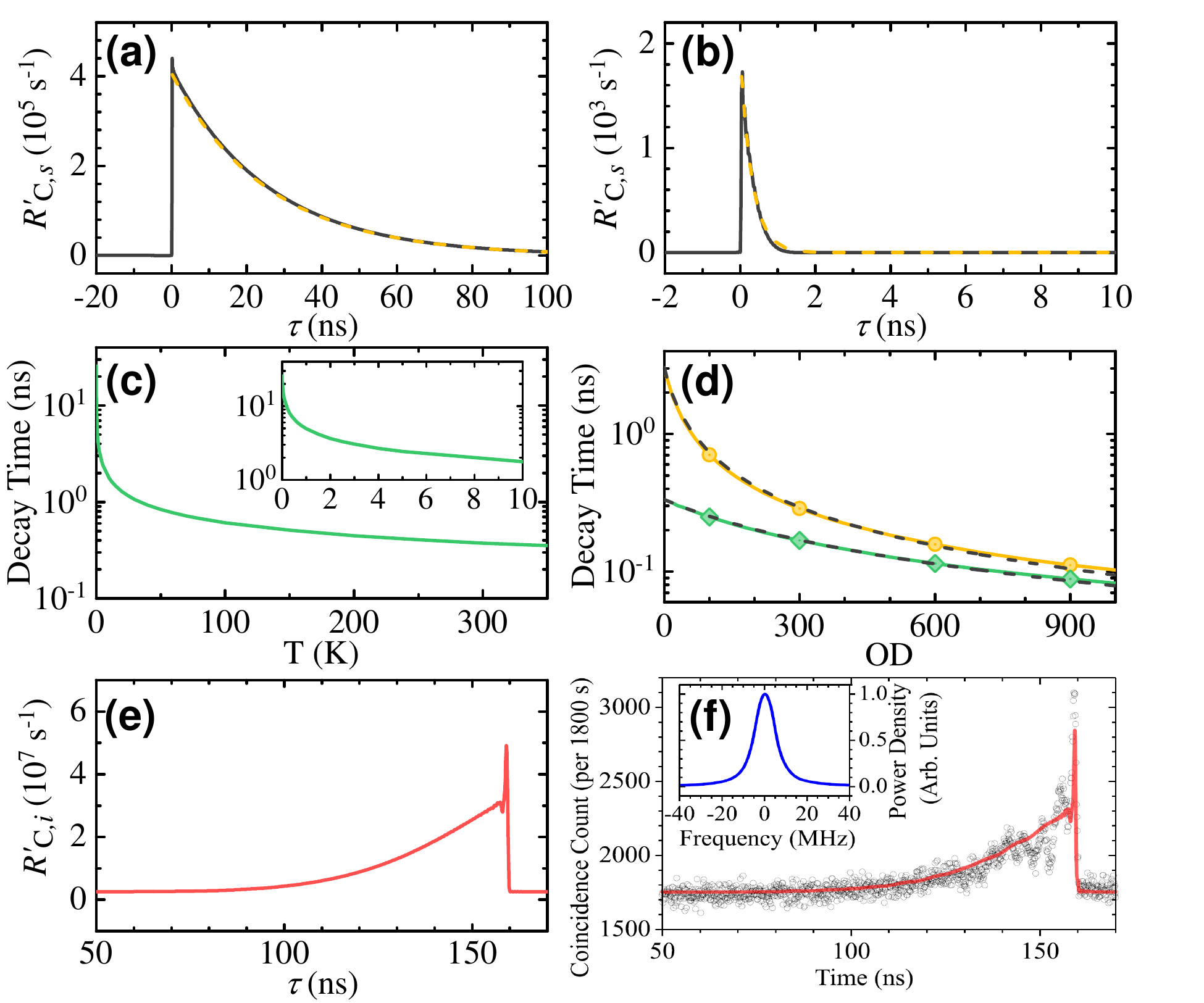}
		\caption{			 
Superradiant signatures in Doppler-broadened biphoton temporal correlations of warm atomic ensembles. 
(a), (b) Signal coincidence rate $R'_{\mathrm{C},s}$ as a function of idler delay for $\mathrm{T}=10^{-4}\,\mathrm{K}$ and $\mathrm{T}=300\,\mathrm{K}$, respectively, with $\mathrm{OD}=0.1$. Yellow dashed curves are exponential fits yielding $1/e$ decay times of approximately $26\,\mathrm{ns}$ and $0.3\,\mathrm{ns}$. 
(c) Biphoton decay time versus temperature for $\mathrm{OD}=0.1$. The inset shows an enlarged view for $0$--$10\,\mathrm{K}$. 
(d) Decay time as a function of OD for $\mathrm{T}=3\,\mathrm{K}$ (yellow circles) and $\mathrm{T}=300\,\mathrm{K}$ (green diamonds). Black dashed curves are fits to $\tau'_d=\tau'_{31}/(1+x\,\mathrm{OD})$, with fitted values of the scaling parameter $x=1/32$ and $1/310$, respectively. 
Unless otherwise stated, parameters in (a)--(d) are $\Omega_c=\Omega_d=5\Gamma$, $\Delta_1=-500\Gamma$, and $\Delta_2=0$. 
(e) Theoretical prediction of $R'_{\mathrm{C},i}$ compared with experimental data (f) from Ref.~\cite{Tu}. 
For panels (e) and (f), the parameters are $\mathrm{OD}=420$, $\Omega_c=17.1\Gamma$, $\Omega_d=78.7\Gamma$, $\Delta_1=353\Gamma$, $\Delta_2=0$, and $\mathrm{T}=328\,\mathrm{K}$. 
Panel (f) is reproduced from Ref.~\cite{Tu} under its Creative Commons license.
}
	\label{fig8}
	\end{figure}
}

\newcommand{\fignine}{
	\begin{figure}[t]
		\centering
		\includegraphics[width = 8.6 cm]{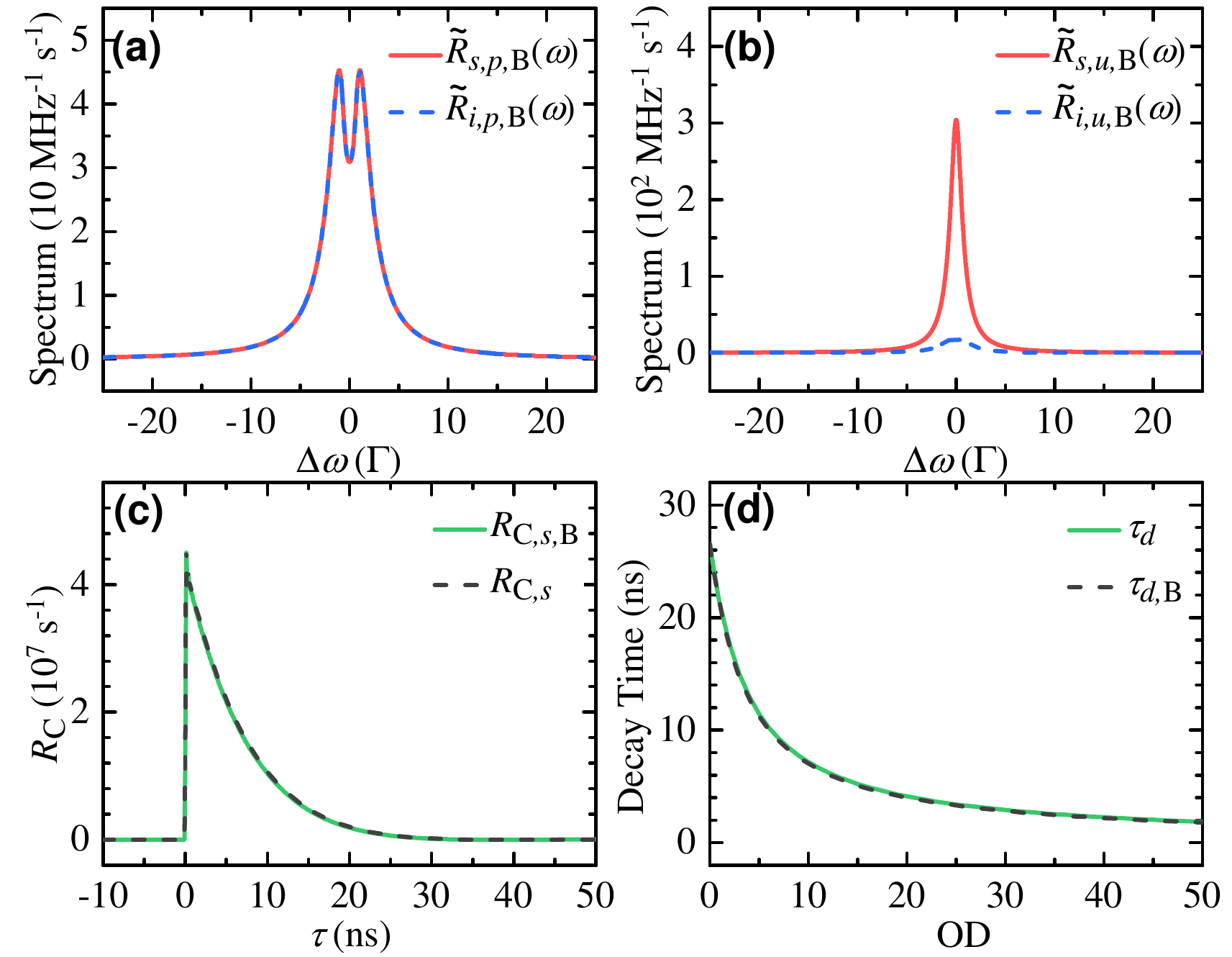}
		\caption{			
Comparison of biphoton generation properties in the backward-field configuration with those in the forward scheme. (a) Pairing spectra of the signal $R_{s,p,{\textrm{B}}}$ and idler $R_{i,p,{\textrm{B}}}$ fields in the backward configuration, which are indistinguishable from those in Fig.~\ref{fig2}(a). (b) Corresponding unpairing spectra for the signal ($R_{s,u,{\textrm{B}}}$) and idler ($R_{i,u,{\textrm{B}}}$) fields. (c) Signal coincidence count rates for the forward ($R_{\mathrm{C},s}$) and backward ($R_{\mathrm{C},s,{\textrm{B}}}$) configurations, exhibiting nearly identical temporal profiles with decay times of approximately $7\,\mathrm{ns}$. In panels (a)–(c), the OD is set to 10. (d) Extracted decay times as a function of OD for the forward ($\tau_d$) and backward ($\tau_{d,{\textrm{B}}}$) schemes, showing excellent agreement between the two curves. The remaining parameters are $\Omega_c=\Omega_d=1\Gamma$, $\Delta_1=-50\Gamma$, $\Delta_2=0$, and $\Delta kL=0$.
}
	\label{fig9}
	\end{figure}
}

%%%%%%%%%%%%%%%%%%%%%%%%%%%%%%%%%%%%%%%%%%%%%%%%%%%%%%%%%%%%%%%%%%%%%%%%%%%%%%%%%%%%%%%%%%%%%%%%%%%%%
%%%%%%%%%%%%%%%%%%%%%%%%%%%%%%%%%%%%%%%%%%%%%%%%%%%%%%%%%%%%%%%%%%%%%%%%%%%%%%%%%%%%%%%%%%%%%%%%%%%%%

\section{Introduction}

Collective spontaneous emission, known as superradiance, refers to a regime in which radiation from different atoms interferes constructively in a phase matched manner, giving rise to an accelerated emission process whose characteristic timescale is governed by the number of participating atoms and the sample geometry~\cite{Dicke,Ernst,Agarwal,Rehler,Gross,Scully}. Beyond its original formulation as a collective modification of independent single-atom spontaneous emission, superradiance can fundamentally reshape the temporal structure of correlated light generated in resonant atomic media. In particular, when applied to biphoton generation, collective emission imprints many body radiative dynamics directly onto the temporal correlations of photon pairs, rather than merely modifying a single atom radiative response. In diamond type atomic configurations driven near resonance, experiments have repeatedly shown that the biphoton correlation time shortens with increasing optical depth (OD), accompanied by a simultaneous increase in the instantaneous biphoton flux~\cite{Chaneliere,Srivathsan,Srivathsan2,Zhang,Park,Jeong2}. These concurrent trends point to a regime in which collective spontaneous emission plays a central role in shaping biphoton temporal properties.

This behavior is conceptually nontrivial because, in many extensively studied narrowband atomic biphoton platforms, most notably double-$\Lambda$ systems, biphoton temporal correlations are predominantly governed by long lived ground-state coherences and dispersion induced slow-light effects, whose physical origins are relatively well understood. In contrast, resonantly driven diamond-type systems exhibit accelerated biphoton correlations in the absence of such slow-light mechanisms, indicating that a qualitatively different physical process is at work. This contrast calls for a microscopic theoretical description in which collective emission, quantum correlations, and dissipation are treated on equal footing, and in which biphoton temporal properties emerge from an open system many body radiative dynamics rather than from single particle or purely dispersive effects.

Nonclassical light sources~\cite{Cirac,Varcoe,Bertet,Walther,Ourjoumtsev,OBrien,Afek,Jeong1,Tan,Jia} are key resources for quantum networks~\cite{Kimble,Wehner,Lodahl,Aguilar,Rad} and integrated photonic technologies~\cite{Flamini,JWang,LLu,Toninelli,Pelucchi}. While long-distance communication benefits from low-loss telecom wavelengths~\cite{Gisin,Muralidharan,Paraiso}, quantum control and metrology typically operate in the visible and optical domains~\cite{Guzik,Slussarenko,Takeda,Giovannetti,Giovannetti2,Polino}. Bridging this spectral gap in a manner compatible with narrowband atomic interfaces remains challenging. Quantum frequency conversion provides one approach~\cite{Ikuta,Zaske,Stolk,Tseng}, but may introduce loss and noise. Alternatively, correlated photon pairs may be generated directly with one photon in the telecom band and the other in the visible. Solid-state platforms have demonstrated two-color biphotons via far-off-resonant nonlinear processes~\cite{Pelton,Arahira,Hentschel,Clausen,Saglamyurek,Stuart,Bussieres2,Dietz,XLu,QWang,Schunk}, typically with terahertz bandwidths that require cavity filtering and reduce usable rates~\cite{Schunk,XLu}.

Operating near atomic resonance provides a complementary route that enables narrowband biphoton generation at the megahertz scale through spontaneous four-wave mixing (SFWM) in atomic media~\cite{Duan, Bussieres, Pu, Covey}. Both double-cascade systems with identical transitions~\cite{Chaneliere, Ding, Jeong2, Lee2, Park, Park3} and diamond-type systems with distinct transitions~\cite{Willis, Willis2, Zhang, Srivathsan, Srivathsan2, Dong, Craddock, Gulati, Whiting, Tu} can place one photon in the telecom band and the other in the visible. In contrast to double-$\Lambda$ configurations, where slow-light effects elongate the biphoton wave packet~\cite{Kolchin, Ooi, LZhao, Kolchin2, Du2, Yan, MZhao, Shiu1, Shiu2, Cui, Zhao, Shiu3}, the correlation time in many cold-atom diamond-type implementations has been observed to shorten with increasing OD, a behavior commonly associated with superradiant emission~\cite{Chaneliere, Srivathsan, Srivathsan2, Zhang, Park}. Increasing OD shortens the correlation time, thereby redistributing the biphoton emission into a narrower temporal window and enhancing the instantaneous biphoton flux and signal-to-background ratio under otherwise comparable conditions~\cite{Srivathsan, Srivathsan2, Zhang, Park, Jeong2}.

Despite experimental progress, existing theories remain incomplete in identifying the microscopic origin of biphoton correlations. Many cascade~\cite{Ding, Jeong2, Jen} and diamond-type~\cite{Willis, Jen2} treatments rely on effective third-order susceptibilities and parametric gain without frequency-resolved separation of correlated biphotons and vacuum-seeded background photons. Consequently, observables such as the pairing ratio~\cite{Shiu1, Shiu2, Shiu3}, which bounds heralding efficiency~\cite{Mosley, Bock}, and accidental backgrounds are not directly accessible. Although numerical phase-space methods have explored superradiant emission in cascade ensembles~\cite{Jen3}, transparent analytical scaling laws linking OD to superradiant correlation time remain limited. In warm vapors~\cite{Ding, Jeong2, Lee2, Park, Park3, Willis, Willis2, Dong, Craddock}, substantially shorter correlation times are observed. Jeong \textit{et al.}~\cite{Jeong2} attributed this primarily to velocity-induced dispersion, leaving the role of genuine collective emission unresolved. While temporal shortening may appear in gain-based models, quantitative prediction of unpaired background photons and pairing ratios requires a self-consistent open-system framework connecting collective emission and vacuum fluctuations.

Here we apply the Heisenberg--Langevin--Maxwell theoretical framework to a four-level diamond-type atomic system and establish a fully quantum microscopic model of biphoton generation through SFWM~\cite{Ford, ScullyBook}. For cold atomic ensembles, we resolve the spectral contributions of paired biphotons and unpaired noise photons seeded by vacuum fluctuations, enabling a direct evaluation of the pairing ratio~\cite{Shiu1, Shiu2, Shiu3}. In the high-OD regime, the coupled Heisenberg--Langevin and Maxwell equations admit simple analytical solutions that reduce the dynamics to an effective collective two-level emission process, making the superradiant origin of the biphoton temporal envelope explicit. The resulting correlation functions reproduce the superradiant time scale observed experimentally~\cite{Chaneliere} and agree with established collective-emission scaling laws~\cite{Rehler}.

Extending the model to Doppler-broadened warm vapors via a Maxwell--Boltzmann velocity distribution~\cite{Yatsenko}, we show that collective emission and Doppler-induced dispersion coexist, with the latter dominating at elevated temperatures. To the best of our knowledge, a closed-form analytical connection between OD and the superradiant biphoton correlation time, together with a frequency-resolved separation of paired and unpaired contributions and a unified extension to warm vapors within the same microscopic framework, has not been reported previously within a single microscopic open-system framework.

The remainder of this paper is organized as follows. Section~\ref{Sec. 2} presents a microscopic theoretical framework and derives analytical solutions for biphoton generation in cold atomic ensembles. Section~\ref{Sec. 3} examines the spectral characteristics and generation rates of the biphotons. Section~\ref{Sec. 4} analyzes the temporal correlation functions and their superradiant scaling behavior. Section~\ref{Sec. 5} extends the theory to Doppler-broadened warm atomic vapors. Section~\ref{Sec. 6} concludes the paper. The Appendices provide technical details supporting the main text, including microscopic derivations and treatments of thermal and noise-induced effects
within the Heisenberg--Langevin framework~\cite{Cheng, Hsu, Liu1, Chen}.

%%%%%%%
\figone
%%%%%%%

%%%%%%%%%%%%%%%%%%%%%%%%%%%%%%%%%%%%%%%%%%%%%%%%%%%%%%%%%%%%%%%%%%%%%%%%%%%%%%%%%%%%%%%%%%%%%%%%%%%%%
%%%%%%%%%%%%%%%%%%%%%%%%%%%%%%%%%%%%%%%%%%%%%%%%%%%%%%%%%%%%%%%%%%%%%%%%%%%%%%%%%%%%%%%%%%%%%%%%%%%%%

\section{Microscopic Theory in the Doppler-Free Regime} \label{Sec. 2}

\subsection{Unified theoretical framework} \label{Sec. 2a}

We consider a diamond-type four-level atomic ensemble consisting of a single ground state $|1\rangle$ and three excited states $|2\rangle$, $|3\rangle$, and $|4\rangle$, as illustrated in Fig.~\ref{fig1}(a). Two strong laser fields, a coupling field with Rabi frequency $\Omega_c$ driving the $|1\rangle \leftrightarrow |2\rangle$ transition and a driving field with Rabi frequency $\Omega_d$ addressing the $|2\rangle \leftrightarrow |4\rangle$ transition, sequentially excite atoms from $|1\rangle$ to $|4\rangle$ through $|2\rangle$. The one-photon and two-photon detunings are defined as $\Delta_1=\omega_c-\omega_{21}$ and $\Delta_2=\omega_c+\omega_d-\omega_{41}$, respectively. Vacuum fluctuations initiate the SFWM process and lead to the generation of weakly correlated signal and idler photons, where the signal mode $\hat{a}_s$ couples $|4\rangle \leftrightarrow |3\rangle$ and the idler mode $\hat{a}_i$ couples $|3\rangle \leftrightarrow |1\rangle$. As the OD increases, collectively enhanced stimulated emission of the idler photons accelerates their decay, resulting in a temporally compressed biphoton wave packet, as shown in Fig.~\ref{fig1}(b). Throughout this subsection, the coupling and driving beams are assumed to be co-propagating. The interaction Hamiltonian $\hat{H}$ in continuous form is:
\begin{align}
	\hat{H} = & -\frac{\hbar N}{2L}\int^L_0  \big[
	\Omega_c\hat{\sigma}_{21}(z,t)+\Omega_d \hat{\sigma}_{42}(z,t) 
	\notag\\
	& + 
	2g_s\hat{\sigma}_{43}(z,t)\hat{a}_s(z,t)+2g_i\hat{\sigma}_{31}(z,t)\hat{a}_i(z,t)
	\notag\\
	& +
	\Delta_1\hat{\sigma}_{22}(z,t)+\Delta_2\hat{\sigma}_{44}(z,t)+\textrm{H.c.} \big]\,dz,
	\label{eq(1)}
\end{align}
where $N$ and $L$ denote the total number of atoms and the ensemble length, respectively. The single-atom coupling constants to the signal and idler fields are $g_{s(i)}=d_{43(31)}E_{s(i)}/\hbar$, where $d_{43(31)}$ is the dipole moment of the $|4\rangle \leftrightarrow |3\rangle$ ($|3\rangle \leftrightarrow |1\rangle$) transition and $E_{s(i)}=\sqrt{\hbar \omega_{s(i)}/(2\epsilon_0 V)}$ represents the single-photon electric-field amplitude. The slowly varying collective atomic operator $\hat{\sigma}_{jk}(z,t)$ describes the coherence between $|j\rangle$ and $|k\rangle$ and evolves according to the Heisenberg--Langevin equations (HLEs)
\begin{eqnarray}
	\frac{\partial}{\partial t}\hat{\sigma}_{jk}=\frac{i}{\hbar}\left[\hat{H},\hat{\sigma}_{jk}\right]+\hat{R}_{jk}+\hat{F}_{jk}.
	\label{eq(2)}
\end{eqnarray}
Here $\hat{R}_{jk}$ and $\hat{F}_{jk}$ represent the relaxation terms and Langevin noise operators. 
In the large-$\Delta_1$ limit, atomic excitation is negligible and both coupling and driving fields remain undepleted, allowing the ground-state approximation (GSA) $\langle \hat{\sigma}_{11}\rangle \approx 1$. Under this condition, the generated signal and idler fields are treated perturbatively. Propagation of the generated signal and idler fields through the atomic medium is described by the coupled Maxwell--Schr\"{o}dinger equations (MSEs):
\begin{align}
	& \left(\frac{1}{c}\frac{\partial}{\partial t}+\frac{\partial}{\partial z}\right)\hat{a}_{s}(z,t)=\frac{ig_{s}N}{c}\hat{\sigma}^{(1)}_{34}(z,t), \label{eq(3)} \\[6pt]
	& \left(\frac{1}{c}\frac{\partial}{\partial t}+\frac{\partial}{\partial z}+i \Delta k\right)\hat{a}^{\dagger}_{i}(z,t)=-\frac{ig_{i}N}{c}\hat{\sigma}^{(1)}_{31}(z,t), 
	\label{eq(4)}
\end{align}
where $c$ is the speed of light in vacuum and $\Delta k =k_c+k_d-k_s-k_i$ denotes the wave-vector mismatch of the SFWM process. Performing a temporal Fourier transform and substituting the first-order atomic coherences $\tilde{\sigma}^{(1)}_{34}(z,\omega)$ and $\tilde{\sigma}^{(1)}_{31}(z,\omega)$ (see Appendix~\ref{ApxA}) yield the coupled frequency-domain equations:
\begin{align}
	& \frac{\partial}{\partial z} \tilde{a}_s
	+G_s \tilde{a}_s +\kappa_s  \tilde{a}^{\dagger}_i =\sum_{jk}\zeta^s_{jk}\tilde{f}_{jk}, 
	\label{eq(5)} \\[6pt]
	& \frac{\partial}{\partial z} \tilde{a}^{\dagger}_i
	+\Gamma_i \tilde{a}^{\dagger}_i +\kappa_i  \tilde{a}_s=\sum_{jk}\zeta^i_{jk}\tilde{f}_{jk}. 
	\label{eq(6)}
\end{align}
Here $G_s$ and $\Gamma_i$ characterize the effective parametric gain and linear response experienced by the signal and idler fields, respectively, while $\kappa_s$ and $\kappa_i$ quantify the parametric coupling between the two fields. Under the GSA and in the large-$\Delta_1$ regime, $G_s$ admits an effective interpretation in terms of a three-photon parametric process involving the $|1\rangle\!\leftrightarrow\!|2\rangle\!\leftrightarrow\!|4\rangle\!\leftrightarrow\!|3\rangle$ manifold. More generally, within the Heisenberg--Langevin framework, these coefficients implicitly incorporate contributions from all higher-order virtual excitation pathways consistent with the full microscopic dynamics. In this regime, the parameters take the forms
\begin{align}
	& \Gamma_i = \frac{\alpha\Gamma_{31}}{2L(\gamma_{31}-2i\omega)} -\frac{i\omega}{c}+i\Delta k, \label{eq(7)}\\[6pt]
	& \kappa_s  =\frac{i\alpha\sqrt{s_\lambda\Gamma_{43}\Gamma_{31}}\Omega_c\Omega_d}{4L\Delta_1(\gamma_{41}-2i\Delta_2)(\gamma_{31}-2i\omega)}, \label{eq(8)}\\[6pt]
	& \kappa_i =  \frac{i\alpha\sqrt{s_\lambda\Gamma_{43}\Gamma_{31}}\Omega^{\ast}_c\Omega^{\ast}_d}{4L\Delta_1(\gamma_{41}+2i\Delta_2)(\gamma_{31}-2i\omega)}
	, \label{eq(9)}\\[6pt]
	& G_s = \frac{-\alpha s_\lambda\Gamma_{43}|\Omega_c|^2|\Omega_d|^2}{8L \Delta^2_1(\gamma_{41}+2i\Delta_2)(\gamma_{31}-2i\omega)[\gamma_{43}-2i(\Delta_2+\omega)]}\notag\\
	& \quad\quad -\frac{i\omega}{c}, \label{eq(10)}
\end{align}
where ${\Gamma}_{jk}$ includes spontaneous emission and $\gamma_{jk}$ denotes the decoherence rate between $|j\rangle$ and $|k\rangle$. Here $s_\lambda = (\lambda_s/\lambda_i)^2$ is the squared wavelength ratio of the signal and idler fields, the terms $-i\omega/c$ in $G_s$ and $\Gamma_i$ account for the phase accumulation due to free propagation, and the relation $g_i^2 N/c=\alpha\Gamma_{31}/4L$ is used, with $\alpha$ denoting the OD on the $|3\rangle\!\leftrightarrow\!|1\rangle$ transition. The renormalized Langevin operators $\tilde{f}_{jk}=\sqrt{N/c}\,\tilde{F}_{jk}$ and their coefficients $\zeta^s_{jk}$ and $\zeta^i_{jk}$ (see Appendix~\ref{ApxB}) yield the general solutions
\begin{align}
	\begin{bmatrix}
		\tilde{a}_{sL} \\
		\tilde{a}^{\dagger}_{iL}
	\end{bmatrix}
	= 
	\begin{bmatrix}
		A_1 & B_1 \\
		C_1 & D_1
	\end{bmatrix}
	\begin{bmatrix}
		\tilde{a}_{s0} \\
		\tilde{a}^{\dagger}_{i0}
	\end{bmatrix}
	+\sum_{jk}\int^L_0
	\begin{bmatrix}
		A_2 & B_2\\
		C_2 & D_2
	\end{bmatrix}
	\begin{bmatrix}
		\zeta^s_{jk} \\
		\zeta^i_{jk}
	\end{bmatrix}
	\tilde{f}_{jk}\,dz. \label{eq(11)}
\end{align}
The subscript $0$ ($L$) denotes the field position for $z=0$ ($z=L$). The first matrix on the right-hand side of Eq.~(\ref{eq(11)}) represents the noiseless parametric evolution of the signal and idler fields, corresponding to the stimulated component of the FWM process. As derived in Appendix~\ref{ApxC}, for sufficiently large one-photon detuning $\Delta_1$, the matrix elements are
\begin{align}
	A_1 & = 1, \label{eq(12)}\\
	B_1 & = \left( \frac{\kappa_s}{\Gamma_i} \right)\left(e^{-\Gamma_i L} -1\right), \label{eq(13)}\\
	C_1 & = \left( \frac{\kappa_i}{\Gamma_i} \right)\left(e^{-\Gamma_i L} -1\right), \label{eq(14)}\\
	D_1 & = e^{-\Gamma_i L}. \label{eq(15)}
\end{align}
Physically, $A_1$ and $D_1$ ($B_1$ and $C_1$) denote the mode-preserved (mode-converted) coefficients for the signal and idler fields. The coefficient $A_1$ equals unity for the signal field because no population occupies states $|3\rangle$ and $|4\rangle$ under the GSA condition. In contrast, the coefficient $D_1$ explicitly indicates that the idler field undergoes absorption via the $|1\rangle \rightarrow |3\rangle$ transition. The second matrix in Eq.~(\ref{eq(11)}), which accounts for the propagation of Langevin noise contributions, has the same functional dependence as Eqs.~(\ref{eq(12)})–(\ref{eq(15)}) with $L$ replaced by $(L - z)$. The spectral and temporal characteristics of the generated biphotons are primarily determined by the structure of the coefficient matrix composed of $A_1$ through $D_1$, which will be analyzed in Sec.~\ref{Sec. 3}.

%%%%%%%%%%%%%%%%%%%%%%%%%%%%%%%%%%%%%%%%%%%%%%%%%%%%%%%%%%%%%%%%%%%%%%%%%%%%%%%%%%%%%%%%%%%%%%%%%%%%%
%%%%%%%%%%%%%%%%%%%%%%%%%%%%%%%%%%%%%%%%%%%%%%%%%%%%%%%%%%%%%%%%%%%%%%%%%%%%%%%%%%%%%%%%%%%%%%%%%%%%%

\subsection{Level scheme and parameters} \label{Sec. 2b}

We consider the diamond-type biphoton generation process using $^{87}$Rb atoms. The selected Zeeman sublevels are 
$|1\rangle = |5S_{1/2}, F=2, m_F=2\rangle$, 
$|2\rangle = |5P_{1/2}, F=1, m_F=1\rangle$, 
$|3\rangle = |5P_{3/2}, F=3, m_F=3\rangle$, and 
$|4\rangle = |4D_{3/2}, F=2, m_F=2\rangle$.  
This choice forms an effectively closed cycling configuration, ensured by the Zeeman selection rules, that confines the atomic population predominantly within the designated four-level manifold~\cite{LCChen}. The signal and idler wavelengths are 1529~nm and 780~nm. The spontaneous decay rates, together with the corresponding squared Clebsch--Gordan (CG) coefficients, are
$\Gamma_{21} = 0.95\Gamma$ (1/2),  
$\Gamma_{31} = 1.00\Gamma$ (1),  
$\Gamma_{42} = 0.30\Gamma$ (1/2), and  
$\Gamma_{43} = 0.05\Gamma$ (1/5),  
where $\Gamma = 2\pi \times 6~\mathrm{MHz}$~\cite{Brink, Balcar}. The decoherence rates used in the simulations are  
$\gamma_{21} = \Gamma_{21}$,
$\gamma_{31} = \Gamma_{31}$,   
$\gamma_{32} = \Gamma_{31}+\Gamma_{21}$,  
$\gamma_{41} = \Gamma_{42} + \Gamma_{43}$,  
$\gamma_{42} = \Gamma_{42} + \Gamma_{43} + \Gamma_{21}$, and  
$\gamma_{43} = \Gamma_{42} + \Gamma_{43} + \Gamma_{31}$~\cite{Tseng}.  
For cold-atom simulations we set $\Delta_1 = -50\Gamma$ and $\Delta_2 = 0$. Throughout the calculations, perfect phase matching ($\Delta k = 0$) is assumed, which can be achieved experimentally by optimizing the relative field geometry.

To benchmark our theoretical model, we compare it with the experimental measurements reported in Ref.~\cite{Chaneliere}.  
In that work, the relevant hyperfine levels of $^{85}$Rb are identified as 
$|1\rangle = |5S_{1/2}, F=3\rangle$, 
$|2\rangle =|3\rangle = |5P_{3/2}, F=4\rangle$, and 
$|4\rangle = |5D_{5/2}, F=5\rangle$.  
Because the Zeeman sublevels were not specified, we adopt a representative cycling configuration,  
$|1\rangle = |5S_{1/2}, F=3, m_F=3\rangle$, 
$|2\rangle = |5P_{3/2}, F=4, m_F=2\rangle$, 
$|3\rangle = |5P_{3/2}, F=4, m_F=4\rangle$, and 
$|4\rangle = |5D_{5/2}, F=5, m_F=3\rangle$.  
Such variations in Zeeman selection are theoretically expected to cause only minor modifications to the biphoton temporal waveform. The spontaneous decay rates with the corresponding squared CG coefficients for the $|5P_{3/2}\rangle \leftrightarrow |5D_{5/2}\rangle$ transitions are
$\Gamma_{21} = 1.00\Gamma$ (1/28),  
$\Gamma_{31} = 1.00\Gamma$ (1),  
$\Gamma_{42} = 0.28\Gamma$ (28/45), and  
$\Gamma_{43} = 0.28\Gamma$ (1/45).  
The associated signal and idler wavelengths are 776~nm and 780~nm. A detailed comparison between theory and experiment is presented in Sec.~\ref{Sec. 4}. 

For completeness, we specify the atomic level configuration used for comparison with the Doppler-broadened experiment in Ref.~\cite{Tu}. In that work, the relevant hyperfine levels of $^{87}$Rb are selected as 
$|1\rangle = |5S_{1/2}, F=2\rangle$, 
$|2\rangle = |5P_{1/2}, F=2\rangle$,
$|3\rangle = |5P_{3/2}, F=3\rangle$, and 
$|4\rangle = |4D_{3/2}, F=3\rangle$ 
with a multi-Zeeman-level configuration. For simplicity, we adopt the following single-Zeeman transition  
$|1\rangle = |5S_{1/2}, F=2, m_F=2\rangle$, 
$|2\rangle = |5P_{1/2}, F=2, m_F=1\rangle$, 
$|3\rangle = |5P_{3/2}, F=3, m_F=3\rangle$, and 
$|4\rangle = |4D_{3/2}, F=3, m_F=2\rangle$. 
Effective single-Zeeman configurations can also be realized in warm vapors through optical pumping and polarization-selective schemes~\cite{Micalizio}. The spontaneous decay rates together with the squared Clebsch–Gordan (CG) coefficients are  
$\Gamma_{21} = 0.95\Gamma$ (1/6),  
$\Gamma_{31} = 1.00\Gamma$ (1),  
$\Gamma_{42} = 0.30\Gamma$ (2/3), and  
$\Gamma_{43} = 0.05\Gamma$ (1/5). 
A detailed discussion is presented in Sec.~\ref{Sec. 5}.

%%%%%%%%%%%%%%%%%%%%%%%%%%%%%%%%%%%%%%%%%%%%%%%%%%%%%%%%%%%%%%%%%%%%%%%%%%%%%%%%%%%%%%%%%%%%%%%%%%%%%
%%%%%%%%%%%%%%%%%%%%%%%%%%%%%%%%%%%%%%%%%%%%%%%%%%%%%%%%%%%%%%%%%%%%%%%%%%%%%%%%%%%%%%%%%%%%%%%%%%%%%

\section{Spectral Decomposition of Biphotons} \label{Sec. 3}

\subsection{Paired and unpaired spectra} \label{Sec. 3a}

The generation rate $R_{s(i)}$ of the signal (idler) field is defined as 
$R_{s(i)}=(c/L)\langle \hat{a}^{\dagger}_{s(i)}(L,t)\hat{a}_{s(i)}(L,t) \rangle$, 
where the prefactor $c/L$ converts the mean photon number at $z=L$ into a photon generation rate, consistent with the vacuum transit time $L/c$ implicit in our field normalization. Applying the inverse Fourier transform to Eq.~\eqref{eq(11)} and using the commutation relation $[\tilde{a}_{s(i)}(L,\omega),\tilde{a}^\dagger_{s(i)}(L,-\omega')] = (L/2\pi c)\,\delta(\omega+\omega')$, the generation rates are obtained as
\begin{align}
	R_s = & 
	\int^{\infty}_{-\infty} \frac{d\omega}{2\pi} \bigg(
	|B_1|^2
	+\sum_{jk,j'k'}
	\int^L_0 dz
	P^{\ast}_{jk}\mathscr{D}_{kj,j'k'}P_{j'k'}
	\bigg)
	\notag\\
	= & \int^{\infty}_{-\infty} d\omega 
	\left[ \tilde{R}_{s,p}(\omega) + \tilde{R}_{s,u}(\omega) \right],
	\label{eq(16)}\\
	R_i = & 
	\int^{\infty}_{-\infty} \frac{d\omega}{2\pi} \bigg(
	|C_1|^2
	+\sum_{jk,j'k'}
	\int^L_0 dz
	Q_{jk}\mathscr{D}_{jk,k'j'}Q^{\ast}_{j'k'} 
	\bigg)
	\notag\\
	= & \int^{\infty}_{-\infty} d\omega 
	\left[ \tilde{R}_{i,p}(\omega) + \tilde{R}_{i,u}(\omega) \right]. 
	\label{eq(17)}
\end{align}

We therefore identify the frequency-resolved quantities $\tilde{R}_{s,p}(\omega)$ and $\tilde{R}_{s,u}(\omega)$ as the paired and unpaired spectral densities of the signal field, with analogous definitions for the idler. Their frequency-integrated counterparts, $R_{s,p}$ and $R_{s,u}$, represent the corresponding photon generation rates. In Eqs.~(\ref{eq(16)}) and~(\ref{eq(17)}), the functions $P=A_2\zeta^s_{jk}+B_2\zeta^i_{jk}$ and $Q=C_2\zeta^s_{jk}+D_2\zeta^i_{jk}$ encode the coupling between field propagation and Langevin noise operators, while $\mathscr{D}_{jk,k'j'}$ and $\mathscr{D}_{kj,j'k'}$ denote the associated diffusion coefficients describing noise correlations (see Appendix~\ref{ApxB}). The two terms in each expression correspond to physically distinct processes. The pairing components $R_{s,p}$ and $R_{i,p}$ are governed by $B_1$ and $C_1$ and describe correlated photon pairs, whereas the unpairing components $R_{s,u}$ and $R_{i,u}$ arise from Langevin noise fluctuations and represent uncorrelated photons~\cite{Shiu1}.

%%%%%%%
\figtwo
%%%%%%%

Figures~\ref{fig2}(a)–\ref{fig2}(c) show the pairing ($R_{s,p}$, $R_{i,p}$), unpairing ($R_{s,u}$, $R_{i,u}$), and total ($R_s$, $R_i$) spectra of the signal and idler photons, calculated from Eqs.~(\ref{eq(16)}) and~(\ref{eq(17)}). All matrix coefficients including $B_{1}$ and $C_{1}$ can be evaluated under the GSA to simplify the analytical structure, while the diffusion coefficients $\mathscr{D}_{kjj'k'}$ and $\mathscr{D}_{jkk'j'}$ are computed using the exact atomic populations since applying the GSA to these terms would not provide an accurate description of diffusion (see Appendix~\ref{ApxB}). The unpairing rates $R_{s,u}$ and $R_{i,u}$ are fully determined by these diffusion coefficients, which encode the Langevin noise correlations of the atomic medium. As shown in Fig.~\ref{fig2}(a), $\tilde{R}_{s,p}(\omega)$ and $\tilde{R}_{i,p}(\omega)$ coincide because weak pumping in the GSA regime ensures that atoms return to $|1\rangle$ after each SFWM cycle. The overall pairing rate is defined as $R_{s,p} = R_{i,p}\equiv R_p$. In contrast, the unpairing spectra exhibit substantial differences as displayed in Fig.~\ref{fig2}(b).

The signal unpairing spectrum $\tilde{R}_{s,u}(\omega)$ is comparable in magnitude to the pairing spectra, whereas $\tilde{R}_{i,u}(\omega)$ is much weaker owing to asymmetric absorption between the two fields.  As illustrated in Fig.~\ref{fig1}(a), idler photons can resonantly drive the $|1\rangle \leftrightarrow |3\rangle$ transition and be reabsorbed by the atoms, followed by spontaneous emission that isotropically scatters photons into the environment. The corresponding signal photons thus lose their partners. In contrast, the medium is nearly transparent to the signal field since $\langle\hat{\sigma}_{33}\rangle \approx \langle\hat{\sigma}_{44}\rangle \approx 0$. This behavior is consistent with the fluctuation–dissipation picture \cite{Kubo, Garrison}, where dissipation of one field is accompanied by quantum noise that couples into its partner, leading to the rate imbalance observed in Fig.~\ref{fig2}(c). 

Figure~\ref{fig2}(d) presents the total generation rates as a function of OD. Under large $\Delta_1$, both $R_s$ and $R_i$ scale linearly with OD and depend on the spontaneous decay rates together with the squared CG coefficients of the relevant transitions.  For the 1529~nm signal and 780~nm idler configuration (main panel), the $|4\rangle \rightarrow |3\rangle$ decay channel is relatively weak ($\Gamma_{43}=0.05\Gamma$, squared CG value 1/5), which limits the parametric gain and therefore the biphoton generation rate. For comparison, the 1367~nm (signal) and 780~nm (idler) configuration shown in the inset of Fig.~\ref{fig2}(d) benefits from a much stronger $|4\rangle \rightarrow |3\rangle$ decay channel ($\Gamma_{43}=0.33\Gamma$, squared CG value 1/2), resulting in substantially higher generation rates. In this case, $R_s = 3.8\times 10^6~\mathrm{s}^{-1}$ and $R_i = 1.9\times 10^6~\mathrm{s}^{-1}$ at $\mathrm{OD}=1000$. The scaling relations $R_p \propto \mathrm{OD}$ and $R_{u,s}, R_{u,i} \propto \sqrt{\mathrm{OD}}$ indicate that higher OD suppresses the fraction of unpaired photons~\cite{Shiu1}. The main-panel simulations use the $^{87}$Rb levels defined in Sec.~\ref{Sec. 2}. For the inset configuration (1367~nm + 780~nm), the relevant sublevels are 
$|1\rangle = |5S_{1/2}, F=2, m_F=2\rangle$,  
$|2\rangle = |5P_{1/2}, F=1, m_F=1\rangle$,  
$|3\rangle = |5P_{3/2}, F=3, m_F=3\rangle$, and  
$|4\rangle = |6S_{1/2}, F=2, m_F=2\rangle$,  
with spontaneous decay rates and squared CG coefficients  
$\Gamma_{21} = 0.95\Gamma$ (1/2),  
$\Gamma_{31} = 1.00\Gamma$ (1),  
$\Gamma_{42} = 0.17\Gamma$ (1/2), and  
$\Gamma_{43} = 0.33\Gamma$ (1/2).  
All simulations adopt the same detunings and decoherence parameters introduced in Sec.~\ref{Sec. 2}.

%%%%%%%%%%%%%%%%%%%%%%%%%%%%%%%%%%%%%%%%%%%%%%%%%%%%%%%%%%%%%%%%%%%%%%%%%%%%%%%%%%%%%%%%%%%%%%%%%%%%%
%%%%%%%%%%%%%%%%%%%%%%%%%%%%%%%%%%%%%%%%%%%%%%%%%%%%%%%%%%%%%%%%%%%%%%%%%%%%%%%%%%%%%%%%%%%%%%%%%%%%%

\subsection{Pairing ratio and uncorrelated background}
\label{Sec. 3b}

%%%%%%%%%
\figthree
%%%%%%%%%

The biphoton pairing rate $R_p$ constitutes the dominant contribution to the detected biphoton signal. Combining Eqs.~(\ref{eq(13)}) and~(\ref{eq(16)}) yields the exact integral expression
\begin{align}
	R_p 
	=\frac{1}{2\pi}
	\int^{\infty}_{-\infty} 
	e^{-\textrm{Re}\left[\Gamma_i\right]L}
	\left|\frac{2\kappa_s}{\Gamma_i}
	\textrm{sinh}\!\left(\frac{\Gamma_i L}{2}\right)\right|^2 d\omega,
	\label{eq(18)}
\end{align}
which contains two distinct physical contributions. The exponential factor describes two-level absorption governed by $\Gamma_i$, while the hyperbolic sine term captures collectively enhanced stimulated emission.

Figures~\ref{fig3}(a) and~\ref{fig3}(b) show the pairing rate spectrum $R_p$ (black solid), together with its absorption (green dash-dotted) and emission (yellow dashed) components, for optical depths of $0.1$ and $10$, respectively. At low OD [Fig.~\ref{fig3}(a)], the medium behaves as an ensemble of independent atoms, and the full width at half maximum (FWHM) of the $R_p$ spectrum is set by the spontaneous decay rate $\Gamma$ associated with the $|3\rangle\rightarrow|1\rangle$ transition. As the OD increases, both absorption and collectively enhanced emission become stronger, resulting in a broadened, double-peaked spectrum, as shown in the inset of Fig.~\ref{fig3}(b).

In the regime $\mathrm{OD}\gg1$, the pairing rate can be further simplified. Rewriting Eq.~(\ref{eq(18)}) in the equivalent form
\begin{align}
	R_p
	= & \frac{1}{2\pi}
	\left|
	\frac{\kappa_s}{\Gamma_i}
	\right|^{2}
	\int_{-\infty}^{\infty}
	\left(
	1-e^{-\Gamma_i L}-e^{-\Gamma_i^{\ast} L}
	+e^{-2\mathrm{Re}[\Gamma_i]L}
	\right)
	\, d\omega,
	\label{eq(19)}
\end{align}
and substituting the explicit frequency dependence of $\Gamma_i$ from Eq.~(\ref{eq(7)}), the integral can be evaluated analytically under the large-OD approximation. The resulting compact analytical expression for the pairing rate is given by
\begin{align}
	R_{p,\mathrm{A}} =
	\sqrt{\frac{\alpha\gamma_{31}\Gamma_{43}}{16\Gamma_{31}}}
	\left(\sqrt{\frac{\alpha\Gamma_{31}}{\gamma_{31}}}-\frac{2}{\sqrt{\pi}}\right)
	\left|\frac{\Omega_c\Omega_d}{\Delta_1(\gamma_{41}+2i\Delta_2)}\right|^2,
	\label{eq(20)}
\end{align}
as derived in Appendix~\ref{ApxD}.

Figure~\ref{fig3}(c) compares the numerical evaluation of Eq.~(\ref{eq(18)}) (black dashed) with the analytical approximation $R_{p,\mathrm{A}}$ (purple solid), showing excellent agreement. For the $1529~\mathrm{nm}$ (signal) and $780~\mathrm{nm}$ (idler) configuration, the pairing rate reaches $2.8\times10^{5}\,\mathrm{s}^{-1}$ at $\mathrm{OD}=1000$. These moderate values arise from the relatively small decay rate $\Gamma_{43}$ of the selected transition. As shown in the inset of Fig.~\ref{fig3}(c), the $1367~\mathrm{nm}$–$780~\mathrm{nm}$ configuration yields substantially larger pairing rates due to the larger $|4\rangle\rightarrow|3\rangle$ decay rate under otherwise identical driving conditions.

The pairing ratio for the signal or idler field is defined as
\begin{align}
	r_{p,s(i)} \equiv \frac{R_{s(i),p}}{R_{s(i)}} = \frac{R_p}{R_{s(i)}},
	\label{eq(21)}
\end{align}
and represents the fraction of correlated photons within the total detected counts. An ideal biphoton source yields $r_{p,s(i)}=100\%$ in the absence of unpaired photons. In realistic diamond-type SFWM systems, dissipation and vacuum fluctuations introduce unavoidable unpaired noise, thereby reducing the pairing ratio. At the same time, collective enhancement at large OD amplifies the paired contribution more rapidly than the unpaired background. As a result, the pairing ratios for both fields increase with OD despite the presence of dissipation, as shown in Fig.~\ref{fig3}(d), consistent with collective enhancement observed in double-$\Lambda$ systems~\cite{Shiu1}.

A pronounced asymmetry nevertheless develops between the signal and idler pairing ratios. This asymmetry does not originate from any difference in the underlying biphoton generation process, but rather from asymmetric propagation loss and detection conditioning~\cite{Shiu3}. The idler photons resonantly couple to the $|1\rangle\leftrightarrow|3\rangle$ transition and are subject to reabsorption during propagation. Consequently, only idler photons that survive this two-level absorption process remain paired with their signal partners. Conditioning on idler detection therefore preferentially selects intact photon pairs, causing $r_{p,i}$ to approach unity as the OD increases.

In contrast, signal photons drive the $|3\rangle\leftrightarrow|4\rangle$ transition, where the population is negligible, and thus experience minimal absorption. Signal photons can therefore be detected even when their idler partners have already been removed by reabsorption, leading to a reduced signal pairing ratio. Under the conditions of Fig.~\ref{fig3}(d), $r_{p,s}$ saturates near $43\%$ because the loss of idler photons breaks the signal–idler correlations.

The same correlated-pair generation therefore appears asymmetric solely due to conditioning, appearing as $r_{p,i}\!\approx\!100\%$ when conditioning on idler detection and as $r_{p,s}\!\approx\!43\%$ when conditioning on signal detection. At $\mathrm{OD}=100$, the idler pairing ratio reaches $r_{p,i}\!\approx\!93\%$, indicating that idler detection heralds a highly pure telecom-band single photon and highlighting the potential of this platform for quantum-network applications.

%%%%%%%%%%%%%%%%%%%%%%%%%%%%%%%%%%%%%%%%%%%%%%%%%%%%%%%%%%%%%%%%%%%%%%%%%%%%%%%%%%%%%%%%%%%%%%%%%%%%%
%%%%%%%%%%%%%%%%%%%%%%%%%%%%%%%%%%%%%%%%%%%%%%%%%%%%%%%%%%%%%%%%%%%%%%%%%%%%%%%%%%%%%%%%%%%%%%%%%%%%%

\section{Superradiant Biphoton Dynamics} \label{Sec. 4}

\subsection{General solution and experimental comparison} \label{Sec. 4a}

%%%%%%%%%
\figfour
%%%%%%%%% 

To characterize the temporal cross correlations, we introduce the second-order
correlation function
$G^{(2)}_{s\textrm{-}i}(\tau)=
\big\langle 
\hat{a}^{\dagger}_s(L,t)
\hat{a}^{\dagger}_i(L,t+\tau)
\hat{a}_i(L,t+\tau)
\hat{a}_s(L,t)
\big\rangle$.
Using the field solutions in Eq.~(\ref{eq(11)}) together with Wick’s theorem \cite{Louisell}, the correlation function becomes
\begin{align}
	G^{(2)}_{s\textrm{-}i}(\tau)
	= &  
	\bigg(\frac{L}{c}\bigg)^2 R_s R_i
	+\bigg|\frac{L}{2\pi c} 
	\int e^{i\omega \tau}d\omega\bigg( B_1D_1^{\ast} \notag\\
	& + \sum_{jk,j'k'}\int dz\, 
	Q^{\ast}_{jk} \mathscr{D}_{jkk'j'} P_{j'k'}\bigg)\bigg|^2,
	\label{eq(22)}
\end{align}
The first term represents the accidental-coincidence background arising from photons belonging to different, uncorrelated pairs. The second term contains both paired and noise-induced contributions arising from the same microscopic evolution. These contributions cannot, in general, be uniquely separated at the level of the correlation function, and together determine the observed temporal profile of the biphoton correlations. The signal-side coincidence rate is $R_{\mathrm{C},s}(\tau)=(c/L)^2 \Delta T G^{(2)}_{s\textrm{-}i}(\tau)$, where $\Delta T$ denotes the coincidence time window. Choosing $\Delta T=1/R_s$ effectively selects events containing a single signal photon, yielding

\begin{align}
	R_{\mathrm{C},s}(\tau)=
	\left(\frac{c}{L}\right)^2 
	\frac{G^{(2)}_{s\textrm{-}i}(\tau)}{R_s}.
	\label{eq(23)}
\end{align}
The corresponding idler-side expression, $R_{\mathrm{C},i}(\tau)=(c/L)^2 G^{(2)}_{i\textrm{-}s}(\tau)/R_i$, is derived in Appendix~\ref{ApxE}.

Figures~\ref{fig4}(a) and~\ref{fig4}(b) show $R_{\mathrm{C},s}$ as functions of the idler delay $\tau$ for $\mathrm{OD}=0.1$ and $\mathrm{OD}=10$. Exponential fits (see captions) indicate that the decay time $\tau_d$ decreases with increasing OD. In the small-OD limit, where independent-atom behavior dominates, $\tau_d \approx 26 \ \mathrm{ns}$, consistent with the inverse spontaneous-emission rate $1/\Gamma$ associated with the $|1\rangle\leftrightarrow|3\rangle$ transition. This result also agrees with the FWHM of the low-OD spectrum in Fig.~\ref{fig3}(a).

As the OD increases [Fig.~\ref{fig4}(b)], $\tau_d$ shortens to $7\ \mathrm{ns}$ because the linear-response parameter $\Gamma_i$, which governs the $|1\rangle\leftrightarrow|3\rangle$ transition under large detuning $\Delta_1$, dominates the biphoton dynamics. In this regime, the SFWM-driven atomic dipoles radiate cooperatively, producing the superradiant behavior illustrated in Fig.~\ref{fig1}(b). Constructive interference among the collectively enhanced stimulated emission amplitudes leads to a brighter and temporally compressed correlation, which also corresponds to broader spectral bandwidths, consistent with the behavior in Fig.~\ref{fig3}(b).

We now demonstrate quantitative agreement between our theory and previous experimental observations. Figure~\ref{fig4}(c) shows the theoretical $R_{\mathrm{C},s}$ at $\mathrm{OD}=25$, which matches the measured waveform shown in Fig.~\ref{fig4}(d). The extracted $1/e$ decay times, 3.3 ns and 3.2 ns, agree closely. The OD dependence of $\tau_d$ in Fig.~\ref{fig4}(e) also reproduces the experimental trend shown in Fig.~\ref{fig4}(f).

When collective effects dominate at large OD, the decay time exhibits the approximate scaling $\tau_d\propto1/\mathrm{OD}$, characteristic of the superradiant regime \cite{Dicke,Rehler,Gross}. In contrast, in the low-OD limit the system approaches independent-atom behavior, with $\tau_d$ tending toward $1/\gamma_{31}$ as $\mathrm{OD}\rightarrow 0$. Figures~\ref{fig4}(c) and~\ref{fig4}(e) are simulated using the backward-field configuration and the corresponding experimental energy-level scheme. For parameters not specified in the experiment, we set $\Omega_c=\Omega_d=1\Gamma$, $\Delta_1=-50\Gamma$, and $\Delta_2=0$. The biphoton waveform and decay time are primarily determined by OD, and differences between forward and backward propagation are discussed in Appendix~\ref{ApxE}.

%%%%%%%%%%%%%%%%%%%%%%%%%%%%%%%%%%%%%%%%%%%%%%%%%%%%%%%%%%%%%%%%%%%%%%%%%%%%%%%%%%%%%%%%%%%%%%%%%%%%%
%%%%%%%%%%%%%%%%%%%%%%%%%%%%%%%%%%%%%%%%%%%%%%%%%%%%%%%%%%%%%%%%%%%%%%%%%%%%%%%%%%%%%%%%%%%%%%%%%%%%%

\subsection{Analytical reduction and physical picture} \label{Sec. 4b}

%%%%%%%%
\figfive
%%%%%%%%

The connection between the biphoton correlation width in the diamond-type system and superradiant dynamics can be understood from the correlation formalism. We introduce the superradiant scaling time $\tau_s=\tau_{31}/(1+N\mu)$, where $\tau_{31}=1/\Gamma_{31}$ is the lifetime of the $|3\rangle\rightarrow|1\rangle$ transition and $\mu$ is the geometric shape factor of the ensemble~\cite{Rehler}.  In the negligible-diffraction limit, $\tau_s$ becomes (see Appendix~\ref{ApxG})
\begin{align}
	\tau_s=\frac{\tau_{31}}{1+\alpha/4},
	\label{eq(24)}
\end{align}
where $\alpha$ is the corresponding OD on the $|3\rangle \leftrightarrow |1\rangle$ transition, as defined in Sec.~\ref{Sec. 2}. According to Ref.~\cite{Rehler}, $\tau_s$ represents the superradiant $1/e$ lifetime in the low-excitation limit. In the present diamond-type biphoton system, this identification is justified under the GSA condition, where the population remains predominantly in the ground state. Figure~\ref{fig5}(a) compares the $1/e$ decay time $\tau_d$ extracted from the full solution of $G^{(2)}_{s\textrm{-}i}(\tau)$ (black curve) with the superradiant prediction $\tau_s$ of Eq.~(\ref{eq(24)}) (red dots). The two results agree closely across a wide range of OD, thereby establishing a quantitative connection between the biphoton decay time and the superradiant timescale. A small deviation remains, as highlighted in the inset of Fig.~\ref{fig5}(a). This deviation likely originates from the more complex FWM dynamics of the diamond-type configuration beyond the simple two-level superradiance model~\cite{Rehler, Gross}, as well as from nonlinear propagation effects included in the Maxwell--Schr\"{o}dinger framework. Here the superradiant lifetime refers to a Dicke-type collective radiative decay governed by geometry and propagation, rather than by near-field dipole--dipole interactions between atoms.

To clarify the physical content of the temporal cross-correlation, it is useful to rewrite Eq.~(\ref{eq(22)}) in a form that explicitly separates the accidental-coincidence background from the correlated contribution. This yields
\begin{align}
	G^{(2)}_{s\textrm{-}i,\mathrm{A}}(\tau)= 
	\left(\frac{L}{c}\right)^2 
	\left[ R_sR_i + \left|\Psi_{s\textrm{-}i}(\tau)\right|^2\right],
	\label{eq(25)}
\end{align}
where $\Psi_{s\textrm{-}i}(\tau)$ denotes the quantity inside the absolute square in Eq.~(\ref{eq(22)}). Its explicit expression is obtained by substituting the field coefficients and is given in Appendix~\ref{ApxE}. 

In the large-OD limit, the biphoton wave function $\Psi_{s\textrm{-}i}(\tau)$ admits a compact analytical form,
\begin{align}
	& \Psi_{s\textrm{-}i}(\tau)
	=
	\sqrt{\frac{i}{2\pi^2|\tau|^3\tau_{c}}}
	\left(\frac{\kappa_s}{\Gamma_i}\right)
	\notag\\
	& \times\bigg[
	\big(-i\tau+|\tau|\big)\mathrm{Kei}_1\!\left(\sqrt{\Theta}\right)+ \big(\tau-i|\tau|\big)\mathrm{Kei}_1\!\left(\sqrt{-\Theta}\right)
	\notag\\
	& +
	\big(-i\tau-|\tau|\big)\mathrm{Ker}_1\!\left(\sqrt{\Theta}\right)+ \big(\tau+i|\tau|\big)\mathrm{Ker}_1\!\left(\sqrt{-\Theta}\right)
	\bigg],
	\label{eq(26)}
\end{align}
with $\Theta(\tau)=4i|\tau|/\tau_{c}$ and $\tau_{c}=4/(\alpha\Gamma_{31})$. Here $\mathrm{Kei}_\mu(x)$ and $\mathrm{Ker}_\mu(x)$ denote Kelvin functions of the second kind (see Appendix~\ref{ApxE}). The characteristic time $\tau_c$ originates from the two-level spectral coefficient $\Gamma_i$ in Eq.~(\ref{eq(7)}), showing that under large detuning $\Delta_1$ the two-level response of the $|1\rangle\leftrightarrow|3\rangle$ transition dominates the biphoton correlation width and sets the superradiant timescale.

Figure~\ref{fig5}(b) shows the coherent time obtained from the general solution $\tau_d$ (black dashed) and from the analytical approximation $\tau_{d,\mathrm{A}}$ (yellow solid). The two agree within $25\%$ for $10\lesssim\alpha\lesssim30$ and within $3\%$ for $\alpha\gtrsim30$, where the curves nearly overlap. In this regime, the decay time decreases approximately linearly with OD and is well approximated by $\tau_d \approx \tau_{d,\mathrm{A}} \approx \tau_c = 4/(\alpha\Gamma_{31})$, consistent with Eq.~(\ref{eq(26)}).

Figures~\ref{fig5}(c) and~\ref{fig5}(d) compare the analytical (yellow solid) and full (black dashed) expressions of $R_{\mathrm{C},s}$ for $\mathrm{OD}=30$ and $\mathrm{OD}=100$, respectively. The temporal decay obtained from the analytical form agrees well with the full solution. An oscillatory structure appears in both results, as shown in the insets. These oscillations become more pronounced and more closely aligned as OD increases, consistent with superradiant ringing~\cite{Burnham, Kaluzny}, in which emitted photons are reabsorbed by trailing atoms and subsequently re-emitted.  Such behavior can only be captured by a nonlinear propagation model that includes coupled field–matter evolution~\cite{Gross}.

%%%%%%%%%%%%%%%%%%%%%%%%%%%%%%%%%%%%%%%%%%%%%%%%%%%%%%%%%%%%%%%%%%%%%%%%%%%%%%%%%%%%%%%%%%%%%%%%%%%%%
%%%%%%%%%%%%%%%%%%%%%%%%%%%%%%%%%%%%%%%%%%%%%%%%%%%%%%%%%%%%%%%%%%%%%%%%%%%%%%%%%%%%%%%%%%%%%%%%%%%%%

\subsection{High-purity biphotons from superradiant enhancement} \label{Sec. 4c}

Superradiant enhancement not only compresses the biphoton correlation time but also enables highly nonclassical photon-pair generation at high pairing rates. To quantify the nonclassicality, we evaluate the normalized second-order correlation function
\begin{align}
	g^{(2)}_{s\textrm{-}i}(\tau)
	= \frac{G^{(2)}_{s\textrm{-}i}(\tau)}{N_s N_i},
	\label{eq(27)}
\end{align}
where $N_{s(i)}=(L/c)R_{s(i)}$ is the mean photon number of the signal (idler) field within the detection time window associated with the transit time $L/c$. Substituting Eq.~(\ref{eq(22)}) into this definition, the correlation function
can be written as
\begin{align}
	g^{(2)}_{s\textrm{-}i}(\tau)
    = & 1 +\frac{1}{R_s R_i}\bigg|\int \frac{d\omega}{2\pi} e^{i\omega \tau}\bigg( B_1D_1^{\ast} \notag\\
    & + \sum_{jk,j'k'}\int dz Q^{\ast}_{jk} \mathscr{D}_{jkk'j'} P_{j'k'}\bigg)\bigg|^2.
\label{eq(28)}
\end{align}
This function satisfies the symmetry relation $g^{(2)}_{s\textrm{-}i}(\tau)=g^{(2)}_{i\textrm{-}s}(-\tau)$. The peak signal-to-background ratio of the biphoton correlation is then $r_{\mathrm{SB}} = g^{(2)}_{s\textrm{-}i}(0)-1$. When $g^{(2)}_{s\textrm{-}i}(0)\gg 1$, one has $r_{\mathrm{SB}}\approx g^{(2)}_{s\textrm{-}i}(0)$.

A complementary measure of nonclassicality is provided by the Cauchy–Schwarz inequality,
\begin{align}
	\left[ g^{(2)}_{s\textrm{-}i}(0)\right]^2 \left[ g^{(2)}_{s\textrm{-}s}(0)
    g^{(2)}_{i\textrm{-}i}(0)\right]^{-1}\leq 1,
	\label{eq(29)}
\end{align}
where $g^{(2)}_{s\textrm{-}s}(0)$ and $g^{(2)}_{i\textrm{-}i}(0)$ are the autocorrelation functions for the signal and idler fields under zero delay, which can be calculated respectively according to 
\begin{align}
	g^{(2)}_{s\textrm{-}s}(\tau)
	= & 1 +R^{-2}_s\left|\int \frac{d\omega}{2\pi} e^{i\omega \tau}\tilde{R}_s(\omega)\right|^2, 
	\label{eq(30)}\\
	g^{(2)}_{i\textrm{-}i}(\tau)
	= & 1 +R^{-2}_i\left|\int \frac{d\omega}{2\pi} e^{i\omega \tau}\tilde{R}_i(\omega)\right|^2.
   \label{eq(31)}
\end{align}
As proven in Appendix~\ref{ApxH}, the signal and idler fields exhibit thermal statistics with $g^{(2)}_{s\textrm{-}s}(0)=g^{(2)}_{i\textrm{-}i}(0)=2$, consistent with the thermal nature of spontaneous fluorescence~\cite{Mandel,Srivathsan}. Violation of Eq.~(\ref{eq(29)}) is quantified by $F_{\mathrm{CSI}}$, the factor by which the inequality is exceeded. Nonclassical behavior is identified when $r_{\mathrm{SB}}>1$ or $F_{\mathrm{CSI}}>1$.

%%%%%%%%
\figsix
%%%%%%%%

Figures~\ref{fig6}(a) and \ref{fig6}(b) show $F_{\mathrm{CSI}}$ as a function of OD for fixed pairing rates of $R_p=1.0\times10^6\,\mathrm{s}^{-1}$ and $R_p=1.0\times10^5\,\mathrm{s}^{-1}$, respectively. The corresponding values of $r_{\mathrm{SB}}$ are presented in the insets. Both quantities increase with OD due to the inverse scaling $\tau_d\propto1/\mathrm{OD}$ and the resulting enhancement of the pairing rate $R_p$. At $\mathrm{OD}=170$, we obtain $r_{\mathrm{SB}}\approx 600$ and $F_{\mathrm{CSI}}\approx 10^{5}$. These values are comparable to, and in some cases slightly exceed, those reported in double-$\Lambda$ biphoton systems when compared at similar biphoton brightness levels of $\sim10^{3}\,\mathrm{s}^{-1}\,\mathrm{MHz}^{-1}$~\cite{Shiu1}.

Figure~\ref{fig6}(c) directly illustrates this effect by isolating the dependence of the nonclassicality on the pairing rate $R_p$ at a fixed OD. By continuously tuning the driving strengths $\Omega_c$ and $\Omega_d$ at $\mathrm{OD}=170$, we find that the Cauchy--Schwarz surpassing factor follows a clear scaling $F_{\mathrm{CSI}}\propto1/R_p^{2}$. This behavior reflects the suppression of uncorrelated background events at lower generation rates, while the superradiantly enhanced correlated contribution remains intact.

Consistent with this scaling, further enhancement is achievable by reducing the pairing rate. At $R_p=1.0\times10^5\,\mathrm{s}^{-1}$ [Fig.~\ref{fig6}(b)], both $r_{\mathrm{SB}}$ and $F_{\mathrm{CSI}}$ increase by one to two orders of magnitude compared with Fig.~\ref{fig6}(a), reaching values comparable to the highest reported results for solid-state biphoton sources ($r_{\mathrm{SB}}\approx12000$)~\cite{Ma}. These results demonstrate the capability of this platform to generate highly nonclassical telecom-optical biphotons. The required OD are readily accessible with current cold-atom technology~\cite{Blatt,Hsiao}.

Finally, beyond the spectral decomposition discussed above, the pairing ratio can be evaluated independently by integrating the correlated area of the coincidence curve. Defining
\begin{align}
	A_{s(i)}
	= \int_{-\infty}^{\infty} d\tau\,
	R_{i(s)}\big[g^{(2)}_{s\textrm{-}i(i\textrm{-}s)}(\tau)-1\big],
	\label{eq(32)}
\end{align}
the quantity $A_{s(i)}$ represents the average number of photons correlated with a post-selected signal (idler) photon.  Figure~\ref{fig6}(d) shows 
$R_{i(s)}[g^{(2)}_{s\textrm{-}i(i\textrm{-}s)}(\tau)-1]$ versus the idler delay
$\tau$. At $\mathrm{OD}=100$, we obtain $A_s=0.41$ and $A_i=0.93$, in close agreement with the pairing ratios $r_{p,s}$ and $r_{p,i}$ in Fig.~\ref{fig3}(d). Thus, when conditioning on a single detected photon, the corresponding correlated partner appears with average probabilities of 41\% (signal) and 93\% (idler), fully consistent with the behavior shown in Fig.~\ref{fig3}(d).

%%%%%%%%%%%%%%%%%%%%%%%%%%%%%%%%%%%%%%%%%%%%%%%%%%%%%%%%%%%%%%%%%%%%%%%%%%%%%%%%%%%%%%%%%%%%%%%%%%%%%
%%%%%%%%%%%%%%%%%%%%%%%%%%%%%%%%%%%%%%%%%%%%%%%%%%%%%%%%%%%%%%%%%%%%%%%%%%%%%%%%%%%%%%%%%%%%%%%%%%%%%
 
\section{Warm Atomic Vapors with Doppler Broadening} \label{Sec. 5}

\subsection{Theoretical framework for Doppler-broadened atomic ensembles} \label{Sec. 5a}

In contrast to the cold-atom model considered in the preceding sections, biphotons generated in a warm atomic vapor are additionally affected by the thermal motion of atoms at finite temperature. The atomic velocities follow a Maxwell--Boltzmann distribution, giving rise to a Doppler detuning $\Delta_{\mathrm{D}}=k v$ for each atom with velocity $v$, where $k$ denotes the wave vector of the relevant optical field. Under this condition, the interaction Hamiltonian of the Doppler-broadened system can be written as
\begin{align}
	& \hat{H}' =  -\frac{\hbar N}{2L}\int^L_0  \big[
	\Omega_c\hat{\sigma}_{21}(z,t)+\Omega_d \hat{\sigma}_{42}(z,t) 
	\notag\\
	& + 
	2g_s\hat{\sigma}_{43}(z,t)\hat{a}_s(z,t)+2g_i\hat{\sigma}_{31}(z,t)\hat{a}_i(z,t) +
	\Delta'_1\hat{\sigma}_{22}(z,t)
	\notag\\
	& +\Delta'_2\hat{\sigma}_{44}(z,t) +\Delta'_{3}\hat{\sigma}_{33}(z,t)+\mathrm{H.c.} \big]dz,
	\label{eq(33)}
\end{align}
where, for a forward-propagating configuration, the detunings are given by $\Delta'_1=\Delta_1+k_c v$, $\Delta'_2=\Delta_2+(k_c+k_d)v$, and $\Delta'_3=k_i v$. Compared with Eq.~(\ref{eq(1)}), the Doppler effect introduces velocity-dependent detunings for all optical transitions. As a result, not only the one-photon and two-photon detunings $\Delta'_1$ and $\Delta'_2$ are modified, but an additional relative detuning $\Delta'_3$ between the signal and idler transitions also appears. This term vanishes identically in the cold-atom limit.

Under the GSA condition, incorporation of the Maxwell--Boltzmann velocity distribution modifies the parametric coefficients of the MSEs, which consequently take the Doppler-broadened forms
\begin{align}
	\Gamma'_{i} = &  
	\frac{\alpha\Gamma_{31}}{2L[\gamma_{31}-2i(\omega+\Delta'_3)]} -\frac{i\omega}{c}+i\Delta k, 
	\label{eq(34)}\\
	\kappa'_{s} = &	    
	\frac{i\alpha\sqrt{s_\lambda\Gamma_{43}\Gamma_{31}}\Omega_c\Omega_d}{4L\Delta'_1(\gamma_{41}-2i\Delta'_2)[\gamma_{31}-2i(\omega+\Delta'_3)]},
	\label{eq(35)}\\
	\kappa'_{i} = & 
	\frac{i\alpha\sqrt{s_\lambda\Gamma_{43}\Gamma_{31}}\Omega^{\ast}_c\Omega^{\ast}_d}{4L\Delta'_1(\gamma_{41}+2i\Delta'_2)[\gamma_{31}-2i(\omega+\Delta'_3)]},
	\label{eq(36)}\\
	G'_{s} = & 
	\frac{-\alpha s_\lambda\Gamma_{43}|\Omega_c|^2|\Omega_d|^2}{8L \Delta'^2_1(\gamma_{41}+2i\Delta'_2)[\gamma_{43}-2i(\Delta'_2+\omega+\Delta'_3)]}\notag\\
	& \times \frac{1}{[\gamma_{31}-2i(\omega+\Delta'_3)]}-\frac{i\omega}{c}.
	\label{eq(37)}
\end{align}
The corresponding Doppler broadened Langevin noise coefficients are summarized in Appendix \ref{ApxI}.

The general solutions for the signal and idler fields retain the same formal structure as Eq.~(\ref{eq(11)}), with all parametric coefficients replaced by their Doppler-broadened counterparts. The elements of the first matrix read
\begin{align}
	A'_{1} & =1, \label{eq(38)}\\
	B'_{1} & =
	\left(\frac{\int\kappa'_{s}}{\int\Gamma'_{i}}\right)\left(e^{-\int\Gamma'_{i} L}-1\right), 
	\label{eq(39)}\\
	C'_{1} & =\left(\frac{\int\kappa'_{i}}{\int\Gamma'_{i}}\right)\left(e^{-\int\Gamma'_{i} L}-1\right),  
	\label{eq(40)}\\
	D'_{1} & =e^{-\int\Gamma'_{i} L}. 
	\label{eq(41)}
\end{align}
Because the Doppler-induced detunings introduce velocity-dependent phase evolution into the field propagation, the contributions associated with different atomic velocity classes acquire rapidly varying relative phases. As a result, cross-velocity interference terms do not survive in experimentally relevant observables. The Doppler effect is therefore fully accounted for by averaging the velocity-dependent field-transfer coefficients over the thermal velocity distribution. In Eqs.~(\ref{eq(38)})--(\ref{eq(41)}), we introduce the compact notation $\int[\bullet]\equiv\int p(v)[\bullet]\,dv$, where $[\bullet]$ denotes the parametric coefficients in Eqs.~(\ref{eq(34)})--(\ref{eq(37)}) and $p(v)=\sqrt{\frac{M}{2\pi k_B \mathrm{T}}}\exp\!\left(-\frac{Mv^2}{2k_B \mathrm{T}}\right)$ is the normalized Maxwell--Boltzmann distribution at temperature $\mathrm{T}$.

In the following, we focus on Doppler-broadened biphoton generation with signal and idler wavelengths of 1529~nm and 780~nm, respectively. The corresponding energy-level configuration is described in Sec.~\ref{Sec. 2b}. The detunings are chosen as $\Delta_1=-500\Gamma$ and $\Delta_2=0$ to ensure that all laser-driven transitions remain far from the Doppler-broadened absorption profile, whose width is approximately $100\Gamma$ at room temperature.

%%%%%%%%%%%%%%%%%%%%%%%%%%%%%%%%%%%%%%%%%%%%%%%%%%%%%%%%%%%%%%%%%%%%%%%%%%%%%%%%%%%%%%%%%%%%%%%%%%%%%
%%%%%%%%%%%%%%%%%%%%%%%%%%%%%%%%%%%%%%%%%%%%%%%%%%%%%%%%%%%%%%%%%%%%%%%%%%%%%%%%%%%%%%%%%%%%%%%%%%%%%

\subsection{Doppler-broadened spectra} \label{Sec. 5b}

%%%%%%%%%
\figseven
%%%%%%%%%

The signal and idler generation rates are derived below, with detailed derivations given in Appendix~\ref{ApxI}.
\begin{align}
	R'_{s} = & \int \frac{d\omega}{2\pi} \bigg[
	|B'_{1}|^2 +\sum_{jk,j'k'} \int^L_0 dz \bigg(A'^\ast_{2} X^{ss}_{S} A'_{2}
	\notag\\
	& 
	+ B'^\ast_{2} X^{is}_{S} A'_{2} 
	+ A'^\ast_{2} X^{si}_{S} B'_{2}
	+ B'^\ast_{2} X^{ii}_{S} B'_{2}
	\bigg)\bigg]
    \notag\\
	\equiv & \int \frac{d\omega}{2\pi}  
	\big[\tilde{R}'_{s,p}(\omega)+\tilde{R}'_{s,u}(\omega)\big],
	\label{eq(42)}
\end{align}
\begin{align}
	R'_{i} = & \int \frac{d\omega}{2\pi} 
	\bigg[
	|C'_{1}|^2
	+\sum_{jk,j'k'} \int^L_0 dz \bigg(C'^\ast_{2} X^{ss}_{I} C'_{2}
	\notag\\
	&
	+ D'^\ast_{2} X^{is}_{I} B'_{2} 
	+ C'^\ast_{2} X^{si}_{I} D'_{2}
	+ D'^\ast_{2} X^{ii}_{I} D'_{2}
	\bigg)\bigg]
    \notag\\
	\equiv & \int \frac{d\omega}{2\pi}   \big[\tilde{R}'_{i,p}(\omega)+\tilde{R}'_{i,u}(\omega)\big],
	\label{eq(43)}
\end{align}
where
\begin{align}
	X^{s(i)s(i)}_{S} = \int  p(v) \zeta'^{s(i)\ast}_{jk}
	\mathscr{D}_{kj,j'k'} \zeta'^{s(i)}_{j'k'} dv,
	\label{eq(44)}\\
	X^{s(i)s(i)}_{I} = \int  p(v) \zeta'^{s(i)}_{jk}
	\mathscr{D}_{jk,k'j'} \zeta'^{s(i)\ast}_{j'k'} dv.
	\label{eq(45)}
\end{align}
The Doppler-broadened signal and idler rates arising from the parametric contribution can be expressed as
\begin{align}
	R'_{s(i),p}
	=  &\;
	\int \frac{d\omega}{2\pi}
	e^{-\textrm{Re}\left[\int\Gamma'_{i}\right]L}
    \left|\frac{2\int\kappa'_{s(i)}}{\int\Gamma'_{i}}
	\textrm{sinh}\left(\frac{\int\Gamma'_{i} L}{2}\right)\right|^2.
	\label{eq(46)}
\end{align}

Figure \ref{fig7}(a) presents the spectra of $R'_{s,p}$ and $R'_{i,p}$ calculated from Eq. (\ref{eq(46)}) at different temperatures with $\textrm{OD}=1000$. As the temperature increases, the complicated spectral structure near the center is gradually smoothed out due to the interference among contributions associated with different Doppler detunings. When $\mathrm{T}=10^{-4}\textrm{ K}$, both Doppler-broadened spectra become identical and recover the predictions of the cold atom model, namely $R'_{s,p}\approx R'_{i,p}\approx R_p$. In contrast, as the temperature increases to $300\textrm{ K}$, the signal rate becomes larger than the idler rate, i.e., $R'_{s,p} > R'_{i,p}$. This asymmetry reflects the fact that Doppler broadening affects the signal and idler generation channels in an intrinsically different manner. Specifically, it originates from the discrepancy between the Doppler-broadened coupling constants $\int\kappa'_s$ and $\int\kappa'_i$. We therefore define the pairing rate in the warm atomic ensemble as $R'_{p}=R'_{i,p}$.

For $\mathrm{T}=10^{-4}\textrm{ K}$, the calculated Doppler-broadened pairing rate recovers the cold atom result obtained from Eq. (\ref{eq(18)}), yielding $R'_{p}\approx R_p \approx 1.5\times 10^6\textrm{ s}^{-1}$. As the temperature increases to $300\textrm{ K}$, the pairing rate decreases drastically by approximately five orders of magnitude to about $20\textrm{ s}^{-1}$. This reduction primarily originates from the Doppler term $(k_c+k_d) v$, which leads to an effective two photon detuning $\Delta'_{2}$. This behavior can be understood from the analytical solution of the cold atom model in Eq. (\ref{eq(20)}), which indicates that the pairing rate is strongly suppressed by the factor $1/(2\Delta_1\Delta_2)^2$ when $\Delta_2\gg \gamma_{41}$. To illustrate this effect, we estimate an effective two photon detuning of approximately $30\Gamma$ based on the full width at half maximum of the Maxwell--Boltzmann distribution at $\mathrm{T}=300\textrm{ K}$ and substitute it into Eq. (\ref{eq(18)}). This yields a pairing rate of $45\textrm{ s}^{-1}$ for $\Delta_2=30\Gamma$, which is consistent with the magnitudes obtained at $3\textrm{ K}$ and $300\textrm{ K}$. These results demonstrate that the Doppler induced $\Delta'_2$ plays a dominant role in reducing the pairing rate of diamond biphoton generation in warm atomic ensembles, even in the cryogenic temperature range, as shown in Fig. \ref{fig7}(b).

The temperature dependence of the signal and idler pairing ratios $r'_{p,s}$ and $r'_{p,i}$ is shown in Fig. \ref{fig7}(c). Since $R'_{s,p} > R'_{i,p}$, the signal pairing ratio is defined as $r'_{p,s}=R'_{i,p}/R'_{s}$, while the definition of the idler pairing ratio remains unchanged. Similar to the pairing rate, both pairing ratios decrease with increasing temperature, reaching approximately $0.1\%$ and $7\%$ for $r'_{p,s}$ and $r'_{p,i}$ at $\mathrm{T}=300\textrm{ K}$. This reduction can be attributed to the effective decrease of OD in the warm atomic vapor, which suppresses collective enhancement and thereby reduces the pairing ratios. Similar behavior has been reported in biphoton generation based on double-$\Lambda$ systems~\cite{Shiu2}. In addition, the interference among velocity dependent phases induced by Doppler detunings degrades the coherence during the SFWM process, further reducing the pairing ratios. Both effects suppress the collective enhancement that underlies high-purity biphoton generation in the diamond-type system.

Although the pairing rate is strongly suppressed by Doppler broadening, it can be enhanced by increasing the coupling and driving powers. Figure \ref{fig7}(d) shows the pairing rates $R'_{p}$ calculated using the GSA model in Eq. (\ref{eq(47)}) shown by the green dashed curve and the model incorporating exact populations shown by the blue solid curve, as detailed in Appendix \ref{ApxI}. As expected for a strongly driven thermal ensemble, we note that due to the increased populations of the excited states, the GSA solution is no longer applicable when both Rabi frequencies exceed $10\Gamma$, as indicated in the figure.

%%%%%%%%%%%%%%%%%%%%%%%%%%%%%%%%%%%%%%%%%%%%%%%%%%%%%%%%%%%%%%%%%%%%%%%%%%%%%%%%%%%%%%%%%%%%%%%%%%%%%
%%%%%%%%%%%%%%%%%%%%%%%%%%%%%%%%%%%%%%%%%%%%%%%%%%%%%%%%%%%%%%%%%%%%%%%%%%%%%%%%%%%%%%%%%%%%%%%%%%%%%

\subsection{Temporal correlations under Doppler broadening} \label{Sec. 5c}

Following the same procedure, the Doppler-broadened biphoton correlation function $G'^{(2)}_{s\textrm{-}i}(\tau)$ and the signal coincidence rate $R'_{\mathrm{C},s}(\tau)$ are obtained as
\begin{align}
	& G'^{(2)}_{s\textrm{-}i}(\tau)
	=   
	\bigg(\frac{L}{c}\bigg)^2 R'_{s} R'_{i}
	+\bigg|\frac{L}{c} \int \frac{d\omega}{2\pi} e^{i\omega \tau}
	\bigg[ B'_{1} D'^{\ast}_{1}
	\notag\\
	&  + \sum_{jk,j'k'} \int_0^L dz \bigg(
	C'^\ast_{2} X^{ss}_{I} A'_{2}
	+ D'^\ast_{2} X^{is}_{I} A'_{2}
	\notag\\
	& \quad
	+ C'^\ast_{2} X^{si}_{I} B'_{2}
	+ D'^\ast_{2} X^{ii}_{I} B'_{2}
	\bigg)\bigg]\bigg|^2,
	\label{eq(47)}
	\\
	& R'_{\mathrm{C},s}(\tau) =
	\left(\frac{c}{L}\right)^2
	\frac{G'^{(2)}_{s\textrm{-}i}(\tau)}{R'_{s}}.
	\label{eq(48)}
\end{align}
The corresponding expressions for $G'^{(2)}_{i\textrm{-}s}(\tau)$ and $R'_{\mathrm{C},i}(\tau)$ can be obtained analogously.

%%%%%%%%%
\figeight
%%%%%%%%%

Figures \ref{fig8}(a) and \ref{fig8}(b) show the biphoton correlation functions $R'_{\mathrm{C},s}(\tau)$ as functions of the idler delay $\tau'_d$ for $\mathrm{T}=10^{-4}\textrm{ K}$ and $\mathrm{T}=300\textrm{ K}$, respectively, with $\textrm{OD}=0.1$. For $\mathrm{T}=10^{-4}\textrm{ K}$, an exponential fit yields a decay time of approximately $26\textrm{ ns}$, which is consistent with the cold atom result shown in Fig. \ref{fig4}(a). In contrast, a substantial reduction of the decay time to about $0.3\textrm{ ns}$ is observed at $\mathrm{T}=300\textrm{ K}$, even though the OD remains unchanged and the system is still within the low OD regime. This result indicates that, in this temperature range, the Doppler effect, rather than superradiance, governs the biphoton dynamics by introducing a broad distribution of detunings and strong dispersion. As the temperature increases, the enhanced dispersion leads to a narrowing of the biphoton correlation time.

The dependence of the decay time $\tau'_d$ on temperature is shown in Fig. \ref{fig8}(c) for $\textrm{OD}=0.1$. Under these conditions, we find that the decay time is primarily inversely proportional to the effective broadening width of $\Delta'_3$, namely $\tau'_d \propto 1/\Delta'_3$, confirming that the Doppler effect governs the temporal profile of the biphoton waveform.

Despite the strong Doppler broadening, signatures of superradiance remain observable in the warm atomic ensemble. Figure \ref{fig8}(d) presents the dependence of the decay time $\tau'_d$ on OD for temperatures of $3\textrm{ K}$ shown by the yellow curve with circles and $300\textrm{ K}$ shown by the green curve with diamonds. In both cases, the decay time decreases with increasing OD, which provides clear evidence of superradiant enhancement. However, the observed trends no longer follow the scaling relation predicted by Eq. (\ref{eq(24)}). From the fitted scaling parameters shown in the figure caption, the extracted values for both $3\textrm{ K}$ and $300\textrm{ K}$ are significantly smaller than the cold atom prediction corresponding to $x=1/4$ in Eq. (\ref{eq(24)}). This result indicates that the OD-dependent superradiant effect is progressively weakened as the temperature increases. Physically, this behavior can be attributed to the increasing phase mismatch among individual atoms induced by Doppler broadening, which degrades the collective atomic coherence and suppresses superradiant emission.

Similar to the cold atom case discussed in Sec. \ref{Sec. 4}, we further compare our theoretical predictions with the experimental results reported by Tu \textit{et al.} \cite{Tu}, as shown in Figs. \ref{fig8}(e) for our theory and \ref{fig8}(f) for the experimental data shown by circles, using the corresponding experimental parameters specified in the figure caption. Apart from the signal to background ratio, which is approximately $16$ in theory and $2$ in the experiment, the calculated biphoton waveform agrees well with the measured temporal profile. The discrepancy in the signal to background ratio may be attributed to the fact that we simulate this biphoton case with a relatively simple single-Zeeman cycling transition in Fig.~\ref{fig7}(e). By contrast, the practical experimental results shown in Fig.~\ref{fig7}(f) are obtained using non-cycling, multi–Zeeman-level transitions, in which the population can decay to $|5S_{1/2}, F=1\rangle$, a state outside the specified transition, thereby reducing the pairing ratio.

In addition, the idler to signal correlation function exhibits a sub nanosecond peak followed by an elongated tail with a decay time of $\tau_d \approx 25\textrm{ ns}$, which is close to $1/\Gamma_{31}$. The rapid decay at the leading edge corresponds to the Doppler broadened component discussed above. When the condition $(\Omega_c\Omega_d/\Delta_1)^2 \ll 1$ is satisfied, the biphoton waveform is dominated by this fast decaying component, as shown in Figs. \ref{fig8}(a) and \ref{fig8}(b). As $\Omega_c$ and $\Omega_d$ are further increased, a slowly decaying component emerges. For larger values of $(\Omega_c\Omega_d/\Delta_1)^2$, the relative amplitude of the elongated tail increases with respect to the sharp leading peak, resulting in a pronounced long lived tail, as illustrated in Figs. \ref{fig8}(c) and \ref{fig8}(d). We note that under these conditions, an accurate description of the biphoton waveform requires solutions that include exact populations.

%%%%%%%%%%%%%%%%%%%%%%%%%%%%%%%%%%%%%%%%%%%%%%%%%%%%%%%%%%%%%%%%%%%%%%%%%%%%%%%%%%%%%%%%%%%%%%%%%%%%%
%%%%%%%%%%%%%%%%%%%%%%%%%%%%%%%%%%%%%%%%%%%%%%%%%%%%%%%%%%%%%%%%%%%%%%%%%%%%%%%%%%%%%%%%%%%%%%%%%%%%%

\FloatBarrier

\section{Conclusion} \label{Sec. 6}

We have established a unified and predictive microscopic description of superradiant biphoton generation in diamond-type atomic media within a Heisenberg--Langevin--Maxwell open-system framework applicable to both cold and warm ensembles. Unlike approaches based on effective nonlinear susceptibilities, the present theory explicitly incorporates vacuum fluctuations and spontaneous emission, enabling a fully quantum and self-consistent treatment of paired and unpaired emission processes.

For cold atomic ensembles, the coupled Heisenberg--Langevin and Maxwell equations admit analytical solutions in the high-OD regime. In this limit, the biphoton dynamics reduce to an effective collective two-level emission process, under which superradiance governs the temporal envelope of the wave packet. The correlation time follows the universal scaling $\tau_d=\tau_{31}(1+\mathrm{OD}/4)^{-1}$, establishing a direct quantitative link between collective emission and OD.

The framework is extended to Doppler-broadened warm vapors, where biphoton correlations arise from the interplay between collective emission and motion-induced spectral broadening. At elevated temperatures, Doppler averaging shortens the emission timescale and suppresses clean superradiant scaling with OD. A residual OD dependence nevertheless persists, indicating that collective emission effects remain present even when partially masked by velocity-induced dispersion.

Although collective temporal shortening can appear in gain-based descriptions, a frequency-resolved account of pairing efficiency and background photons requires an open-system treatment. By explicitly connecting optical depth, collective emission, and biphoton correlation time, this work shows that superradiant biphoton dynamics emerge from many-body radiative interactions rather than from phenomenological gain modifications. The resulting analytical scaling relations provide a first-principles basis for predicting and engineering atomic biphoton sources with tailored temporal and spectral properties. More broadly, our results identify the physical conditions under which superradiant biphoton behavior becomes experimentally accessible and clarify the complementary roles of cold and warm ensembles in shaping correlation time, bandwidth, and pairing efficiency. Within this framework, narrowband and hybrid quantum light sources interfacing atomic systems with visible and telecom-band platforms can be systematically optimized. The formalism can be extended to incorporate entanglement generation protocols, including mappings between orbital angular momentum and polarization~\cite{Lin}, enabling scalable quantum networking architectures bridging remote frequency domains.

%%%%%%%%%%%%%%%%%%%%%%%%%%%%%%%%%%%%%%%%%%%%%%%%%%%%%%%%%%%%%%%%%%%%%%%%%%%%%%%%%%%%%%%%%%%%%%%%%%%%%
%%%%%%%%%%%%%%%%%%%%%%%%%%%%%%%%%%%%%%%%%%%%%%%%%%%%%%%%%%%%%%%%%%%%%%%%%%%%%%%%%%%%%%%%%%%%%%%%%%%%%

\section*{ACKNOWLEDGMENTS}

We thank C.-C. Huang for helpful discussions. This work was supported by the National Science and Technology Council of Taiwan under Grant Nos. 113-2811-M-006-022 and 114-2112-M-006-007. We also acknowledge support from the Center for Quantum Science and Technology (CQST) within the framework of the Higher Education Sprout Project by the Ministry of Education (MOE) in Taiwan.

%%%%%%%%%%%%%%%%%%%%%%%%%%%%%%%%%%%%%%%%%%%%%%%%%%%%%%%%%%%%%%%%%%%%%%%%%%%%%%%%%%%%%%%%%%%%%%%%%%%%%
%%%%%%%%%%%%%%%%%%%%%%%%%%%%%%%%%%%%%%%%%%%%%%%%%%%%%%%%%%%%%%%%%%%%%%%%%%%%%%%%%%%%%%%%%%%%%%%%%%%%%

\appendix

\label{Apx}

\FloatBarrier

\section{DERIVATION OF THE COUPLED PROPAGATION EQUATIONS} \label{ApxA}

To derive the signal and idler field solutions, we first solve the zeroth-order HLEs using Eqs. (\ref{eq(1)}) and (\ref{eq(2)}). Assuming that the signal and idler field operators $\hat{a}_s$ and $\hat{a}_i$ vanish at zeroth order, the relevant zeroth-order HLEs are
\begin{align}
	\frac{\partial}{\partial t} &  \hat{\sigma}^{(0)}_{11} = 
	\hat{F}_{11}+\Gamma_{21}\hat{\sigma}^{(0)}_{22}+\Gamma_{31}\hat{\sigma}^{(0)}_{33}
	\notag\\
	& \quad\quad\quad\quad\quad\quad\quad\quad
	+\frac{i}{2} \left(\Omega^{\ast}_c\hat{\sigma}^{(0)}_{12}-\Omega_c\hat{\sigma}^{(0)}_{21}\right), \label{eq(A1)}\\
	\frac{\partial}{\partial t} &  \hat{\sigma}^{(0)}_{22} =  
	\hat{F}_{22}+ \Gamma_{42}\hat{\sigma}^{(0)}_{44}-\Gamma_{21}\hat{\sigma}^{(0)}_{22}\notag\\
	& \quad
	+\frac{i}{2} \left(\Omega_c\hat{\sigma}^{(0)}_{21}-\Omega^{\ast}_c\hat{\sigma}^{(0)}_{12}-
	\Omega_d\hat{\sigma}^{(0)}_{42}+\Omega^{\ast}_d\hat{\sigma}^{(0)}_{24}\right), 
	\label{eq(A2)}\\
	\frac{\partial}{\partial t} &  \hat{\sigma}^{(0)}_{33} = 
	\hat{F}_{33}+\Gamma_{43}\hat{\sigma}^{(0)}_{44}-\Gamma_{31}\hat{\sigma}^{(0)}_{33}, 
    \label{eq(A3)}\\
	\frac{\partial}{\partial t} &  \hat{\sigma}^{(0)}_{44} = 
	\hat{F}_{44}-(\Gamma_{42}+\Gamma_{43})\hat{\sigma}^{(0)}_{44}\notag\\
	& \quad\quad\quad\quad\quad\quad\quad\quad
	+\frac{i}{2} \left(\Omega_d\hat{\sigma}^{(0)}_{42}-\Omega^{\ast}_d\hat{\sigma}^{(0)}_{24}\right), 
	\label{eq(A4)}\\
	\frac{\partial}{\partial t} &  \hat{\sigma}^{(0)}_{21} = 
	\hat{F}_{21}-\frac{1}{2}\gamma_{21}\hat{\sigma}^{(0)}_{21} \notag\\
	& +\frac{i}{2}\left[\Omega^{\ast}_c\left(\hat{\sigma}^{(0)}_{22}-\hat{\sigma}^{(0)}_{11}\right)-\Omega_d\hat{\sigma}^{(0)}_{41}-2\Delta_1\hat{\sigma}^{(0)}_{21}\right],  
	\label{eq(A5)}\\
	\frac{\partial}{\partial t} &  \hat{\sigma}^{(0)}_{41} = 
	\hat{F}_{41}-\frac{1}{2}\gamma_{41}\hat{\sigma}^{(0)}_{41} \notag\\
	& \quad\quad\quad\quad -\frac{i}{2}\left[2\Delta_2\hat{\sigma}^{(0)}_{41}-\Omega^{\ast}_c\hat{\sigma}^{(0)}_{42}+\Omega^{\ast}_d\hat{\sigma}^{(0)}_{21}\right],
	\label{eq(A6)}\\
	\frac{\partial}{\partial t} &  \hat{\sigma}^{(0)}_{42} = 
	\hat{F}_{42}-\frac{1}{2}\gamma_{42}\hat{\sigma}^{(0)}_{42} \notag\\ 
	+ & \frac{i}{2}\left[\Omega_c\hat{\sigma}^{(0)}_{41}+\Omega^{\ast}_d\left(\hat{\sigma}^{(0)}_{44}-\hat{\sigma}^{(0)}_{22}\right)+2\left(\Delta_1-\Delta_2\right)\hat{\sigma}^{(0)}_{42}\right]. 
	\label{eq(A7)} 
\end{align}
Together with their adjoint terms $\hat{\sigma}^{(0)}_{12}$, $\hat{\sigma}^{(0)}_{14}$, and $\hat{\sigma}^{(0)}_{24}$, and using  $\langle\hat{\sigma}^{(0)}_{11}\rangle +\langle\hat{\sigma}^{(0)}_{22}\rangle +\langle\hat{\sigma}^{(0)}_{33}\rangle +\langle\hat{\sigma}^{(0)}_{44}\rangle =1$, the steady-state solutions can be formally written as
\begin{align}
	\hat{\sigma}^{(0)}_{jk}=
	\big\langle \hat{\sigma}^{(0)}_{jk} \big\rangle
	+\sum_{mn}\epsilon^{mn}_{jk}\hat{F}_{mn}.
	\label{eq(A8)}
\end{align}

From the steady-state solutions, we find that $\langle \hat{\sigma}^{(0)}_{11} \rangle > 96\%$ for all parameter sets considered in this work. The exact analytical expressions are lengthy and therefore not reproduced here. By contrast, the solutions become significantly simplified in the large-$\Delta_1$ limit. Retaining only the leading-order terms in $1/\Delta_1$, we find that $\langle \hat{\sigma}^{(0)}_{jk} \rangle = 0$ for most combinations of $(j,k)$. The remaining nonzero expectation values are given by
\begin{align}
	&
	\big\langle \hat{\sigma}^{(0)}_{11} \big\rangle
	=1, 	
	\label{eq(A9)}\\
	&
	\big\langle \hat{\sigma}^{(0)}_{12} \big\rangle
	=
	\big\langle \hat{\sigma}^{(0)}_{21} \big\rangle^{\ast}
	=-\frac{\Omega_c}{2\Delta_1}, 	
	\label{eq(A10)}\\
	&
	\big\langle \hat{\sigma}^{(0)}_{14} \big\rangle
	=
	\big\langle \hat{\sigma}^{(0)}_{41} \big\rangle^{\ast}
	=-\frac{i\Omega_c\Omega_d}{2\gamma_{41}\Delta_1-4i\Delta_1\Delta_2}. 	
	\label{eq(A11)}
\end{align}
Next, we obtain the first-order HLEs as follows
\begin{align}
	& \frac{\partial}{\partial t} \hat{\sigma}^{(1)}_{31}  =
	\hat{F}_{31}-\frac{1}{2}\gamma_{31}\hat{\sigma}^{(1)}_{31} \notag\\
	& -i\left[g_s\hat{a}_s \hat{\sigma}^{(0)}_{41} - g_i\hat{a}^{\dagger}_i\left(\hat{\sigma}^{(0)}_{33}-\hat{\sigma}^{(0)}_{11}\right)- \frac{\Omega^{\ast}_c}{2}\hat{\sigma}^{(1)}_{32}\right], \label{eq(A12)} \\
	\notag\\
	& \frac{\partial}{\partial t}  \hat{\sigma}^{(1)}_{32}  =
	\hat{F}_{32}-\frac{1}{2}\gamma_{32}\hat{\sigma}^{(1)}_{32} \notag\\
	& -i\left(g_i\hat{a}^{\dagger}_i\hat{\sigma}^{(0)}_{12}+g_s\hat{a}_s\hat{\sigma}^{(0)}_{42}-\Delta_1\hat{\sigma}^{(1)}_{32}  -\frac{\Omega_c}{2}\hat{\sigma}^{(1)}_{31}-\frac{\Omega^{\ast}_d}{2}\hat{\sigma}^{(1)}_{34}\right), 
	\label{eq(A13)} \\
	& \frac{\partial}{\partial t}  \hat{\sigma}^{(1)}_{34}  =
	\hat{F}_{34}-\frac{1}{2}\gamma_{43}\hat{\sigma}^{(1)}_{34} \notag\\
	& -i\left[	g_i\hat{a}^{\dagger}_i\hat{\sigma}^{(0)}_{14} -g_s\hat{a}_s\left(\hat{\sigma}^{(0)}_{33}-\hat{\sigma}^{(0)}_{44}\right)
	-\Delta_2\hat{\sigma}^{(1)}_{34}
	-\frac{\Omega_d}{2}\hat{\sigma}^{(1)}_{32}  \right]. 
	\label{eq(A14)}
\end{align}

Substituting the results of Eqs. (\ref{eq(A8)})-(\ref{eq(A11)}) into Eqs. (\ref{eq(A12)})-(\ref{eq(A14)}) and neglecting higher-order terms involving products of the signal or idler field operators with Langevin noise operators, i.e., $\epsilon^{mn}_{jk}\hat{a}^{\dagger}_s\hat{F}_{mn}$ ($\epsilon^{mn}_{jk}\hat{a}_s\hat{F}_{mn}$) and $\epsilon^{mn}_{jk}\hat{a}^{\dagger}_i\hat{F}_{mn}$ ($\epsilon^{mn}_{jk}\hat{a}_i\hat{F}_{mn}$), the first-order HLEs for the signal and idler fields are
\begin{align}
	\frac{\partial}{\partial t} &  \hat{\sigma}^{(1)}_{31}  =
	\hat{F}_{31}-\frac{1}{2}\gamma_{31}\hat{\sigma}^{(1)}_{31} \notag\\
	& -i\bigg[g_s\hat{a}_s \left(
	\frac{i\Omega^\ast_c\Omega^\ast_d}{2\gamma_{41}\Delta_1+4i\Delta_1 \Delta_2} \right) + g_i\hat{a}^{\dagger}_i- \frac{\Omega^{\ast}_c}{2}\hat{\sigma}^{(1)}_{32}\bigg], \label{eq(A15)} \\
	\frac{\partial}{\partial t} &  \hat{\sigma}^{(1)}_{32}  =
	\hat{F}_{32}-\frac{1}{2}\gamma_{32}\hat{\sigma}^{(1)}_{32} \notag\\
	& -i\bigg[ g_i\hat{a}^{\dagger}_i\left(\frac{-\Omega_c}{2\Delta_1} \right)-\Delta_1\hat{\sigma}^{(1)}_{32}  -\frac{\Omega_c}{2}\hat{\sigma}^{(1)}_{31}-\frac{\Omega^{\ast}_d}{2}\hat{\sigma}^{(1)}_{34} \bigg], 
	\label{eq(A16)} \\
	\frac{\partial}{\partial t} &  \hat{\sigma}^{(1)}_{34}  =
	\hat{F}_{34}-\frac{1}{2}\gamma_{43}\hat{\sigma}^{(1)}_{34} \notag\\
	& -i\bigg[ g_i\hat{a}^{\dagger}_i\left(
	\frac{-i\Omega_c\Omega_d}{2\gamma_{41}\Delta_1-4i\Delta_1 \Delta_2} \right) -\frac{\Omega_d}{2}\hat{\sigma}^{(1)}_{32}  -\Delta_2\hat{\sigma}^{(1)}_{34}\bigg]. \label{eq(A17)}
\end{align}
Transforming Eqs.~(\ref{eq(A15)})--(\ref{eq(A17)}) to the frequency domain, we obtain the solutions for $\tilde{\sigma}^{(1)}_{31}(z,\omega)$ and $\tilde{\sigma}^{(1)}_{34}(z,\omega)$. Substituting back into the frequency-domain MSEs, we recover the coupled propagation equations given in Eqs. (\ref{eq(5)}) and (\ref{eq(6)}) of Sec. \ref{Sec. 2a}.

%%%%%%%%%%%%%%%%%%%%%%%%%%%%%%%%%%%%%%%%%%%%%%%%%%%%%%%%%%%%%%%%%%%%%%%%%%%%%%%%%%%%%%%%%%%%%%%%%%%%%
%%%%%%%%%%%%%%%%%%%%%%%%%%%%%%%%%%%%%%%%%%%%%%%%%%%%%%%%%%%%%%%%%%%%%%%%%%%%%%%%%%%%%%%%%%%%%%%%%%%%%

\section{LANGEVIN NOISE OPERATORS AND DIFFUSION COEFFICIENTS} \label{ApxB}

The Langevin noise coefficients $\zeta^s_{jk}$ and $\zeta^i_{jk}$ that enter the frequency-domain coupled propagation equations governing the signal and idler fields in Eqs. (\ref{eq(5)}) and (\ref{eq(6)}), and that originate from the quantum Langevin operators associated with dissipative atomic transitions in the Heisenberg--Langevin formalism, thereby ensuring the preservation of commutation relations and consistency with the fluctuation--dissipation theorem, are explicitly given by
\begin{align}
	\zeta^s_{31} & =
	\sqrt{\frac{\alpha s_\lambda\Gamma_{43}}{4L}}
	\left(
	\frac{\Omega_c\Omega_d}{\Delta_1(\gamma_{31}-2i\omega)[\gamma_{43}-2i(\Delta_2+\omega)]}
	\right),
	\label{eq(B1)}\\
	\zeta^s_{32} & =
	\sqrt{\frac{\alpha s_\lambda\Gamma_{43}}{L}}
	\left(
	\frac{-i \Omega_d}{2\Delta_1[\gamma_{43}-2i(\Delta_2+\omega)]}
	\right),
	\label{eq(B2)}\\
	\zeta^s_{34} & =
	\sqrt{\frac{\alpha s_\lambda\Gamma_{43}}{L}}
	\left[
	\frac{i}{\gamma_{43}-2i(\Delta_2+\omega)}
	\right],
	\label{eq(B3)}\\
	\zeta^i_{31} & =
	\sqrt{\frac{\alpha\Gamma_{31}}{L}}
	\left(
	\frac{-i}{\gamma_{31}-2i\omega}
	\right),
	\label{eq(B4)}\\
	\zeta^i_{32} & =
	\sqrt{\frac{\alpha\Gamma_{31}}{L}}
	\left(
	\frac{i\Omega^{\ast}_c}{2\Delta_1(\gamma_{31}-2i\omega)}
	\right),
	\label{eq(B5)}\\
	\zeta^i_{34} & =
	\sqrt{\frac{\alpha\Gamma_{31}}{L}}
	\left(
	\frac{-\Omega^{\ast}_c\Omega^{\ast}_d}{2\Delta_1(\gamma_{31}-2i\omega)[\gamma_{43}-2i(\Delta_2+\omega)]}
	\right).
	\label{eq(B6)}
\end{align}
Using these noise coefficients, the parameters $P_{jk}$ and $Q_{jk}$ entering the generation rates can be obtained. Next, we assume that the Langevin noise operators are delta-correlated,
\begin{align}
	\langle \tilde{f}^\dagger_{jk}(z,-\omega)\tilde{f}_{j'k'}(z',\omega')\rangle 
	=\frac{L}{2\pi c}\mathscr{D}_{kj,j'k'} \delta(z-z')\delta(\omega+\omega'), \label{eq(B7)}
\end{align}
where the diffusion coefficients $\mathscr{D}_{kj,j'k'}$ are evaluated using the Einstein relation
\begin{align}
	\mathscr{D}_{kj,j'k'} = & \frac{\partial}{\partial{t}}\langle \hat{\sigma}_{kj}\hat{\sigma}_{j'k'} \rangle 
	- \left\langle \left( \frac{\partial}{\partial{t}} \hat{\sigma}_{kj}-\hat{F}_{kj} \right) \hat{\sigma}_{j'k'} \right\rangle 
	\notag\\
	& - \left\langle \hat{\sigma}_{kj} \left(   \frac{\partial}{\partial {t}} \hat{\sigma}_{j'k'}-\hat{F}_{j'k'} \right) \right\rangle
	. \label{eq(B8)}
\end{align}
With the help of Eq.~(\ref{eq(B8)}), the diffusion coefficients $\mathscr{D}_{kj,j'k'}$ can be written in matrix form as
\begin{align}
	\mathscr{D}_{kj,j'k'} 
	& =
	\begin{bmatrix}
		\mathscr{D}_{1331} & \mathscr{D}_{1332} & \mathscr{D}_{1334} \\
		\mathscr{D}_{2331} & \mathscr{D}_{2332} & \mathscr{D}_{2334} \\
		\mathscr{D}_{4331} & \mathscr{D}_{4332} & \mathscr{D}_{4334} 
	\end{bmatrix}
	,
	\label{eq(B9)}
\end{align}
with nonzero elements given by
\begin{align}
	\mathscr{D}_{1331} & =\gamma_{31} \langle\hat{\sigma}_{11}\rangle
	+\Gamma_{21}\langle\hat{\sigma}_{22}\rangle+\Gamma_{31}\langle\hat{\sigma}_{33}\rangle,
	\label{eq(B10)}\\
    \mathscr{D}_{2332} & =(\gamma_{32}-\Gamma_{21})\langle\hat{\sigma}_{22}\rangle
    +\Gamma_{42}\langle\hat{\sigma}_{44}\rangle,
	\label{eq(B11)}\\
	\mathscr{D}_{4334} & =(\gamma_{43}-\Gamma_{42}-\Gamma_{43})\langle\hat{\sigma}_{44}\rangle,
	\label{eq(B12)}\\
	\mathscr{D}_{2331} & =
	\mathscr{D}^{\ast}_{1332}=
	\frac{1}{2}(\gamma_{31}+\gamma_{32}-\gamma_{21})\langle\hat{\sigma}_{21}\rangle,
	\label{eq(B13)}\\
	\mathscr{D}_{4331} & =
	\mathscr{D}^{\ast}_{1334}=
	\frac{1}{2}(\gamma_{31}+\gamma_{43}-\gamma_{41})\langle\hat{\sigma}_{41}\rangle,
	\label{eq(B14)}\\
	\mathscr{D}_{4332} & =
	\mathscr{D}^{\ast}_{2334}=
	\frac{1}{2}(\gamma_{32}+\gamma_{43}-\gamma_{42})\langle\hat{\sigma}_{42}\rangle.
	\label{eq(B15)}
\end{align}
In the same manner, the remaining diffusion coefficients $\mathscr{D}_{jk,k'j'}$ are obtained as
\begin{align}
	\mathscr{D}_{jk,k'j'} 
	& =
	\begin{bmatrix}
		\mathscr{D}_{3113} & \mathscr{D}_{3123} & \mathscr{D}_{3143} \\
		\mathscr{D}_{3213} & \mathscr{D}_{3223} & \mathscr{D}_{3243} \\
		\mathscr{D}_{3413} & \mathscr{D}_{3423} & \mathscr{D}_{3443} 
	\end{bmatrix}
	,
	\label{eq(B16)}
\end{align}
where the nonzero elements are given by
\begin{align}
	\mathscr{D}_{3113} & =
	\Gamma_{43}\langle\hat{\sigma}_{44}\rangle
	+(\gamma_{31}-\Gamma_{31})\langle\hat{\sigma}_{33}\rangle,
	\label{eq(B17)}\\
	\mathscr{D}_{3223} & =
	\Gamma_{43}\langle\hat{\sigma}_{44}\rangle
	+(\gamma_{32}-\Gamma_{31})\langle\hat{\sigma}_{33}\rangle,
	\label{eq(B18)}\\
	\mathscr{D}_{3443} & =
	\Gamma_{43}\langle\hat{\sigma}_{44}\rangle
	+(\gamma_{43}-\Gamma_{31})\langle\hat{\sigma}_{33}\rangle.
	\label{eq(B19)}
\end{align}
All remaining elements vanish. The diffusion coefficients $\mathscr{D}_{kj,j'k'}$ and $\mathscr{D}_{jk,k'j'}$ are evaluated using the exact expectation values of the zeroth-order atomic operators. We find that these coefficients are sensitive to the population distributions, and that the GSA yields inaccurate predictions, even when $\langle \hat{\sigma}_{11} \rangle \approx 1$. Using these diffusion coefficients, the unpaired spectral densities $\tilde{R}_{s,u}(\omega)$ and $\tilde{R}_{i,u}(\omega)$ in Eqs.~(\ref{eq(16)}) and~(\ref{eq(17)}), as well as $\tilde{R}'_{s,u}(\omega)$ and $\tilde{R}'_{i,u}(\omega)$ in Eqs.~(\ref{eq(42)}) and~(\ref{eq(43)}), are evaluated.

%%%%%%%%%%%%%%%%%%%%%%%%%%%%%%%%%%%%%%%%%%%%%%%%%%%%%%%%%%%%%%%%%%%%%%%%%%%%%%%%%%%%%%%%%%%%%%%%%%%%%
%%%%%%%%%%%%%%%%%%%%%%%%%%%%%%%%%%%%%%%%%%%%%%%%%%%%%%%%%%%%%%%%%%%%%%%%%%%%%%%%%%%%%%%%%%%%%%%%%%%%%

\section{MODEL BEYOND THE GROUND-STATE APPROXIMATION} \label{ApxC}

In this Appendix, we derive the solutions of the signal and idler fields $\tilde{a}_{sL}$ and $\tilde{a}_{iL}$ by retaining the full steady-state atomic populations, thereby going beyond the GSA adopted in the main text. Following the same procedure as in Appendix~\ref{ApxA}, the exact zeroth-order steady-state solutions are substituted directly into the first-order HLEs, Eqs.~(\ref{eq(A12)})--(\ref{eq(A14)}). Under these conditions, the coefficients appearing in the frequency-domain coupled propagation equations, Eqs.~(\ref{eq(5)}) and~(\ref{eq(6)}), take the following forms, where we define $\langle \hat{\sigma}^{(0)}_{jk} \rangle \equiv \bar{\sigma}_{jk}$.
\begin{widetext}
	\begin{align}
	\Gamma_{i,{\textrm{E}}} & = \frac{\alpha\Gamma_{31}}{2LM}
    \bigg(\bar{\sigma}_{12}\Omega^{\ast}_c\big[i\gamma_{43}+2(\Delta_2+\omega)\big]-\bar{\sigma}_{14}\Omega^{\ast}_c\Omega^{\ast}_d+(\bar{\sigma}_{11}-\bar{\sigma}_{33})
    \left[(\gamma_{32}-2i\Delta_1-2i\omega)(\gamma_{43}-2i\Delta_2-2i\omega)+|\Omega_d|^2
    \right] \bigg) \notag\\
	& \quad -\frac{i\omega}{c}+i\Delta k, 
	\label{eq(C1)}\\
	\kappa_{s,{\textrm{E}}} & =\frac{-\alpha\sqrt{s_\lambda\Gamma_{43}\Gamma_{31}}}{2LM}
    \bigg[\bar{\sigma}_{14}
    \left(\gamma_{32}-2i\Delta_1-2i\omega
    \right)
    (\gamma_{31}-2i\omega)+\bar{\sigma}_{14}|\Omega_c|^2+\bar{\sigma}_{12}\Omega_d(i\gamma_{31}+2\omega)+(\bar{\sigma}_{33}-\bar{\sigma}_{11})\Omega_c\Omega_d
    \bigg],
	\label{eq(C2)}\\
	\kappa_{i,{\textrm{E}}} & =\frac{\alpha\sqrt{s_\lambda\Gamma_{43}\Gamma_{31}}}{2LM}
    \bigg(
    \bar{\sigma}_{41}\big[(-i\gamma_{32}-2\Delta_1-2\omega)+\bar{\sigma}_{42}\Omega^{\ast}_c\big](i\gamma_{43}+2\Delta_2+2\omega)+(\bar{\sigma}_{33}-\bar{\sigma}_{44})\Omega^{\ast}_c\Omega^{\ast}_d+\bar{\sigma}_{41}|\Omega_d|^2
    \bigg),
	\label{eq(C3)}\\
	G_{s,{\textrm{E}}} & = \frac{\alpha s_\lambda\Gamma_{43}}{2LM}
    \bigg((\bar{\sigma}_{33}-\bar{\sigma}_{44})\big[    (\gamma_{32}-2i\Delta_1-2i\omega)(\gamma_{31}-2i\omega)+|\Omega_c|^2\big]
    +\bar{\sigma}_{42}(-2\omega-i\gamma_{31})\Omega_d+\bar{\sigma}_{41}\Omega_c\Omega_d\bigg)-\frac{i\omega}{c}, 
	\label{eq(C4)}
    \end{align}
    \begin{align}
	& \zeta^s_{31,{\textrm{E}}} =
	-i\sqrt{\frac{\alpha s_\lambda\Gamma_{43}}{LM^2}}
	\Omega_c\Omega_d,
	\label{eq(C5)}\\
	& \zeta^s_{32,{\textrm{E}}} =
	-\sqrt{\frac{\alpha s_\lambda\Gamma_{43}}{LM^2}}
	(\gamma_{31}-2i\omega)\Omega_d,
	\label{eq(C6)}\\
	& \zeta^s_{34,{\textrm{E}}} =
	i\sqrt{\frac{\alpha s_\lambda\Gamma_{43}}{LM^2}}
	\bigg[
	(\gamma_{32}-2i\Delta_1-2i\omega)(\gamma_{31}-2i\omega)+|\Omega_c|^2
	\bigg],
	\label{eq(C7)}\\
	& \zeta^i_{31,{\textrm{E}}} =
	-i\sqrt{\frac{\alpha\Gamma_{31}}{LM^2}}
    \bigg(	[\gamma_{32}-2i(\Delta_1+\omega)]\left[\gamma_{43}-2i(\Delta_2+\omega)\right]+|\Omega_d|^2\bigg),
	\label{eq(C8)}\\
	& \zeta^i_{32,{\textrm{E}}} =
	\sqrt{\frac{\alpha\Gamma_{31}}{LM^2}}
	\left(
	\gamma_{43}-2i\Delta_2-2i\omega
	\right)
	\Omega^{\ast}_c,
	\label{eq(C9)}\\
	& \zeta^i_{34,{\textrm{E}}} =
	i\sqrt{\frac{\alpha\Gamma_{31}}{LM^2}}
	\Omega^{\ast}_c\Omega^{\ast}_d,
	\label{eq(C10)}
    \end{align}
	\begin{align}
	\intertext{where}
		M =
		(\gamma_{43}-2i\Delta_2-2i\omega)\big[(\gamma_{32}-2i\Delta_1-2i\omega)(\gamma_{31}-2i\omega)+|\Omega_c|^2\big]+(\gamma_{31}-2i\omega)|\Omega_d|^2.
		\label{eq(C11)}
	\end{align}
Equations~(\ref{eq(C1)})--(\ref{eq(C11)}) therefore provide the exact coefficients of the coupled MSEs when the full atomic populations are retained. The general solutions for the signal and idler fields have the same functional form as Eq.~(\ref{eq(11)}) in the main text, with the corresponding coefficients and matrix elements replaced by their exact-population counterparts. The first transfer matrix becomes
	\begin{align}
		\begin{bmatrix}
			A_{1,{\textrm{E}}} & B_{1,{\textrm{E}}} \\
			C_{1,{\textrm{E}}} & D_{1,{\textrm{E}}} \\
		\end{bmatrix}
		= -e^{-(G_{s,{\textrm{E}}}+\Gamma_{i,{\textrm{E}}})L/2}
		\begin{bmatrix}
			\left(\frac{G_{s,{\textrm{E}}}-\Gamma_{i,{\textrm{E}}}}{2\Phi_{\textrm{E}}}\right)\cosh(\Phi_{\textrm{E}} L)-\sinh(\Phi_{\textrm{E}} L) &     \left(\frac{\kappa_{s,{\textrm{E}}}}{\Phi_{\textrm{E}}}\right)\sinh(\Phi_{\textrm{E}} L) \\
			\left(\frac{\kappa_{i,{\textrm{E}}}}{\Phi_{\textrm{E}}}\right)\sinh(\Phi_{\textrm{E}} L) &     -\left(\frac{G_{s,{\textrm{E}}}-\Gamma_{i,{\textrm{E}}}}{2\Phi_{\textrm{E}}}\right)\sinh(\Phi_{\textrm{E}} L)-\cosh(\Phi_{\textrm{E}} L) \\
		\end{bmatrix}
		,
		\label{eq(C12)}
	\end{align}
\end{widetext}
where $\Phi_{\mathrm{E}} = \frac{1}{2}\sqrt{(G_{s,\mathrm{E}}-\Gamma_{i,\mathrm{E}})^2 +4\kappa_{s,\mathrm{E}}\kappa_{i,\mathrm{E}}}$. The second transfer matrix has the same functional dependence as Eq.~(\ref{eq(C12)}) with $L$ replaced by $(L-z)$. In the large-$\Delta_1$ limit, the coefficient $\Gamma_{i,\mathrm{E}}$ dominates the dynamics, such that $\Gamma_{i,\mathrm{E}} \gg G_{s,\mathrm{E}}, \kappa_{s,\mathrm{E}}, \kappa_{i,\mathrm{E}}$. Under this condition, the approximations $(G_{s,\mathrm{E}}+\Gamma_{i,\mathrm{E}})\approx \Gamma_i$, $(G_{s,\mathrm{E}}-\Gamma_{i,\mathrm{E}})\approx -\Gamma_i$, and $\sqrt{(G_{s,\mathrm{E}}-\Gamma_{i,\mathrm{E}})^2 +4\kappa_{s,\mathrm{E}}\kappa_{i,\mathrm{E}}}\approx \Gamma_i$ are valid. As a result, the exact-population solutions reduce to those obtained under the GSA in the main text, establishing the GSA as the large-detuning limit of the full theory.

%%%%%%%%%%%%%%%%%%%%%%%%%%%%%%%%%%%%%%%%%%%%%%%%%%%%%%%%%%%%%%%%%%%%%%%%%%%%%%%%%%%%%%%%%%%%%%%%%%%%%
%%%%%%%%%%%%%%%%%%%%%%%%%%%%%%%%%%%%%%%%%%%%%%%%%%%%%%%%%%%%%%%%%%%%%%%%%%%%%%%%%%%%%%%%%%%%%%%%%%%%%

\section{DERIVATION OF THE ANALYTICAL PAIRING RATE} \label{ApxD}

In this Appendix, we derive the analytical approximation for the pairing rate given in Eq.~(\ref{eq(20)}), starting from the exact integral expression in Eq.~(\ref{eq(18)}). Rewriting Eq.~(\ref{eq(18)}) in the equivalent form of Eq.~(\ref{eq(19)}), the integrand is expressed explicitly in terms of the frequency-dependent parameter $\Gamma_i(\omega)$. Substituting the expression of $\Gamma_i(\omega)$ from Eq.~(\ref{eq(7)}), the pairing rate can be simplified under the large-OD condition. In this regime, collective spectral broadening causes the contribution of the $\gamma_{31}$ term in the denominator of Eq.~(\ref{eq(7)}) to become subdominant over the frequency range that provides the dominant contribution to the integral. Under this approximation, the pairing rate reduces to the following integral form,
\begin{align}
	R_p
	\approx & \frac{1}{2\pi}
	\left|
	\frac{\kappa_s}{\Gamma_i}
	\right|^{2}
	\int_{-\infty}^{\infty}
	\bigg[
	1-e^{-\left(\frac{\alpha\Gamma_{31}}{2}\right)\left(\frac{2i\omega}{4\omega^2}\right)}
	\notag\\
	& \quad
	-e^{-\left(\frac{\alpha\Gamma_{31}}{2}\right)\left(\frac{\gamma_{31}-2i\omega}{4\omega^2}\right)}
	+e^{-(\alpha\Gamma_{31})\left(\frac{\gamma_{31}}{4\omega^2}\right)}
	\bigg]
	\, d\omega .
	\label{eq(D1)}
\end{align}
To evaluate the frequency integral in Eq.~(\ref{eq(D1)}), we define the parameters $u=\alpha\Gamma_{31}/4$ and $\nu=\gamma_{31}/2$. With these definitions, the integral can be expressed in terms of the error function as
\begin{align}
	& \int_{-\infty}^{\infty}
	\left[
	1
	-2\exp\!\left(-\frac{u\nu}{\omega^{2}}\right)
	\cos\!\left(\frac{u}{\omega}\right)
	+\exp\!\left(-\frac{2u\nu}{\omega^{2}}\right)
	\right]
	\, d\omega
	\notag\\
	&\quad =
	e^{-u/4\nu}\sqrt{4\pi u\nu}
	\left(
	2
	+e^{u/4\nu}
	\left[
	\sqrt{\frac{\pi u}{\nu}}\,
	\mathrm{Erf}\!\left(\frac{u}{4\nu}\right)
	-\sqrt{2}
	\right]
	\right),
	\label{eq(D2)}
\end{align}
where $\mathrm{Erf}(x)=\frac{2}{\sqrt{\pi}}\int_{0}^{x} e^{-t^{2}}\,dt$ denotes the error function. Substituting Eq.~(\ref{eq(D2)}) into Eq.~(\ref{eq(D1)}), the pairing rate can be written as
\begin{align}
	R_p
	& =
	\sqrt{\frac{\alpha\Gamma_{31}\gamma_{31}}{2\pi}}
	\left(\frac{s_\lambda\Gamma_{43}}{\Gamma_{31}}\right)
	\left|
	\frac{\Omega_c\Omega_d}
	{2\Delta_1(\gamma_{41}+2i\Delta_2)}
	\right|^{2}
	\notag\\
	&\quad \times
	\left[
	\exp(-K)
	+\sqrt{\pi K}\,\mathrm{Erf}(K)
	-\frac{1}{\sqrt{2}}
	\right],
	\label{eq(D3)}
\end{align}
where $K\equiv \alpha\Gamma_{31}/(8\gamma_{31})$. Finally, in the asymptotic large-OD limit ($\alpha\gg1$), one has $\mathrm{Erf}(K)\simeq1$ and $\exp(-K)\simeq0$. Under this condition, Eq.~(\ref{eq(D3)}) reduces to the compact analytical expression $R_{p,\mathrm{A}}$ given in Eq.~(\ref{eq(20)}) of Sec.~\ref{Sec. 3b}.

%%%%%%%%%%%%%%%%%%%%%%%%%%%%%%%%%%%%%%%%%%%%%%%%%%%%%%%%%%%%%%%%%%%%%%%%%%%%%%%%%%%%%%%%%%%%%%%%%%%%%
%%%%%%%%%%%%%%%%%%%%%%%%%%%%%%%%%%%%%%%%%%%%%%%%%%%%%%%%%%%%%%%%%%%%%%%%%%%%%%%%%%%%%%%%%%%%%%%%%%%%%

\section{DERIVATION OF THE ANALYTICAL CORRELATION FUNCTION} \label{ApxE}

In this Appendix, we derive the analytical expression for the biphoton wave function $\Psi_{s\textrm{-}i}(\tau)$ appearing in Eq.~(\ref{eq(25)}), whose large-OD limit yields the compact analytical form given in Eq.~(\ref{eq(26)}). Starting from Eq.~(\ref{eq(22)}), we substitute the coefficients $B_1$ and $D_1$ obtained from Eqs.~(\ref{eq(13)}) and~(\ref{eq(15)}), and focus on the quantity inside the absolute square, which we define as the biphoton wave function $\Psi_{s\textrm{-}i}(\tau)$. This yields
\begin{align}
	\Psi_{s\textrm{-}i}(\tau)
	=
	\frac{\kappa_s L}{\Gamma_i c}
	\int_{-\infty}^{\infty}
	\frac{d\omega}{2\pi}
	e^{i\omega\tau}
	\left(
	e^{-2\mathrm{Re}[\Gamma_i]L}
	-
	e^{-\Gamma_i^{\ast}L}
	+
	\mathcal{N}_{si}
	\right),
	\label{eq(E1)}
\end{align}
where the factor $\kappa_s/\Gamma_i$ is frequency independent under the large-OD condition, and we have introduced $\mathcal{N}_{si}\equiv \sum_{jk,j'k'}\int dz\, Q^{\ast}_{jk}\mathscr{D}_{jk,k'j'}P_{j'k'}$ to denote the contribution from Langevin noise terms.

In the large-OD limit ($\alpha\gg1$), using the expression for $\Gamma_i$ given in Eq.~(\ref{eq(7)}) and defining $u=\alpha\Gamma_{31}/4$, the frequency integral in Eq.~(\ref{eq(E1)}) can be simplified. Numerical evaluation shows that the leading contribution arises from the phase-dependent term, while the noise contribution $\mathcal{N}_{si}$ and the Gaussian damping factors $\exp(-u\gamma_{31}/\omega^2)$ are negligible at the level of the dominant temporal envelope for $\alpha\gtrsim30$. Under these conditions, Eq.~(\ref{eq(E1)}) can be approximated as
\begin{align}
	\Psi_{s\textrm{-}i}(\tau)
	\approx
	\int_{-\infty}^{\infty}
	\frac{d\omega}{2\pi}
	e^{i\omega\tau}
	\left[
	1-
	\exp\!\left(i\frac{u}{\omega}\right)
	\right].
	\label{eq(E2)}
\end{align}

The integral in Eq.~(\ref{eq(E2)}) can be evaluated analytically, yielding
\begin{align}
	& \Psi_{s\textrm{-}i}(\tau)
	=
	\sqrt{\frac{i u}{2\pi^2|\tau|^3}}
	\left(\frac{\kappa_s}{\Gamma_i}\right)
	\notag\\
	& \times\bigg[
	\big(-i\tau+|\tau|\big)\mathrm{Kei}_1\!\left(\sqrt{\Theta}\right)+ \big(\tau-i|\tau|\big)\mathrm{Kei}_1\!\left(\sqrt{-\Theta}\right)
	\notag\\
	& +
	\big(-i\tau-|\tau|\big)\mathrm{Ker}_1\!\left(\sqrt{\Theta}\right)+ \big(\tau+i|\tau|\big)\mathrm{Ker}_1\!\left(\sqrt{-\Theta}\right)
	\bigg],
	\label{eq(E3)}
\end{align}
where $\Theta=4iu|\tau|$. Here $\mathrm{Kei}_\mu(x)$ and $\mathrm{Ker}_\mu(x)$ denote the imaginary and real parts of the Kelvin function of the second kind, respectively, defined as~\cite{Abramowitz,Olver}
\begin{align}
	\mathrm{Kei}_\mu(x)
	&=
	\mathrm{Im}\!\left[
	e^{-i\mu\pi/2}
	K_\mu\!\left(e^{i\pi/4}x\right)
	\right],
	\label{eq(E4)}\\
	\mathrm{Ker}_\mu(x)
	&=
	\mathrm{Re}\!\left[
	e^{-i\mu\pi/2}
	K_\mu\!\left(e^{i\pi/4}x\right)
	\right],
	\label{eq(E5)}
\end{align}
with $K_\mu(x)$ the modified Bessel function of the second kind.

Substituting $\Psi_{s\textrm{-}i}(\tau)$ into Eq.~(\ref{eq(22)}) in Sec.~\ref{Sec. 4a}, we obtain the analytical form of the biphoton correlation function given in Eq.~(\ref{eq(26)}). Numerical comparisons confirm that this approximation is accurate for $\alpha\gtrsim30$, while for $10\lesssim\alpha\lesssim30$ the relative error in the extracted decay time remains below $25\%$. Similarly, the idler coincident count rate is given by $R_{\mathrm{C},i}(\tau)=(c/L)^2 G^{(2)}_{i\textrm{-}s}(\tau)/R_i$, where $G^{(2)}_{i\textrm{-}s}(\tau) =(L/c)^2 R_s R_i + |\Psi_{i\textrm{-}s}(\tau)|^2$, and
\begin{align}
	\Psi_{i\textrm{-}s}(\tau)
	= &  \frac{ L}{2\pi c}
	\int d\omega e^{i\omega \tau}\bigg( A_1C_1^{\ast} \notag\\
	& + \sum_{jk,j'k'}\int dz P_{jk} \mathscr{D}_{kjj'k'} Q^{\ast}_{j'k'}\bigg).
	\label{eq(E6)}
\end{align}
We numerically verify that $\Psi_{s\textrm{-}i}(\tau)=\Psi_{i\textrm{-}s}(-\tau)$. Under the large-OD condition, the analytical biphoton wave function for the idler field can be obtained in the same manner.

%%%%%%%%%%%%%%%%%%%%%%%%%%%%%%%%%%%%%%%%%%%%%%%%%%%%%%%%%%%%%%%%%%%%%%%%%%%%%%%%%%%%%%%%%%%%%%%%%%%%%
%%%%%%%%%%%%%%%%%%%%%%%%%%%%%%%%%%%%%%%%%%%%%%%%%%%%%%%%%%%%%%%%%%%%%%%%%%%%%%%%%%%%%%%%%%%%%%%%%%%%%

\section{BIPHOTON GENERATION IN A BACKWARD CONFIGURATION} \label{ApxF}

Here we derive the biphoton solutions in a backward-field arrangement and compare them with those obtained in the forward scheme. In this configuration, the coupling laser in Fig.~\ref{fig1} propagates in the opposite direction to the driving laser. In this scenario, the HLEs remain identical to those in the forward scheme, while momentum conservation requires the idler field to be counter-propagating with respect to the signal field. Therefore, the coupled MSEs become
\begin{align}
	& \bigg(\frac{1}{c}\frac{\partial}{\partial t}+\frac{\partial}{\partial z}\bigg)\hat{a}_{s}(z,t)=\frac{ig_{s}N}{c}\hat{\sigma}^{(1)}_{34}(z,t), \label{eq(F1)} \\
	& \bigg(\frac{1}{c}\frac{\partial}{\partial t}-\frac{\partial}{\partial z}-i\Delta k\bigg)\hat{a}^{\dagger}_{i}(z,t)=-\frac{ig_{i}N}{c}\hat{\sigma}^{(1)}_{31}(z,t).
	\label{eq(F2)}
\end{align}
In this case, the corresponding coefficients for the signal field are
\begin{align}
	\kappa_{s,{\textrm{B}}}	& = \frac{i\alpha\sqrt{s_\lambda\Gamma_{43}\Gamma_{31}}\Omega_c\Omega_d}{4L\Delta_1(\gamma_{41}-2i\Delta_2)(\gamma_{31}-2i\omega)},
	\label{eq(F3)}\\
	G_{s,{\textrm{B}}} 	& =  \frac{-\alpha s_\lambda\Gamma_{43}|\Omega_c|^2|\Omega_d|^2}{8L \Delta^2_1(\gamma_{41}+2i\Delta_2)(\gamma_{31}-2i\omega)[\gamma_{43}-2i(\Delta_2+\omega)]}\notag\\
	& \quad -\frac{i\omega}{c},
	\label{eq(F4)}\\
	\zeta^s_{31,{\textrm{B}}} & = \sqrt{\frac{\alpha s_\lambda\Gamma_{43}}{4L}} 
	\left(
	\frac{\Omega_c\Omega_d}{\Delta_1(\gamma_{31}-2i\omega)[\gamma_{43}-2i(\Delta_2+\omega)]}
	\right),
	\label{eq(F5)}\\
	\zeta^s_{32,{\textrm{B}}} & = \sqrt{\frac{\alpha s_\lambda\Gamma_{43}}{L}}
	\left(
	\frac{-i\Omega_d}{2\Delta_1[\gamma_{43}-2i(\Delta_2+\omega)]}
	\right),
	\label{eq(F6)}\\
	\zeta^s_{34,{\textrm{B}}} 	&  = \sqrt{\frac{\alpha s_\lambda\Gamma_{43}}{L}}\left[\frac{i}{\gamma_{43}-2i(\Delta_2+\omega)}\right].
	\label{eq(F7)}
\end{align}
These terms are formally identical to those obtained in the forward scheme. Similarly, the coefficients for the idler field are
\begin{align}
	&\Gamma_{i,{\textrm{B}}} =  -\frac{\alpha\Gamma_{31}}{2L(\gamma_{31}-2i\omega)}-\frac{i\omega}{c}+i\Delta k, \label{eq(F8)}\\
	&\kappa_{i,{\textrm{B}}} =  -\frac{i\alpha\sqrt{s_\lambda\Gamma_{43}\Gamma_{31}}\Omega^{\ast}_c\Omega^{\ast}_d}{4L\Delta_1(\gamma_{41}+2i\Delta_2)(\gamma_{31}-2i\omega)},
	\label{eq(F9)}\\
	&\zeta^i_{31,{\textrm{B}}}  = \sqrt{\frac{\alpha\Gamma_{31}}{L}}\left(\frac{i}{\gamma_{31}-2i\omega}\right),
	\label{eq(F10)}\\
	&\zeta^i_{32,{\textrm{B}}}  = \sqrt{\frac{\alpha\Gamma_{31}}{L}}\left[\frac{-i\Omega^{\ast}_c}{2\Delta_1(\gamma_{31}-2i\omega)}\right],
	\label{eq(F11)}\\
	&\zeta^i_{34,{\textrm{B}}}  =  \sqrt{\frac{\alpha\Gamma_{31}}{L}} 
	\left(
	\frac{\Omega^{\ast}_c\Omega^{\ast}_d}{2\Delta_1(\gamma_{31}-2i\omega)[\gamma_{43}-2i(\Delta_2+\omega)]}
	\right).
	\label{eq(F12)}
\end{align}
Equations~(\ref{eq(F8)})--(\ref{eq(F12)}) have the same functional form as those in the forward scheme, differing only by an overall minus sign. The general solutions of the signal and idler fields have the same functional form as those for the forward scheme in Eq. (\ref{eq(11)}) with the corresponding coefficients replaced. However, because the fields are counter-propagating, the idler output is taken at $z=0$. Therefore, we rearrange Eq. (\ref{eq(11)}) as follows:
\begin{align}
	\begin{bmatrix}
		\tilde{a}_{sL} \\
		\tilde{a}^\dagger_{i0} \\
	\end{bmatrix}
	=  &
	\begin{bmatrix}
		A_{1,{\textrm{B}}} & B_{1,{\textrm{B}}} \\
		C_{1,{\textrm{B}}} & D_{1,{\textrm{B}}} \\
	\end{bmatrix}
	\begin{bmatrix}
		\tilde{a}_{s0} \\
		\tilde{a}^{\dagger}_{iL} \\
	\end{bmatrix}
	\notag\\
	& +\sum_{jk}\int^L_0
	\begin{bmatrix}
		A_{2,{\textrm{B}}} & B_{2,{\textrm{B}}}\\
		C_{2,{\textrm{B}}} & D_{2,{\textrm{B}}} \\
	\end{bmatrix}
	\begin{bmatrix}
		\zeta^s_{jk,{\textrm{B}}} \\
		\zeta^i_{jk,{\textrm{B}}} \\
	\end{bmatrix}
	\tilde{f}_{jk}dz, 
	\label{eq(F13)}
\end{align}
where the linearity of Eq. (\ref{eq(F13)}) yields that 
\begin{align}
	A_{1,{\textrm{B}}} = &  A_1 -\frac{B_1 C_1}{D_1}  \notag\\
	= & 1+2\bigg(\frac{\kappa_{s,{\textrm{B}}}\kappa_{i,{\textrm{B}}}}{\Gamma^2_{i,{\textrm{B}}}}\bigg)
	\bigg(1-\frac{e^{L\Gamma_{i,{\textrm{B}}}}+e^{-L\Gamma_{i,{\textrm{B}}}}}{2}\bigg),
	\label{eq(F14)}\\
	B_{1,{\textrm{B}}} = & \frac{B_1}{D_1}
	= 
    \frac{\kappa_{s,{\textrm{B}}}}{\Gamma_{i,{\textrm{B}}}}\left(
    1-e^{L\Gamma_{i,{\textrm{B}}}}
    \right),
	\label{eq(F15)}\\
	C_{1,{\textrm{B}}} = & -\frac{C_1}{D_1}
	=     \frac{\kappa_{i,{\textrm{B}}}}{\Gamma_{i,{\textrm{B}}}}\left(
	e^{L\Gamma_{i,{\textrm{B}}}}-1
	\right),
	\label{eq(F16)}\\
	D_{1,{\textrm{B}}} = & \frac{1}{D_1}
	=  e^{L\Gamma_{i,{\textrm{B}}}},
	\label{eq(F17)}
\end{align}
and
\begin{align}
	A_{2,{\textrm{B}}} = &  A_2 -\frac{B_1 C_2}{D_1}  \notag\\
	= & 1+\bigg(\frac{\kappa_{s,{\textrm{B}}}\kappa_{i,{\textrm{B}}}}{\Gamma_{i,{\textrm{B}}}^2}\bigg)
    \left(e^{L\Gamma_{i,{\textrm{B}}}}-1\right) \left[e^{(z-L)\Gamma_{i,{\textrm{B}}}}-1\right],
	\label{eq(F18)}\\
	B_{2,{\textrm{B}}} = & B_2-\frac{B_1 D_2}{D_1}
	=  
    \frac{\kappa_{s,{\textrm{B}}}}{\Gamma_{i,{\textrm{B}}}}\left(
    e^{z\Gamma_{i,{\textrm{B}}}}-1
    \right),
	\label{eq(F19)}\\
	C_{2,{\textrm{B}}} = & -\frac{C_2}{D_1}
	=     \frac{\kappa_{i,{\textrm{B}}}}{\Gamma_{i,{\textrm{B}}}}\left(e^{L\Gamma_{i,{\textrm{B}}}}-
	e^{z\Gamma_{i,{\textrm{B}}}}
	\right),
	\label{eq(F20)}\\
	D_{2,{\textrm{B}}}
    = & -\frac{D_2}{D_1}
	= -e^{z\Gamma_{i,{\textrm{B}}} }.
	\label{eq(F21)}
\end{align}
Note that the coefficients $B_{1,{\textrm{B}}}$ and $C_{1,{\textrm{B}}}$ are identical to those in the forward scheme [Eqs. (\ref{eq(13)}) and (\ref{eq(14)})]. 

%%%%%%%%%%
\fignine
%%%%%%%%%%

In the following comparison, we consider the phase-matched condition $\Delta k \approx 0$, which can be realized experimentally in both forward and backward configurations through appropriate geometrical alignment and detuning adjustments, such that $\Delta kL \ll 1$. Figs. \ref{fig9}(a) depicts the backward pairing spectra of the signal $\tilde{R}_{s,p,{\textrm{B}}}(\omega)$ and idler $\tilde{R}_{i,p,{\textrm{B}}}(\omega)$. Since $B_{1,{\textrm{B}}}=B_1$ and $C_{1,{\textrm{B}}}=C_1$, both profiles are identical to the forward results in Fig. \ref{fig2}(a). The presented unpairing spectra in Figs. \ref{fig9}(b) also show no significant differences to those in in Fig. \ref{fig2}(b). As shown in Fig. \ref{fig9}(c), the forward ($R_{\mathrm{C},s}$) and backward ($R_{\mathrm{C},s,{\textrm{B}}}$) signal coincidence count rates agree quantitatively with each other. The consistency of the biphoton correlation function decay time versus the OD in Fig. \ref{fig9}(d) between two schemes also indicates that the superradiance still dominates the biphoton profile under the backward-field arrangement.

%%%%%%%%%%%%%%%%%%%%%%%%%%%%%%%%%%%%%%%%%%%%%%%%%%%%%%%%%%%%%%%%%%%%%%%%%%%%%%%%%%%%%%%%%%%%%%%%%%%%%
%%%%%%%%%%%%%%%%%%%%%%%%%%%%%%%%%%%%%%%%%%%%%%%%%%%%%%%%%%%%%%%%%%%%%%%%%%%%%%%%%%%%%%%%%%%%%%%%%%%%%

\section{SUPERRADIANT DECAY TIME} \label{ApxG}

The characteristic superradiant decay time can be expressed as $\tau_s=\tau_{31}/(1+N\mu)$, where $\tau_{31}=1/\Gamma_{31}$ is the natural radiative lifetime of the $\lvert 3\rangle\!\to\!\lvert 1\rangle$ transition, and $\mu$ is a geometrical factor determined by the spatial mode of the atomic ensemble~\cite{Rehler}. In the limit $A\gg L\lambda_{31}$, where $A$ denotes the transverse cross-sectional area of the medium and $\lambda_{31}$ is the transition wavelength, the geometrical factor is given by~\cite{Friedberg}
\begin{align}
	\mu=\frac{3\lambda_{31}^{2}}{8\pi A}.
	\label{eq(G1)}
\end{align}
This condition corresponds to a large Fresnel number, so that diffraction can be neglected within the plane-wave approximation~\cite{Gross}. Substituting Eq.~(\ref{eq(G1)}) into the expression for $\tau_s$, we obtain
\begin{align}
	\tau_s
	&=\frac{\tau_{31}}
	{1+nAL\,(3\lambda_{31}^{2}/8\pi A)}
	=\frac{\tau_{31}}{1+\alpha/4},
	\label{eq(G2)}
\end{align}
where $N=nAL$ is the total number of atoms in the interaction volume and $\alpha=n\sigma L$ is the resonant OD, with $\sigma = 3\lambda_{31}^2/(2\pi)$ the single-atom resonant scattering (absorption) cross section.

%%%%%%%%%%%%%%%%%%%%%%%%%%%%%%%%%%%%%%%%%%%%%%%%%%%%%%%%%%%%%%%%%%%%%%%%%%%%%%%%%%%%%%%%%%%%%%%%%%%%%
%%%%%%%%%%%%%%%%%%%%%%%%%%%%%%%%%%%%%%%%%%%%%%%%%%%%%%%%%%%%%%%%%%%%%%%%%%%%%%%%%%%%%%%%%%%%%%%%%%%%%

\section{STATISTICAL PROPERTIES OF THE SIGNAL AND IDLER FIELDS} \label{ApxH}

We consider the signal and idler fields initially uncorrelated with a vacuum reservoir. The initial density operator is taken as $\rho(0)=\rho^s(0)\otimes\rho^i(0)\otimes\rho^R(0)$, and the total state at time $t$ is given by $U(t)\rho(0)U^{\dagger}(t)$, where $U(t)$ denotes the evolution operator. The reduced density matrix of the output signal field is therefore $\rho^s(t)=\mathrm{Tr}_{i,R}[U(t)\rho(0)U^{\dagger}(t)]$. To evaluate the Fock-basis matrix elements of $\rho^s(t)$, we employ the vacuum projector
\begin{align}
	|0\rangle\langle 0|
	= :\exp(-\hat{a}^\dagger\hat{a}):
	= \sum_{l=0}^{\infty}\frac{(-1)^l}{l!}
	(\hat{a}^\dagger)^l \hat{a}^l,
	\label{eq(H1)}
\end{align}
where $:\cdots:$ denotes normal ordering. The output matrix elements of the signal field can then be written as~\cite{Cheng,Hsu,Liu1}
\begin{align}
	\rho^s_{mn}(t)
	=\mathrm{Tr}\!\left\{
	\hat{\rho}^{\,s}_{mn}(t)\,\rho(0)
	\right\},
	\label{eq(H2)}
\end{align}
with
\begin{align}
	\hat{\rho}^{\,s}_{mn}(t)
	=
	\sum_{l=0}^{\infty}
	\frac{(-1)^l}{l!\sqrt{m!n!}}
	\left\langle
	\big[\hat{a}^{\dagger}_{sL}(t)\big]^{n+l}
	\big[\hat{a}_{sL}(t)\big]^{m+l}
	\right\rangle .
	\label{eq(H3)}
\end{align}
The output field operator $\hat{a}_{sL}(t)$ follows from Eq.~(\ref{eq(11)}) via the inverse Fourier transform,
\begin{align}
	\hat{a}_{sL}(t)=
	\mathcal{IFT}\!\left(
	A_1\tilde{a}_{s0}
	+ B_1\tilde{a}^\dagger_{i0}
	+ \sum_s\tilde{\mathcal{F}}_s
	\right),
	\label{eq(H4)}
\end{align}
where $\sum_s \tilde{\mathcal{F}}_s$ denotes the collection of Langevin noise operators that couple into the signal mode.

Assuming vacuum inputs for both signal and idler fields, $\rho(0)=|0_s\rangle\langle 0_s|\otimes|0_i\rangle\langle 0_i|\otimes\rho^R(0)$, only diagonal density-matrix elements ($m=n$) remain nonzero. Upon substituting Eq.~(\ref{eq(H4)}) into Eq.~(\ref{eq(H3)}), the resulting expression naturally separates into contributions originating from the idler-vacuum term, which produces paired photons, and from the Langevin noise term, which produces unpaired photons.

Under the Gaussian-noise assumption~\cite{Milonni} and by applying Wick's theorem~\cite{Louisell}, the contribution from the Langevin noise term can be evaluated as
\begin{align}
	\left\langle
	\left(
	\sum_{ss'}\hat{\mathcal{F}}^\dagger_s\hat{\mathcal{F}}_{s'}
	\right)^k
	\right\rangle
	=
	k!\left(\frac{L}{c}R_{s,u}\right)^k,
	\label{eq(H5)}
\end{align}
which is directly determined by the unpaired signal generation rate $R_{s,u}$ defined in Eq.~(\ref{eq(16)}).

The contribution associated with paired photons arises from the idler-vacuum term. Using the commutation relation
$[\hat{a}_{i0}(t),\hat{a}^\dagger_{i0}(t')]=(L/c)\delta(t-t')$,
one finds
\begin{align}
	\left(
	|\mathcal{B}_1|^2\,
	\langle \hat{a}_{i0}\hat{a}^\dagger_{i0}\rangle
	\right)^q
	=
	\left(\frac{L}{c}R_{s,p}\right)^q,
	\label{eq(H6)}
\end{align}
where $\mathcal{B}_1\equiv\mathcal{IFT}(B_1)$ and $R_{s,p}$ denotes the paired signal generation rate defined in Eq.~(\ref{eq(16)}).

Combining Eqs.~(\ref{eq(H5)}) and~(\ref{eq(H6)}), the diagonal elements of the reduced density matrix take the form
\begin{align}
	\rho^s_{nn}(t)
	=
	\sum_{l=0}^{\infty}
	\frac{(-1)^l(l+n)!}{l!\,n!}
	\left[\frac{L}{c}(R_{s,p}+R_{s,u})\right]^{l+n}.
	\label{eq(H7)}
\end{align}
Using the series identity $\sum_{l=0}^\infty \frac{(n)_l}{l!}x^l=(1-x)^{-n}$ with the Pochhammer symbol $(n)_l=n(n+1)\cdots(n+l-1)$~\cite{Olver}, the photon-number distribution of the signal field is obtained as the thermal (geometric) distribution
\begin{align}
	\rho^s_{nn}(t)
	=
	\frac{\langle n_s\rangle^n}{(1+\langle n_s\rangle)^{n+1}},
	\qquad
	\langle n_s\rangle \equiv \frac{L}{c}R_s,
	\label{eq(H8)}
\end{align}
where $R_s\equiv R_{s,p}+R_{s,u}$. An analogous result holds for the idler field,
\begin{align}
	\rho^i_{nn}(t)
	=
	\frac{\langle n_i\rangle^n}{(1+\langle n_i\rangle)^{n+1}},
	\qquad
	\langle n_i\rangle \equiv \frac{L}{c}R_i.
	\label{eq(H9)}
\end{align}

%%%%%%%%%%%%%%%%%%%%%%%%%%%%%%%%%%%%%%%%%%%%%%%%%%%%%%%%%%%%%%%%%%%%%%%%%%%%%%%%%%%%%%%%%%%%%%%%%%%%%
%%%%%%%%%%%%%%%%%%%%%%%%%%%%%%%%%%%%%%%%%%%%%%%%%%%%%%%%%%%%%%%%%%%%%%%%%%%%%%%%%%%%%%%%%%%%%%%%%%%%%

\section{DOPPLER-BROADENED MODEL} \label{ApxI}

\subsection{Derivation of the signal and idler spectra and the biphoton wave function} \label{ApxIa}

In this Appendix, we derive the analytical expressions for the signal and idler spectra, as well as the biphoton wave function, in the presence of Doppler broadening. By solving the first-order HLEs with the Hamiltonian given in Eq.~(\ref{eq(33)}) in Sec.~\ref{Sec. 5} under the GSA condition, and substituting the solutions into the coupled MSEs, we obtain
\begin{align}
	\frac{\partial}{\partial z} \tilde{a}_{s,\textrm{D}}
	& + \int G'_{s} \tilde{a}_{s,\textrm{D}}
	+ \int \kappa'_{s} \tilde{a}^{\dagger}_{i,\textrm{D}}
	= \sum_{jk} \int \zeta'^s_{jk} \tilde{f}_{jk,\textrm{D}},
	\label{eq(I1)} \\
	\frac{\partial}{\partial z} \tilde{a}^{\dagger}_{i,\textrm{D}}
	& + \int \Gamma'_{i} \tilde{a}^{\dagger}_{i,\textrm{D}}
	+ \int \kappa'_{i} \tilde{a}_{s,\textrm{D}}
	= \sum_{jk} \int \zeta'^i_{jk} \tilde{f}_{jk,\textrm{D}}.
	\label{eq(I2)}
\end{align}
On the left-hand side of Eqs.~(\ref{eq(I1)}) and (\ref{eq(I2)}), we define $\int[\bullet] \equiv \int p(v)[\bullet]\,dv$, where $[\bullet]\in\{G'_{s},\kappa'_{s},\Gamma'_{i},\kappa'_{i}\}$ denotes the Doppler-dependent coefficients given in Eqs.~(\ref{eq(34)})--(\ref{eq(37)}), and $p(v)$ is the normalized Maxwell--Boltzmann velocity distribution. On the right-hand side, we define $\int[\ast] \equiv \int \sqrt{p(v)}[\ast]\,dv$, where $[\ast]\in\{\zeta'^s_{jk},\zeta'^i_{jk}\}$. This definition ensures that the effective Langevin noise operators retain their delta-correlation after Doppler averaging. The parameter $\tilde{f}_{jk,\textrm{D}}\equiv \sqrt{p(\textrm{v})N/c}\,\tilde{F}_{jk,\textrm{D}}$ is the renormalized Doppler-broadened Langevin noise operator. The corresponding noise coefficients are given by
\begin{align}
	\zeta'^s_{31} =  &
	\sqrt{\frac{\alpha s_\lambda\Gamma_{43}}{L}}
	\bigg(
	\frac{\Omega_c\Omega_d}{2\Delta'_1[\gamma_{31}-2i(\omega+\Delta'_3)]}
	\notag\\
	& \times \frac{1}{\gamma_{43}-2i(\Delta'_2+\omega+\Delta'_3)}\bigg),
	\label{eq(I3)}\\
	\zeta'^s_{32} = &
	\sqrt{\frac{\alpha s_\lambda\Gamma_{43}}{L}}
	\left(
	\frac{-i\Omega_d}{2\Delta'_1[\gamma_{43}-2i(\Delta'_2+\omega+\Delta'_3)]}
	\right),
	\label{eq(I4)}\\
	\zeta'^s_{34} = & \sqrt{\frac{\alpha s_\lambda\Gamma_{43}}{L}}\left[\frac{i}{\gamma_{43}-2i(\Delta'_2+\omega+\Delta'_3)}\right],
	\label{eq(I5)}\\
	\zeta'^i_{31} =  & \sqrt{\frac{\alpha\Gamma_{31}}{L}}\left[\frac{-i}{\gamma_{31}-2i(\omega+\Delta'_3)}\right],
	\label{eq(I6)}\\
	\zeta'^i_{32} =  & \sqrt{\frac{\alpha\Gamma_{31}}{L}}
	\left(
	\frac{i\Omega^{\ast}_c}{2\Delta'_1[\gamma_{31}-2i(\omega+\Delta'_3)]}
	\right),
	\label{eq(I7)}\\
	\zeta'^i_{34} =  & \sqrt{\frac{\alpha\Gamma_{31}}{L}}
	\bigg(
	\frac{-\Omega^{\ast}_c\Omega^{\ast}_d}{2\Delta'_1[\gamma_{31}-2i(\omega+\Delta'_3)]}
	\notag\\
	& \times \frac{1}{\gamma_{43}-2i(\Delta'_2+\omega+\Delta'_3)}
	\bigg).
	\label{eq(I8)}
\end{align}
Here $jk\in\{31,32,34\}$ labels the relevant atomic transitions, and the Doppler-shifted detunings $\Delta'_1$, $\Delta'_2$, and $\Delta'_3$ are defined in Sec.~\ref{Sec. 5a}.

Next, we derive the signal and idler generation rates $R'_{s}$ and $R'_{i}$. Using the general expression in Eq.~(\ref{eq(42)}), the signal generation rate is obtained as
\begin{align}
	& R'_{s}
	= \frac{c}{L} \int \int
	\frac{d\omega d\omega'}{2\pi}
	\bigg[
	|B'_{1}|^2
	\notag\\
	& + \sum_{jk,j'k'}
	\int \int dzdz' \left(A'^\ast_{2}\int\zeta'^{s\ast}_{jk}+B'^\ast_{2}\int\zeta'^{i\ast}_{jk}\right)
	\notag\\
	& \times \langle \tilde{f}^\dagger_{jk,\textrm{D}}\tilde{f}_{j'k',\textrm{D}}\rangle
	\left(\int\zeta'^s_{j'k'}A'_{2} +\int\zeta'^i_{j'k'}B'_{2}\right)
	\bigg]
	\notag\\
	\equiv & \int \frac{d\omega}{2\pi}  
	\big[\tilde{R}'_{s,p}(\omega)+\tilde{R}'_{s,u}(\omega)\big].
	\label{eq(I9)}
\end{align}
In the Doppler-broadened system, the Langevin noise correlations associated with different atoms are delta-correlated not only in frequency $\omega$ and position $z$, but also in atomic velocity $v$~\cite{Yatsenko},
\begin{align}
	& \langle \tilde{f}^\dagger_{jk,\textrm{D}}(z,-\omega,v)\tilde{f}_{j'k',\textrm{D}}(z',\omega',v')\rangle \notag\\
	= & \frac{L}{2\pi c}\mathscr{D}_{kj,j'k'}(v) \delta(z-z')\delta(\omega+\omega')\delta(v-v'). \label{eq(I10)}
\end{align}
Substituting Eq.~(\ref{eq(I10)}) into Eq.~(\ref{eq(I9)}), the unpairing contribution $R'_{s,u}$ can be explicitly evaluated, since it originates from the Langevin noise correlations. By contrast, the pairing contribution $R'_{s,p}$ arises solely from the coherent field-transmission term $|B'_1|^2$ and can be directly obtained without invoking Eq.~(\ref{eq(I10)}). Combining these two contributions yields the total signal generation rate $R'_s = R'_{s,p} + R'_{s,u}$. The idler generation rate $R'_i$ follows from the same procedure. The resulting expressions for both $R'_{s,p}$ and $R'_{i,p}$ can then be cast into the compact form given in Eq.~(\ref{eq(46)}) in Sec.~\ref{Sec. 5b}.

%%%%%%%%%%%%%%%%%%%%%%%%%%%%%%%%%%%%%%%%%%%%%%%%%%%%%%%%%%%%%%%%%%%%%%%%%%%%%%%%%%%%%%%%%%%%%%%%%%%%%

\subsection{Model with exact populations or backward scheme} \label{ApxIb}

In the main text, we present the Doppler-broadened coefficients under the GSA for clarity. If one wishes to include the exact steady-state populations and coherences, the Doppler-broadened expressions can be constructed in a straightforward way from the corresponding Doppler-free coefficients in Eqs.~(\ref{eq(C1)})--(\ref{eq(C11)}). Specifically, for each velocity class $v$ one replaces the detunings by their Doppler-shifted counterparts and evaluates the velocity average, i.e.,
\begin{equation}
	\Delta_1 \rightarrow \Delta'_1,\qquad
	\Delta_2 \rightarrow \Delta'_2,\qquad
	\omega \rightarrow \omega+\Delta'_3,
\end{equation}
where $\Delta'_1$, $\Delta'_2$, and $\Delta'_3$ are defined in Sec.~\ref{Sec. 5a}. With these substitutions, the structure of the coupled MSEs remains unchanged, while the susceptibility-like coefficients become velocity dependent. Performing the Maxwell--Boltzmann average as in Eqs.~(\ref{eq(I1)}) and~(\ref{eq(I2)}) then yields the Doppler-broadened model with exact populations. The resulting pairing rate is used to produce the curve with exact populations shown in Fig.~\ref{fig7}(d).

The Doppler-broadened biphoton properties in the backward configuration can be obtained by combining the backward-field formulation in Appendix~\ref{ApxF} with the Doppler averaging procedure described in Appendix~\ref{ApxIa}. In the backward geometry, the coupling and driving fields counter-propagate, so that the Doppler contribution to the two-photon detuning changes from $(k_c+k_d)v$ (forward) to $(k_c-k_d)v$ (backward). Accordingly, the effective two-photon detuning becomes $\Delta'_2=\Delta_2+(k_c-k_d)v$, whereas the one- and three-photon detunings $\Delta'_1$ and $\Delta'_3$ remain unchanged because they are associated with single-photon and three-photon resonance conditions set by the optical frequencies. As a consequence, the velocity-induced spread of $\Delta'_2$ is significantly reduced in the backward configuration, which suppresses the photon generation rates (both pairing and unpairing) through the reduced effective two-photon response. By contrast, the biphoton correlation time is only weakly affected by the field geometry, since it is primarily governed by the effective three-photon detuning $\Delta'_3$ (and its Doppler-broadened distribution), as discussed in Sec.~\ref{Sec. 5}.

%%%%%%%%%%%%%%%%%%%%%%%%%%%%%%%%%%%%%%%%%%%%%%%%%%%%%%%%%%%%%%%%%%%%%%%%%%%%%%%%%%%%%%%%%%%%%%%%%%%%%
%%%%%%%%%%%%%%%%%%%%%%%%%%%%%%%%%%%%%%%%%%%%%%%%%%%%%%%%%%%%%%%%%%%%%%%%%%%%%%%%%%%%%%%%%%%%%%%%%%%%%

%%%%%%%%%%%%%%%%%%%%%%%%%%%%%%%%%%%%%%%%%%%%%%%%%%%%%%%%%%%%%%%%%%%%%%%%%%%%%%%%%%%%%%%%%%%%%%%%%%%%%
%%%%%%%%%%%%%%%%%%%%%%%%%%%%%%%%%%%%%%%%%%%%%%%%%%%%%%%%%%%%%%%%%%%%%%%%%%%%%%%%%%%%%%%%%%%%%%%%%%%%%


\begin{thebibliography}{1}
	
	\bibitem{Dicke}
	R. H. Dicke,
	Coherence in Spontaneous Radiation Processes,
	Phys. Rev. \textbf{93}, 99 (1954).
	
	\bibitem{Ernst}
	V. Ernst and P. Stehle,
	Emission of Radiation from a System of Many Excited Atoms,
	Phys. Rev. \textbf{176}, 1456 (1968).
	
	\bibitem{Agarwal}
	G. S. Agarwal,
	Master-Equation Approach to Spontaneous Emission,
	Phys. Rev. A \textbf{2}, 2038 (1970).
	
	\bibitem{Rehler}
	N. E. Rehler and J. H. Eberly,
	Superradiance,
	Phys. Rev. A \textbf{3}, 1735 (1971).
	
	\bibitem{Gross}
	M. Gross and S. Haroche,
	Superradiance: An essay on the theory of collective spontaneous emission,
	Phys. Rep. \textbf{93}, 301 (1982).
	
	\bibitem{Scully}
	M. O. Scully, E. S. Fry, C. H. R. Ooi, and K. W\'{o}dkiewicz,
	Directed Spontaneous Emission from an Extended Ensemble of
	N Atoms: Timing Is Everything,
	Phys. Rev. Lett. \textbf{96}, 010501 (2006).
	
	%\bibitem{Scully2}
	%M. O. Scully and A. A. Svidzinsky,
	%The Super of Superradiance,
	%Science \textbf{325}, 1510 (2009).
	
	\bibitem{Chaneliere}
	T. Chaneli\`{e}re, D. N. Matsukevich, S. D. Jenkins, T. A. B. Kennedy, M. S. Chapman, and A. Kuzmich,
	Quantum Telecommunication Based on Atomic Cascade Transitions,
	Phys. Rev. Lett. \textbf{96}, 093604 (2006).
	
	\bibitem{Srivathsan}
	B. Srivathsan, G. K. Gulati, B. Chng, G. Maslennikov, D. Matsukevich, and C. Kurtsiefer,
	Narrow Band Source of Transform-Limited Photon Pairs via Four-Wave Mixing in a Cold Atomic Ensemble,
	Phys. Rev. Lett. \textbf{111}, 123602 (2013).
	
	\bibitem{Srivathsan2}
	B. Srivathsan, G. K. Gulati, A. Cer\`{e}, B. Chng, and C. Kurtsiefer,
	Reversing the Temporal Envelope of a Heralded Single Photon using a Cavity,
	Phys. Rev. Lett. \textbf{113}, 163601 (2014).
	
	\bibitem{Zhang}
	W. Zhang, D.-S. Ding, S. Shi, Y. Li, Z.-Y. Zhou, B.-S. Shi, and G.-C. Guo,
	Storing a single photon as a spin wave entangled with a flying photon in the telecommunication bandwidth,
	Phys. Rev. A \textbf{93}, 022316 (2016).
	
	\bibitem{Park}
	J. Park, T. Jeong, H. Kim, and H. S. Moon,
	Time-Energy Entangled Photon Pairs from Doppler-Broadened Atomic Ensemble via Collective Two-Photon Coherence,
	Phys. Rev. Lett. \textbf{121}, 263601 (2018).
	
	\bibitem{Jeong2}
	H. Jeong, H. Kim, and H. S. Moon,
	High‐Performance Telecom‐Wavelength Biphoton Source from a Hot Atomic Vapor Cell,
	Adv. Quantum Technol. \textbf{7}, 2300108 (2023).
	
	\bibitem{Cirac}
	J. I. Cirac, R. Blatt, A. S. Parkins, and P. Zoller,
	Preparation of Fock states by observation of quantum jumps in an ion trap,
	Phys. Rev. Lett. \textbf{70}, 762 (1993).
	
	\bibitem{Varcoe}
	B. T. H. Varcoe, S. Brattke, M. Weidinger, and H. Walther,
	Preparing pure photon number states of the radiation field,
	Nature \textbf{403}, 743 (2000).
	
	\bibitem{Bertet}
	P. Bertet, S. Osnaghi, P. Milman, A. Auffeves, P. Maioli, M. Brune, J. M. Raimond, and S. Haroche,
	Generating and Probing a Two-Photon Fock State with a Single Atom in a Cavity,
	Phys. Rev. Lett. \textbf{88}, 143601 (2002).
	
	\bibitem{Walther}
	P. Walther, J.-W. Pan, M. Aspelmeyer, R. Ursin, S. Gasparoni, and A. Zeilinger,
	De Broglie wavelength of a non-local four-photon state,
	Nature \textbf{429}, 158 (2004).
	
	\bibitem{Ourjoumtsev}
	A. Ourjoumtsev, H. Jeong, R. Tualle-Brouri, and P. Grangier,
	Generation of optical ‘Schr\"{o}dinger cats’ from photon number states,
	Nature \textbf{448}, 784 (2007).
	
	\bibitem{OBrien}
	J. L. O'Brien, A. Furusawa, and J. Vu\v{c}kovi\'{c},
	Photonic quantum technologies,
	Nat. Photon. \textbf{3}, 687 (2009).
	
	\bibitem{Afek}
	I. Afek, O. Ambar, Y. Silberberg,
	High-NOON States by Mixing Quantum and Classical Light,
	Science \textbf{328}, 879 (2010).
	
	\bibitem{Jeong1}
	H. Jeong, A. Zavatta, M. Kang, S.-W. Lee, L. S. Costanzo, S. Grandi, T. C. Ralph, and M. Bellini,
	Generation of hybrid entanglement of light,
	Nat. Photon. \textbf{8}, 564 (2014).
	
	\bibitem{Tan}
	K. C. Tan and H. Jeong,
	Nonclassical light and metrological power: An introductory review,
	AVS Quantum Sci. \textbf{1}, 014701 (2019).
	
	\bibitem{Jia}
	J.-M. Chen, T. Peters, P.-H. Hsieh, and I. A. Yu,
	Review of Biphoton Sources Based on the Double-$\Lambda$
	Spontaneous Four-Wave Mixing Process,
	Adv. Quantum Technol. \textbf{7}, 2400138 (2024).
	
	\bibitem{Kimble}
	H. J. Kimble,
	The quantum internet,
	Nature \textbf{453}, 1023 (2008).
	
	\bibitem{Wehner}
	S. Wehner, D. Elkouss, and R. Hanson,
	Quantum internet: A vision for the road ahead,
	Science \textbf{362}, eaam9288 (2018).
	
	\bibitem{Lodahl}
	P. Lodahl,
	Quantum-dot based photonic quantum networks,
	Quantum Sci. Technol. \textbf{3}, 013001 (2018).
	
	\bibitem{Aguilar}
	L. Villegas-Aguilar, E. Polino, F. Ghafari, M. T. Quintino, K. T. Laverick, I. R. Berkman, S. Rogge, L. K. Shalm, N. Tischler, E. G. Cavalcanti, S. Slussarenko, and G. J. Pryde,
	Nonlocality activation in a photonic quantum network,
	Nat. Commun. \textbf{15}, 3112 (2024).
	
	\bibitem{Rad}
	H. A. Rad \textit{et al.},
	Scaling and networking a modular photonic quantum computer,
	Nature \textbf{638}, 912 (2025).
	
	\bibitem{Flamini}
	F. Flamini, N. Spagnolo, and F. Sciarrino,
	Photonic quantum information processing: a review,
	Rep. Prog. Phys. \textbf{82}, 016001 (2019).
	
	\bibitem{JWang}
	J. Wang, F. Sciarrino, A. Laing, and M. G. Thompson,
	Integrated photonic quantum technologies,
	Nat. Photon. \textbf{14}, 273 (2020).
	
	\bibitem{LLu}
	L. Lu, X. Zheng, Y. Lu, S. Zhu, X.-S. Ma,
	Advances in Chip-Scale Quantum Photonic Technologies,
	Adv. Quantum Technol. \textbf{4}, 2100068 (2021).
	
	\bibitem{Toninelli}
	C. Toninelli, I. Gerhardt, A. S. Clark, A. Reserbat-Plantey, S. G\"{o}tzinger, Z. Ristanovi\'{c}, M. Colautti, P. Lombardi, K. D. Major, I. Deperasi\'{n}ska, W. H. Pernice, F. H. L. Koppens, B. Kozankiewicz, A. Gourdon, V. Sandoghdar, and M. Orrit ,
	Single organic molecules for photonic quantum technologies,
	Nat. Mater. \textbf{20}, 1615 (2021).
	
	\bibitem{Pelucchi}
	E. Pelucchi, G. Fagas, I. Aharonovich, D. Englund, E. Figueroa, Q. Gong, H. Hannes, J. Liu, C.-Y. Lu, N. Matsuda, J.-W. Pan, F. Schreck, F. Sciarrino, C. Silberhorn, J. Wang, and K. D. J\"{o}ns,
	The potential and global outlook of integrated photonics for quantum technologies,
	Nat. Rev. Phys. \textbf{4}, 194 (2022).
	
	\bibitem{Gisin}
	N. Gisin and R. Thew,
	Quantum communication,
	Nat. Photon. \textbf{1}, 165 (2007).
	
	\bibitem{Muralidharan}
	S. Muralidharan, L. Li, J. Kim, N. L\"{u}tkenhaus, M. D. Lukin, and L. Jiang,
	Optimal architectures for long distance quantum communication,
	Sci. Rep. \textbf{6}, 20463 (2016).
	
	\bibitem{Paraiso}
	T. K. Para\"{i}so, T. Roger, D. G. Marangon, I. D. Marco, M. Sanzaro, R. I. Woodward, J. F. Dynes, Z. Yuan, and Andrew J. Shields,
	A photonic integrated quantum secure communication system,
	Nat. Photon. \textbf{15}, 850 (2021).
	
	\bibitem{Guzik}
	A. Aspuru-Guzik and P. Walther,
	Photonic quantum simulators,
	Nat. Phys. \textbf{8}, 285 (2012).
	
	\bibitem{Slussarenko}
	S. Slussarenko and G. J. Pryde,
	Photonic quantum information processing: A concise review,
	Appl. Phys. Rev. \textbf{6}, 041303 (2019).
	
	\bibitem{Takeda}
	S. Takeda and A. Furusawa,
	Toward large-scale fault-tolerant universal photonic quantum computing,
	APL Photonics \textbf{4}, 060902 (2019).
	
	\bibitem{Giovannetti}
	V. Giovannetti, S. Lloyd, and L. Maccone,
	Quantum metrology,
	Phys. Rev. Lett. \textbf{96}, 010401 (2006).
	
	\bibitem{Giovannetti2}
	V. Giovannetti, S. Lloyd, and L. Maccone,
	Advances in quantum metrology,
	Nat. Photon. \textbf{5}, 222 (2011).
	
	\bibitem{Polino}
	E. Polino, M. Valeri, N. Spagnolo, and F. Sciarrino,
	Photonic quantum metrology,
	AVS Quantum Sci. \textbf{2}, 024703 (2020).
	
	\bibitem{Ikuta}
	R. Ikuta, Y. Kusaka, T. Kitano, H. Kato, T. Yamamoto, M. Koashi, and N. Imoto,
	Wide-band quantum interface for visible-to-telecommunication wavelength conversion,
	Nat. Commun. \textbf{2}, 537 (2011).
	
	\bibitem{Zaske}
	S. Zaske, A. Lenhard, C. A. Keßler, J. Kettler, C. Hepp, C. Arend, R. Albrecht, W.-M. Schulz, M. Jetter, P. Michler, and C. Becher,
	Visible-to-Telecom Quantum Frequency Conversion of Light from a Single Quantum Emitter,
	Phys. Rev. Lett.  \textbf{109}, 147404 (2012).
	
	\bibitem{Stolk}
	A. J. Stolk, K. L. van der Enden, M.-C. Roehsner, A. Teepe, S. O. J. Faes, C. E. Bradley, S. Cadot, J. van Rantwijk, I. te Raa, R. A. J. Hagen, A. L. Verlaan, J. J. B. Biemond, A. Khorev, R. Vollmer, M. Markham, A. M. Edmonds, J. P. J. Morits, T. H. Taminiau, E. J. van Zwet, and R. Hanson,
	Telecom-Band Quantum Interference of Frequency-Converted Photons from Remote Detuned NV Centers,
	PRX Quantum \textbf{3}, 020359 (2022).
	
	\bibitem{Tseng}
	P.-H. Tseng, L.-C. Chen, J.-S. Shiu, and Y.-F. Chen,
	Quantum interface for telecom frequency conversion based on diamond-type atomic ensembles,
	Phys. Rev. A \textbf{109}, 043716 (2024).
	
	\bibitem{Pelton}
	M. Pelton, P. Marsden, D. Ljunggren, M. Tengner, A. Karlsson, A. Fragemann, C. Canalias, and F. Laurell,
	Bright, single-spatial-mode source of frequency non-degenerate, polarization-entangled photon pairs using periodically poled KTP,
	Opt. Express \textbf{12}, 3573 (2004).
	
	\bibitem{Arahira}
	S. Arahira, N. Namekata, T. Kishimoto, H. Yaegashi, and S. Inoue,
	Generation of polarization entangled photon pairs at telecommunication wavelength using cascaded $\chi^{(2)}$ processes in a periodically poled LiNbO$_3$ ridge waveguide,
	Opt. Express \textbf{19}, 16032 (2011).
	
	\bibitem{Hentschel}
	M. Hentschel, H. H\"{u}bel, A. Poppe, and A. Zeilinger,
	Three-color Sagnac source of polarization-entangled photon pairs,
	Opt. Express \textbf{17}, 23153 (2009).
	
	\bibitem{Clausen}
	C. Clausen, I. Usmani, F\'{e}lix Bussi\`{e}res, N. Sangouard, M. Afzelius, H. d. Riedmatten, and N. Gisin,
	Quantum storage of photonic entanglement in a crystal,
	Nature \textbf{469}, 508 (2011).
	
	\bibitem{Saglamyurek}
	E. Saglamyurek, N. Sinclair, J. Jin, J. A. Slater, D. Oblak, F\'{e}lix Bussi\`{e}res, M. George,
	R. Ricken, W. Sohler, and W. Tittel,
	Broadband waveguide quantum memory for entangled photons,   Nature \textbf{469}, 512 (2011).
	
	\bibitem{Stuart}
	T. E. Stuart, J. A. Slater, F. Bussi\`{e}res, and W. Tittel,
	Flexible source of nondegenerate entangled photons based on a two-crystal Sagnac interferometer,
	Phys. Rev. A \textbf{88}, 012301 (2013).
	
	\bibitem{Bussieres2}
	F. Bussi\`{e}res, C. Clausen, A. Tiranov, B. Korzh, V. B. Verma,
	S. W. Nam, F. Marsili, A. Ferrier, P. Goldner, H. Herrmann, C. Silberhorn, W. Sohler, M. Afzelius, and N. Gisin,
	Quantum teleportation from a telecom-wavelength photon to a solid-state quantum memory,
	Nat. Photon. \textbf{8}, 775 (2014).
	
	\bibitem{Dietz}
	O. Dietz, C. M\"{u}ller, T. Kreißl, U. Herzog, T. Kroh, A. Ahlrichs, and O. Benson,
	A folded-sandwich polarization-entangled two-color photon pair source with large tuning capability for applications in hybrid quantum systems,
	Appl. Phys. B \textbf{122}, 33 (2016).
	
	\bibitem{XLu}
	X. Lu, Q. Li, D. A. Westly, G. Moille, A. Singh, V. Anant, and K. Srinivasan,
	Chip-integrated visible–telecom entangled photon pair source for quantum communication,
	Nat. Phys. \textbf{15}, 373 (2019).
	
	\bibitem{QWang}
	J.-Q. Wang, Y.-H. Yang, M. Li, H. Zhou, X.-B. Xu, J.-Z. Zhang, C.-H. Dong, G.-C. Guo, and C.-L. Zou,
	Synthetic five-wave mixing in an integrated microcavity for visible-telecom entanglement generation,
	Nat. Commun. \textbf{13}, 6223 (2022).
	
	\bibitem{Schunk}
	G. Schunk, U. Vogl, D. V. Strekalov, M. F\"{o}rtsch, F. Sedlmeir, H. G. L. Schwefel, M. G\"{o}belt, S. Christiansen, G. Leuchs, and C. Marquardt,
	Interfacing transitions of different alkali atoms and telecom bands using one narrowband photon pair source,
	Optica \textbf{2}, 773 (2015).
	
	\bibitem{Duan}
	L.-M. Duan, M. D. Lukin, J. I. Cirac, and P. Zoller,
	Long-distance quantum communication with atomic ensembles and linear optics,
	Nature \textbf{414}, 413 (2001).
	
	\bibitem{Bussieres}
	F. Bussi\`{e}res, N. Sangouard, M. Afzelius, H. d. Riedmatten, C. Simon, and W. Tittel,
	Prospective applications of optical quantum memories,
	J. Mod. Opt. \textbf{60}, 1519 (2013).
	
	\bibitem{Pu}
	Y. Pu, Y. Wu, N. Jiang, W. Chang, C. Li, S. Zhang, and Luming Duan,
	Experimental entanglement of 25 individually accessible atomic quantum interfaces,
	Sci. Adv. \textbf{4}, eaar3931 (2018).
	
	\bibitem{Covey}
	J. P. Covey, H. Weinfurter, and H. Bernien,
	Quantum networks with neutral atom processing nodes,
	npj. Quantum Inf. \textbf{9}, 90 (2023).
	
	\bibitem{Ding}
	D.-S. Ding, Z.-Y. Zhou, B.-S. Shi, X.-B. Zou, and G.-C. Guo,
	Generation of non-classical correlated photon pairs via a ladder-type atomic configuration: theory and experiment,
	Opt. Express \textbf{20}, 11433 (2012).
	
	\bibitem{Lee2}
	Y.-S. Lee, S. M. Lee, H. Kim, and H. S. Moon,
	Single-photon superradiant beating from a Doppler-broadened ladder-type atomic ensemble,
	Phys. Rev. A \textbf{96}, 063832 (2017).
	
	\bibitem{Park3}
	J. Park, H. Kim, and H. S. Moon,
	Polarization-Entangled Photons from a Warm Atomic Ensemble Using a Sagnac Interferometer,
	Phys. Rev. Lett. \textbf{122}, 143601 (2019).
	
	\bibitem{Willis}
	R. T. Willis, F. E. Becerra, L. A. Orozco, and S. L. Rolston,
	Correlated photon pairs generated from a warm atomic ensemble,
	Phys. Rev. A  \textbf{82}, 053842  (2010).
	
	\bibitem{Willis2}
	R. T. Willis, F. E. Becerra, L. A. Orozco, and S. L. Rolston,
	Photon statistics and polarization correlations at telecommunications wavelengths from a warm atomic ensemble,
	Opt. Express \textbf{19}, 14632 (2011).
	
	\bibitem{Dong}
	M.-X. Dong, W. Zhang, S. Shi, K. Wang, Z.-Y. Zhou, S.-L. Liu, D.-S. Ding, and B.-S. Shi,
	Two-color hyper-entangled photon pairs generation in a cold $^{85}$Rb atomic ensemble,
	Opt. Express \textbf{25}, 10145 (2017).
	
	\bibitem{Craddock}
	A. N. Craddock, Y. Wang, F. Giraldo, R. Sekelsky, M. Flament, and M. Namazi,
	High-rate subgigahertz-linewidth bichromatic entanglement source for quantum networking,
	Phys. Rev. Appl. \textbf{21}, 034012 (2024).
	
	\bibitem{Gulati}
	G. K. Gulati, B. Srivathsan, B. Chng, A. Cer\`{e}, and C. Kurtsiefer,
	Polarization entanglement and quantum beats of photon pairs from four-wave mixing in a cold $^{87}$Rb ensemble,
	New. J. Phys. \textbf{17}, 093034 (2015).
	
	\bibitem{Whiting}
	D. J. Whiting, N. \u{S}ibali\'c, J. Keaveney, C. S. Adams, and I. G. Hughes,
	Single-Photon Interference due to Motion in an Atomic Collective Excitation,
	Phys. Rev. Lett. \textbf{118}, 253601 (2017).
	
	\bibitem{Tu}
	P.-Y. Tu, C.-Y. Hsu, W.-K. Huang, T.-Y. Lin, C.-S. Chuu, and I. A. Yu,
	Temporally long C-band heralded single photons generated from hot atoms,
	APL Photon. \textbf{10}, 106104 (2025).
	
	\bibitem{Kolchin}
	P. Kolchin,
	Electromagnetically-induced-transparency-based paired photon generation,
	Phys. Rev. A \textbf{75}, 033814 (2007).
	
	\bibitem{Ooi}
	C. H. R. Ooi, Q. Sun, M. S. Zubairy, and M. O. Scully,
	Correlation of photon pairs from the double Raman amplifier: Generalized analytical quantum Langevin theory,
	Phys. Rev. A \textbf{75}, 013820 (2007).
	
	\bibitem{LZhao}
	L. Zhao, Y. Su, and S. Du,
	Narrowband biphoton generation in the group delay regime,
	Phys. Rev. A \textbf{93}, 033815 (2016).
	
	\bibitem{Kolchin2}
	P. Kolchin, S. Du, C. Belthangady, G. Y. Yin, and S. E. Harris,
	Generation of Narrow-Bandwidth Paired Photons: Use of a Single Driving Laser,
	Phys. Rev. Lett. \textbf{97}, 113602 (2006).
	
	\bibitem{Du2}
	S. Du, P. Kolchin, C. Belthangady, G. Y. Yin, and S. E. Harris,
	Subnatural Linewidth Biphotons with Controllable Temporal Length,
	Phys. Rev. Lett. \textbf{100}, 183603 (2008).
	
	\bibitem{Yan}
	K. Liao, H. Yan, J. He, S. Du, Z.-M. Zhang, and S.-L. Zhu,
	Subnatural-Linewidth Polarization-Entangled Photon Pairs with Controllable Temporal Length,
	Phys. Rev. Lett. \textbf{112}, 243602 (2014).
	
	\bibitem{MZhao}
	T.-M. Zhao, Y. S. Ihn, and Y.-H. Kim,
	Direct Generation of Narrow-band Hyperentangled Photons,
	Phys. Rev. Lett. \textbf{122}, 123607 (2019).
	\bibitem{Cui}
	K.-S. Cui, X.-J. Zhang, and J.-H. Wu,
	Enhanced photon-pair generation under coherent control,
	Phys. Rev. A \textbf{109}, 063701 (2024).	
	\bibitem{Zhao}
	H.-M. Zhao, X.-J. Zhang, M. Artoni, G. C. La Rocca, and J.-H. Wu,
	Nonlocal Rydberg enhancement for four-wave-mixing biphoton generation,
	Phys. Rev. A \textbf{109}, 043711 (2024).
	
	\bibitem{Shiu1}
	J.-S. Shiu, Z.-Y. Liu, C.-Y. Cheng, Y.-C. Huang, I. A. Yu, Y.-C. Chen, C.-S. Chuu, C.-M. Li, S.-Y. Wang, and Y.-F. Chen,
	Observation of highly correlated ultrabright biphotons through increased atomic ensemble density in spontaneous four-wave mixing,
	Phys. Rev. Res. \textbf{6}, L032001 (2024).
	
	\bibitem{Shiu2}
	J.-S. Shiu, C.-W. Lin, Y.-C. Huang, M.-J. Lin, I-C. Huang, T.-H. Wu, P.-C. Kuan, and Y.-F. Chen,
	Frequency-tunable biphoton generation via spontaneous four-wave mixing,
	Phys. Rev. A \textbf{110}, 063723 (2024).
	
	\bibitem{Shiu3}
	J.-S. Shiu, C.-W. Lin, and Y.-F. Chen,
	Asymmetric Biphoton Generation under Ground-State Decoherence and Phase Mismatch in a Cold Atomic Ensemble,
	Adv. Quantum Technol. \textbf{8}, 2500052 (2025).
	
	\bibitem{Jen}
	H. H. Jen,
	Spectral analysis for cascade-emission-based quantum communication in atomic ensembles,
	J. Phys. B: At. Mol. Opt. Phys. \textbf{45}, 165504 (2012).
	
	\bibitem{Jen2}
	H. H. Jen and Y.-C. Chen,
	Spectral shaping of cascade emissions from multiplexed cold atomic ensembles,
	Phys. Rev. A \textbf{93}, 013811 (2016).
	
	\bibitem{Mosley}
	P. J. Mosley, J. S. Lundeen, B. J. Smith, P. Wasylczyk, A. B. U’Ren, C. Silberhorn, and A. I. Walmsley,
	Heralded Generation of Ultrafast Single Photons in Pure Quantum States,
	Phys. Rev. Lett. \textbf{100}, 133601 (2008).
	
	\bibitem{Bock}
	M. Bock, A. Lenhard, C. Chunnilall, and C. Becher,
	Highly efficient heralded single-photon source for telecom wavelengths based on a PPLN waveguide,
	Opt. Express \textbf{24}, 23992 (2016).
	
	\bibitem{Jen3}
	H. H. Jen,
	Positive-$P$ phase-space-method simulation of superradiant emission from a cascade atomic ensemble,
	Phys. Rev. A \textbf{85}, 013835 (2012).
	
	\bibitem{Ford}
	G. W. Ford, J. T. Lewis, and R. F. O’Connell,
	Quantum Langevin equation,
	Phys. Rev. A \textbf{37}, 4419 (1988).
	
	\bibitem{ScullyBook}
	M. O. Scully and M. S. Zubairy,
	\textit{Quantum Optics}
	(Cambridge University Press, Cambridge, England, 1997).
	
	\bibitem{Yatsenko}
	L. Yatsenko, M. Cordier, L. Pache, M. Schemmer,
	Photon Transport in a Gas of Two-Level Atoms: Unveiling Quantum Light Creation,
	New J. Phys. \textbf{27}, 104505 (2025).
	
	\bibitem{Cheng}
	C.-Y. Cheng, J.-J. Lee, Z.-Y. Liu,  J.-S. Shiu, and Y.-F. Chen,
	Quantum frequency conversion based on resonant four-wave mixing,
	Phys. Rev. A \textbf{103}, 023711 (2021).
	
	\bibitem{Hsu}
	H. Hsu, C.-Y. Cheng, J.-S. Shiu, L.-C. Chen, and Y.-F. Chen,
	Quantum fidelity of electromagnetically induced transparency: the full quantum theory,
	Opt. Express \textbf{30}, 2097 (2022).
	
	\bibitem{Liu1}
	Z.-Y. Liu, J.-S. Shiu, C.-Y. Cheng, and Y.-F. Chen,
	Controlling frequency-domain Hong-Ou-Mandel interference via electromagnetically induced transparency,
	Phys. Rev. A \textbf{108}, 013702 (2023).
	
	\bibitem{Chen}
	P.-H. Tseng and Y.-F. Chen,
	Entanglement Preservation and Clauser-Horne Nonlocality in Electromagnetically Induced Transparency Quantum Memories,
	arXiv:2507.15453 (2025).
     
    \bibitem{LCChen}
     L.-C. Chen, M.-Y. Lin, J.-S. Shiu, X.-Q. Zhong, P.-H. Tseng, and Y.-F. Chen, 
     High-efficiency telecom frequency conversion via a diamond-type atomic ensemble,
     Phys. Rev. A \textbf{112}, 013709 (2025).     

    \bibitem{Brink}
    D. M. Brink and G. R. Satchler,
    \textit{Angular Momentum},
    2nd ed. (Oxford University Press, Oxford, 1968). 

    \bibitem{Balcar}
    E. Balcar and S. W. Lovesey,
    \textit{Introduction to the Graphical Theory of Angular Momentum: Case Studies},
    Springer Tracts in Modern Physics Vol. 234 (Springer, Berlin, 2009).
    
    \bibitem{Micalizio}
    S. Micalizio, A. Godone, F. Levi, and C. Calosso,
    Multistep preparation into a single Zeeman sublevel in a $^{87}$Rb vapor cell: Theory and experiment,
    Phys. Rev. A \textbf{80}, 023419 (2009).
	
    \bibitem{Kubo}
	R. Kubo,
	The fluctuation-dissipation theorem,
	Rep. Prog. Phys. \textbf{29}, 255 (1966).
	
	\bibitem{Garrison}
	J. C. Garrison and R. Y. Chiao,
	\textit{Quantum Optics}
	(Oxford University Press, Oxford, 2008).
	
	\bibitem{Louisell}
	W. H. Louisell,
	\textit{Quantum Statistical Properties of Radiation}
	(John Wiley \& Sons, New York, 1973).
	
    \bibitem{Burnham}
	D. C. Burnham and R. Y. Chiao,
	Coherent Resonance Fluorescence Excited by Short Light Pulses,
	Phys. Rev. \textbf{188}, 667 (1969).
	
	\bibitem{Kaluzny}
	Y. Kaluzny, P. Goy, M. Gross, J. M. Raimond, and S. Haroche,
	Observation of Self-Induced Rabi Oscillations in Two-Level Atoms Excited Inside a Resonant Cavity: The Ringing Regime of Superradiance,
	Phys. Rev. Lett. \textbf{51}, 1175  (1983).
	
	\bibitem{Mandel}
	L. Mandel and E. Wolf,
	\textit{Optical Coherence and Quantum Optics}
	(Cambridge University Press, Cambridge, England, 1995).
	
	\bibitem{Ma}
	C. Ma, X. Wang, V. Anant, A. D. Beyer, M. D. Shaw, and S. Mookherjea,  
	Silicon photonic entangled photon-pair and heralded single photon generation with CAR $> 12,000$ and $g^{(2)}(0) < 0.006$,
	Opt. Express \textbf{25}, 32995 (2017).
	
    \bibitem{Blatt}
	F. Blatt, T. Halfmann, and T. Peters,
	One-dimensional ultracold medium of extreme optical depth,
	Opt. Lett. \textbf{39}, 446 (2014).
	
    \bibitem{Hsiao}
	Y.-F. Hsiao, H.-S. Chen, P.-J. Tsai, and Y.-C. Chen,
	Cold atomic media with ultrahigh optical depths,
	Phys. Rev. A \textbf{90}, 055401 (2014).
	
	\bibitem{Abramowitz}
	M. Abramowitz and I. A. Stegun,
	\textit{Handbook of Mathematical Functions with Formulas, Graphs, and Mathematical Tables}
	(U.S. Government Printing Office, Washington, DC, 1968).
	
	\bibitem{Olver}
	F. W. J. Olver \textit{et al.},
	\textit{NIST Handbook of Mathematical Functions}
	(Cambridge University Press, Cambridge, England, 2010).
	
    \bibitem{Friedberg}
	R. Friedberg and S. R. Hartmann,
	Superradiant lifetime: Its definitions and relation to absorption length,
	Phys. Rev. A \textbf{13}, 495 (1976).
	
    \bibitem{Milonni}
	 P. W. Milonni,
	\textit{An Introduction to Quantum Optics and Quantum Fluctuations}
	(Oxford University Press, Oxford, 2019).
	
    \bibitem{Lin}
	C.-W. Lin, Y.-T. Ma, J.-S. Shiu, and Y.-F. Chen,
	Polarization Entanglement in Atomic Biphotons via OAM-to-Spin Mapping,
	arXiv:2512.11625 (2025).
	
\end{thebibliography}
\end{document}